\PassOptionsToPackage{svgnames}{xcolor}
\documentclass[twocolumn]{aastex62}
\usepackage{graphicx}
\usepackage[flushleft]{threeparttable}
\usepackage{blindtext}
\usepackage{amsmath}
\usepackage{mathtools}
\usepackage{multirow}
\usepackage{comment}
\usepackage{booktabs}

\makeatletter
\DeclareRobustCommand{\HI}{%
  \mbox{H\check@mathfonts\fontsize\sf@size\z@\selectfont I}%
}
\makeatother

\makeatletter
\DeclareRobustCommand{\HII}{%
  \mbox{H\check@mathfonts\fontsize\sf@size\z@\selectfont II}%
}
\makeatother

\newcommand{\beq}{\begin{equation}}
\newcommand{\eeq}{\end{equation}}
\newcommand{\bea}{\begin{eqnarray}}
\newcommand{\eea}{\end{eqnarray}}

\begin{document}
\shortauthors{Park et al.}
\def\nar{New Astron.}
\def\na{New Astron.}
\title{\large \textbf{Constraining Reionization with Ly$\alpha$ Damping-Wing Absorption in Galaxy Spectra: A Machine Learning Model Based on Reionization Simulations}}

\correspondingauthor{Hyunbae Park}
\email{hyunbae.park@gmail.com}

\author[0000-0002-7464-7857]{Hyunbae Park}
\affil{Center for Computational Sciences, University of Tsukuba, 1-1-1 Tennodai, Tsukuba, Ibaraki 305-8577, Japan}
\author[0000-0003-1187-4240]{Intae Jung}
\affil{Space Telescope Science Institute, 3700 San Martin Drive, Baltimore, MD 21218, USA}
\author[0000-0002-1319-3433]{Hidenobu Yajima}
\affil{Center for Computational Sciences, University of Tsukuba, 1-1-1 Tennodai, Tsukuba, Ibaraki 305-8577, Japan}
\author{Jenny G. Sorce}
\affil{Univ. Lille, CNRS, Centrale Lille, UMR 9189 CRIStAL, F-59000 Lille, France}
\affil{Universit\'e Paris-Saclay, CNRS, Institut d'Astrophysique Spatiale, 91405, Orsay, France}
\affil{Leibniz-Institut f\"{u}r Astrophysik, An der Sternwarte 16, 14482 Potsdam, Germany}
\author[0000-0002-0410-3045]{Paul R. Shapiro}
\affil{Department of Astronomy, University of Texas, Austin, TX 78712-1083, USA}
\author[0000-0003-3974-1239]{Kyungjin Ahn}
\affil{Department of Earth Sciences, Chosun University, Gwangju 61452, Republic of Korea}
\author{Pierre Ocvirk}
\affil{Observatoire Astronomique de Strasbourg, Université de Strasbourg, CNRS UMR 7550, 11 rue de l’Université, 67000 Strasbourg, France}
\author{Romain Teyssier}
\affil{Institute for Theoretical Physics, University of Zurich, Winterthurerstrasse 190, CH-8057 Zürich, Switzerland}
\author[0000-0001-5031-7936]{Gustavo Yepes}
\affil{Departamento de Física Teórica M-8, Universidad Autónoma de Madrid, Cantoblanco, 28049, Madrid, Spain}
\affil{Centro de Investigación Avanzada en Física Fundamental (CIAFF), Universidad Autónoma de Madrid, 28049 Madrid, Spain}
\author[0000-0002-5174-1365]{Ilian T. Iliev}
\affil{Astronomy Centre, Department of Physics \& Astronomy, Pevensey III Building, University of Sussex, Falmer, Brighton, BN1 9QH, United Kingdom}
\author{Joseph S. W. Lewis}
\affil{Institut d’Astrophysique de Paris, UMR 7095, CNRS, UPMC Univ. Paris VI, 98 bis boulevard Arago, 75014 Paris, France}

\begin{abstract} 
Recently, NIRSpec PRISM/CLEAR observations by JWST have begun providing rest-frame UV continuum measurements of galaxies at $z\gtrsim7$, revealing signatures of Ly$\alpha$ damping-wing (DW) absorption by the intergalactic medium (IGM). We develop a methodology to constrain the global ionization fraction of the IGM $(Q_{\rm HII})$ using low-resolution spectra, employing the random forest classification (RFC) method. We construct mock spectra using the simulated galaxies and the IGM from the Cosmic Dawn II simulation and train RFC models to estimate $Q_{\rm HII}$ at the redshift of the source and to detect the presence of a damped Ly$\alpha$ absorber (DLA). We find that individual galaxy spectra with spectral bins between 1220 and 1270 Å and with signal-to-noise ratios greater than 20 can place tight constraints on $Q_{\rm HII}$, provided the UV continuum is accurately modeled. This method is particularly effective for the early phase of reionization ($Q_{\rm HII}<50\%$), when the IGM opacity is high in the DW. As a demonstration, we apply our model to existing NIRSpec PRISM/CLEAR spectra, placing upper bounds of $Q_{\rm HII}=59.6\%$, $5.6\%$, and $18.5\%$ at $z=7.7,~9.4,$ and $10.6$, respectively, with $68\%$ confidence, though several modeling uncertainties remain to be discussed. These constraints favor late-starting reionization models, where $\gtrsim 80\%$ of the IGM is ionized after $z=8$. We conclude that high SNR observations of carefully selected targets around $z\sim7-9$ can effectively constrain reionization models.
\end{abstract}

\section{Introduction}

One of the pivotal epochs in the quest to understand cosmic evolution is the reionization era, during which early galaxies and black holes ionized the neutral hydrogen in intergalactic space. The ionizing efficiency of these sources is a key unknown in star formation physics, essential for understanding the subsequent evolution of the cosmos. In recent years, advances in both observational and computational capabilities have enabled us to probe this era in diverse ways.

Galactic radiation near Ly$\alpha$ is subject to attenuation by the neutral intergalactic medium (IGM) and is, therefore, considered an effective probe of the HII regions created by early galaxies \citep[see reviews by e.g.,][]{2020ARA&A..58..617O}. The equivalent width (EW) of the Ly$\alpha$ emission line has been widely used in recent studies to estimate the IGM opacity to Ly$\alpha$, and considerable progress has been made in understanding the ionization structure in the IGM up to $z\sim8$ with Ly$\alpha$ EW statistics and Ly$\alpha$ luminosity function \citep{2011ApJ...730L..35V,2012ApJ...744...83O,2012ApJ...752..114S,2012ApJ...747...27T,2013ApJ...775L..29T,2013Natur.502..524F,2014ApJ...795...20S,2015ApJ...810L..12Z,2016ApJ...826..113S,2018ApJ...856....2M,2018A&A...619A.147P,2018ApJ...864..103J,2019ApJ...878...12H,2020MNRAS.495.3602W,2020ApJ...891L..10T,2020ApJ...904..144J,2021ApJ...919..120M,2022MNRAS.511.6042E,2022ApJ...933...87J,2022ApJ...927...36W,2022ApJ...930..104L,2022MNRAS.517.3263B,2023ApJ...949L..40B,2023ApJ...947L..24M,2024ApJ...967...28N}. 

A challenge in utilizing Ly$\alpha$ EW is that both the intrinsic emission strength and IGM opacity can vary significantly from galaxy to galaxy and sight line to sight line due to the complexity of the internal structure of galaxies and density/velocity fluctuations in the IGM \citep[e.g., ][]{2014MNRAS.441.2861H,2021ApJ...922..263P,2022MNRAS.512.3243S}. The sight line variation causes a large scatter in the relationship between Ly$\alpha$ EW and global ionization fraction, requiring a large number of galaxies to accurately constrain the global ionization fraction. Additionally, the emergent Ly$\alpha$ emission strength before the IGM attenuation is highly uncertain and is expected to depend on the viewing angle, further increasing the scatter \citep{2022MNRAS.517....1S,2023MNRAS.523.3749B}. For instance, the JADES and CEERS surveys by the James Webb Space Telescope \citep[JWST;][]{2006SSRv..123..485G} found significantly different fractions of strong Ly$\alpha$ emitters (LAEs; EW>$25$\r{A}) from 25 galaxies each ($20\%$ vs. $52\%$). This discrepancy can be explained by the statistical variation in Ly$\alpha$ opacity calculated from numerically simulated IGM \citep{2024A&A...688A.106N}. It has been suggested that the number of galaxies required to constrain IGM opacity to within 10\% accuracy is larger than 100 \citep{2021ApJ...922..263P}.

In this work, we focus on the Ly$\alpha$ damping-wing (DW) absorption in the continuum of Lyman break galaxies (LBGs) as an alternative probe that is free from some of the limitations of the Ly$\alpha$ EW statistics. When the Universe was highly neutral, the Ly$\alpha$ DW can produce a significant absorption feature up to $\sim100$ \r{A} redward of the resonance \citep{1998ApJ...501...15M}. The DW feature from the high-$z$ Universe was typically searched for in bright active galactic nuclei (AGN) spectra, as their strong flux provides high signal-to-noise ratio (SNR) measurements \citep{2011Natur.474..616M,2018Natur.553..473B,2018ApJ...864..142D,2020ApJ...896...23W,2020ApJ...904...26Y,2024arXiv240612071K,2024arXiv240612070H}. Despite the lower flux, DWs from LBGs is potentially useful for constraining the reionization era as they are much more numerous than quasars, and their narrow emission lines, no broader than a few hundred kilometers per second, allow easier and more accurate fitting of the continuum component to extract the IGM transmissivity \citep{2025A&A...693A..60H}.

A clean measurement of galactic continuum around Ly$\alpha$ is challenging for galaxies at $z>6$ with ground-based telescopes, as their spectra mostly fall outside the visible window of Earth's atmosphere. It is, therefore, advantageous to utilize space instruments such as JWST to acquire damping-wing absorption data from the reionization era. Recent observations by NIRSpec/PRISM on JWST have yielded clean measurements of Ly$\alpha$ DW from high-$z$ galaxies \citep{2023A&A...677A..88B,2023ApJ...951L..22A,2024NatAs...8..657B,2024arXiv240816608W}, and several studies have produced constraints on the reionization history using these data \citep{2023ApJ...949L..34H,2023NatAs...7..622C,2024ApJ...971..124U,2025arXiv250111702M,2025A&A...693A..60H}. A significant portion of these high-$z$ observations exhibits a strong damped Ly$\alpha$ absorber (DLA) component added to the IGM component, presenting a challenge in separating the IGM opacity.

In this study, we develop a methodology to constrain the reionization history using LBG spectra, combining a reionization simulation with a machine learning (ML) algorithm. Cosmological simulations capture the complexity of density/velocity structures in the IGM, providing a realistic database of mock DW spectra. Using reionization simulations is advantageous over purely analytic models, as simulated ionization fields yield the possible range of HII bubble sizes around the source galaxy at a given global IGM ionization fraction \citep{2024MNRAS.531L..34K,2024MNRAS.528L..33C}, providing a prior that can tighten the constraint.

The rest of the paper is as follows. In Section~\ref{sec:LyDW}, we describe how we prepare the mock Ly$\alpha$ damping-wing spectra from a reionization simulation. In Section~\ref{sec:mlModel}, we present our ML model for constraining the volume-averaged ionization fraction of the IGM, $Q_{\rm HII}$, from input galactic spectra. In Section~\ref{sec:application}, we apply our ML model to existing JWST observation data to examine our methodology. In Section~\ref{sec:summary}, we summarize and discuss our results. 

\section{Mock Ly$\alpha$ Damping-wing Spectra} \label{sec:LyDW}

\subsection{Cosmic Dawn II Simulation}

The Ly$\alpha$ opacity of the IGM is influenced by density and velocity fluctuations on subMpc scales, making numerical simulations essential for accurately modeling the complexity of individual sight lines. To this end, we utilize data from the Cosmic Dawn II (CoDa II) simulation to generate mock Ly$\alpha$ damping-wing (DW) spectra from galaxies during the reionization era. CoDa II is a cosmological simulation designed to replicate the formation of galaxies and the evolution of the IGM during cosmic reionization, in a cubic box of $94.4~{\rm Mpc}$ or $64~h^{-1}~{\rm Mpc}$. The simulation computes the density, velocity, ionization fraction, and gas temperature on a $4096^3$ mesh, resolving structures with the cell size of $23.6~{\rm kpc}$. Dark matter is represented using $4096^3$ collisionless particles. Subgrid prescriptions are used to convert gas into stars, represented by collisionless star particles, and to estimate the ionizing radiation produced by these stars. These prescriptions are calibrated to reproduce the observed galaxy luminosity function at $z \sim 6$ \citep{2015ApJ...810...71F,2015ApJ...811..140B,2017ApJ...843..129B,2018MNRAS.479.5184A} and match existing constraints on reionization history, with reionization ending at $z = 6.2$. The simulation fully couples radiative transfer calculations with hydrodynamics, allowing for a realistic modeling of HII regions expanding from star-forming sites and their feedback on the IGM. The cosmology parameters used in the simulation are $\Omega_{m,0}=0.307$, $\Omega_{b,0}=0.048$, and $h=0.678$, which are based on \cite{2014A&A...571A..16P}. Further details of the simulation can be found in \cite{2020MNRAS.496.4087O}.

CoDa II provides a robust dataset for calculating Ly$\alpha$ opacity, capturing hundreds of HII bubbles from star-forming galaxies while resolving the IGM down to pressure-smoothing scales uniformly across the simulation volume. The Ly$\alpha$ opacity is sensitive to compact structures below the atomic-cooling mass ($\lesssim 10^8 M_\odot$) \citep{2021ApJ...922..263P, 2022MNRAS.512.3243S,2024ApJ...969...46P} and requires a spatial resolution element of 20 kpc or smaller for accurate convergence \citep{2015MNRAS.446.3697L}. The CoDa II simulation’s cell size of $23.6~{\rm kpc}$ is sufficiently small to resolve these small-scale structures. However, we note that the simulation box size of $94.4~{\rm Mpc}$ is limited in sampling rare, bright galaxies with $M_{\rm UV} \lesssim -22$ at $z \gtrsim 6$.

For this study, we use snapshots at $z = 6,~6.5,~7,~8,$ and $10$ to compute the mock spectra. The corresponding global ionization fractions in these snapshots are $Q_{\rm HII} = 100,~80,~50,~13.2,$ and $1.6\%$, respectively. Although we use a limited number of snapshots for this proof-of-concept study, it is preferable to utilize more snapshots in practical applications to more accurately capture the redshift evolution of Ly$\alpha$ opacity. In future work, we plan to employ the CoDa III simulation data, which offers higher resolution at $8192^3$ and updated reionization prescriptions that better align with recent constraints,\footnote{The main paper for CoDa III is in preparation. However, some details of CoDa III can be found in \cite{2022MNRAS.516.3389L} and \cite{2024arXiv240905946O}.} and to incorporate more than 20 snapshots to enhance the robustness of our results.

\subsection{Ly$\alpha$ Opacity Calculation}

\begin{figure}
  \begin{center}
    \includegraphics[scale=0.55]{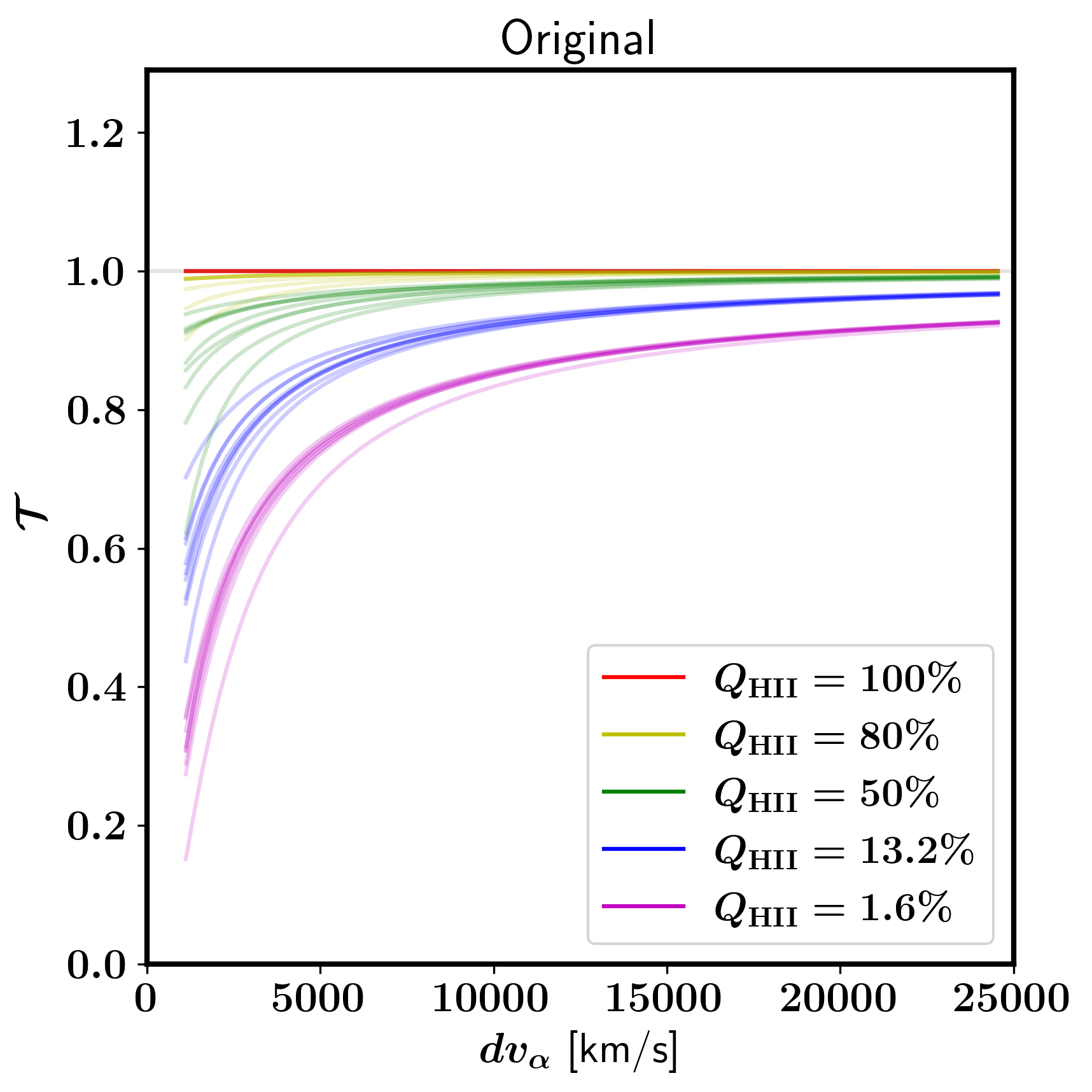}
  \caption{A subset of Ly$\alpha$ transmission curves calculated from the CoDa II simulation data. Red, yellow, green, blue, and magenta curves are from the snapshots with the global ionization fraction of $Q_{\rm HII}=100,~80,~50,~13.2,$ and $1.6\%$, respectively. }
   \label{fig:original}
  \end{center}
\end{figure}

\begin{figure*}
  \begin{center}
    \includegraphics[scale=0.4]{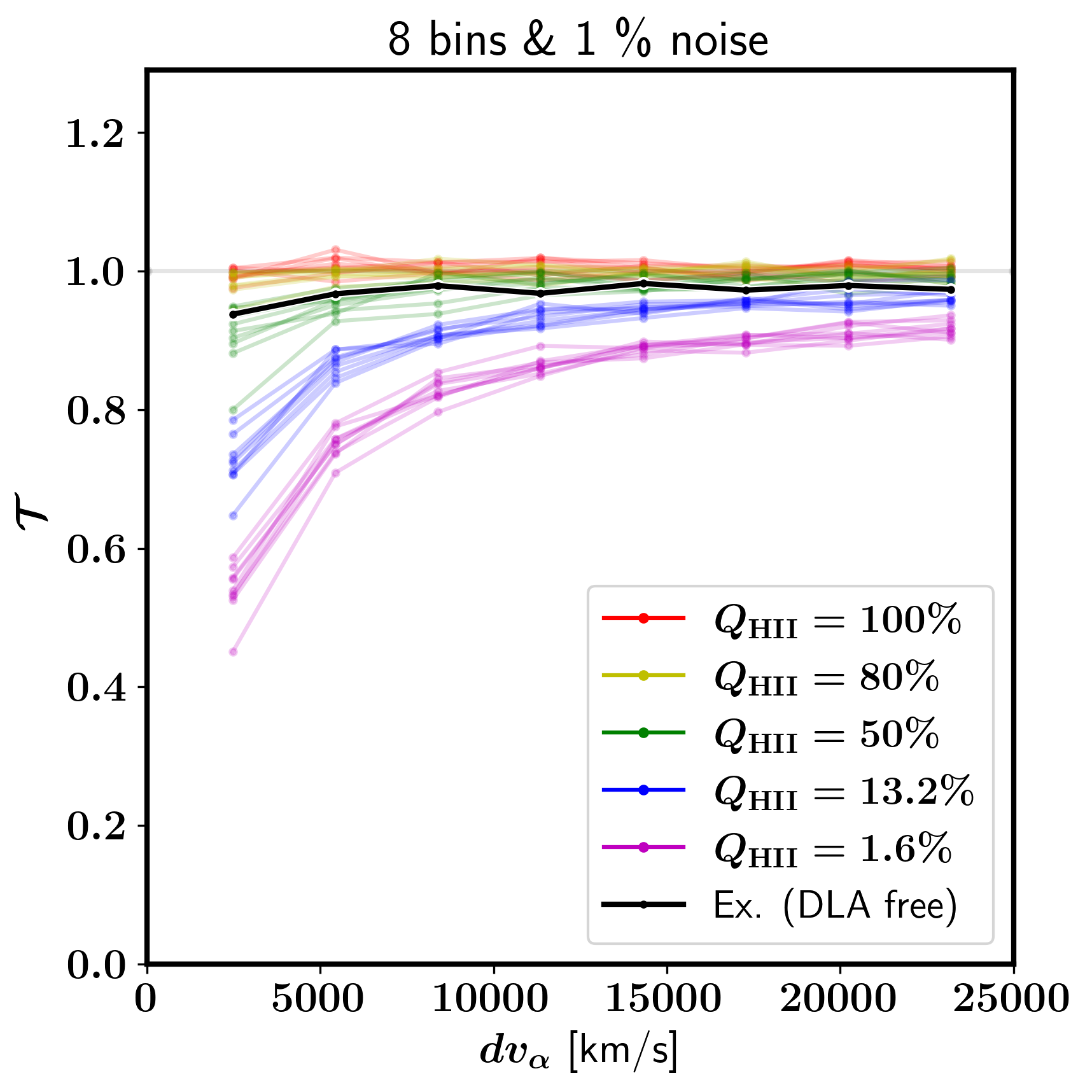}
    \includegraphics[scale=0.4]{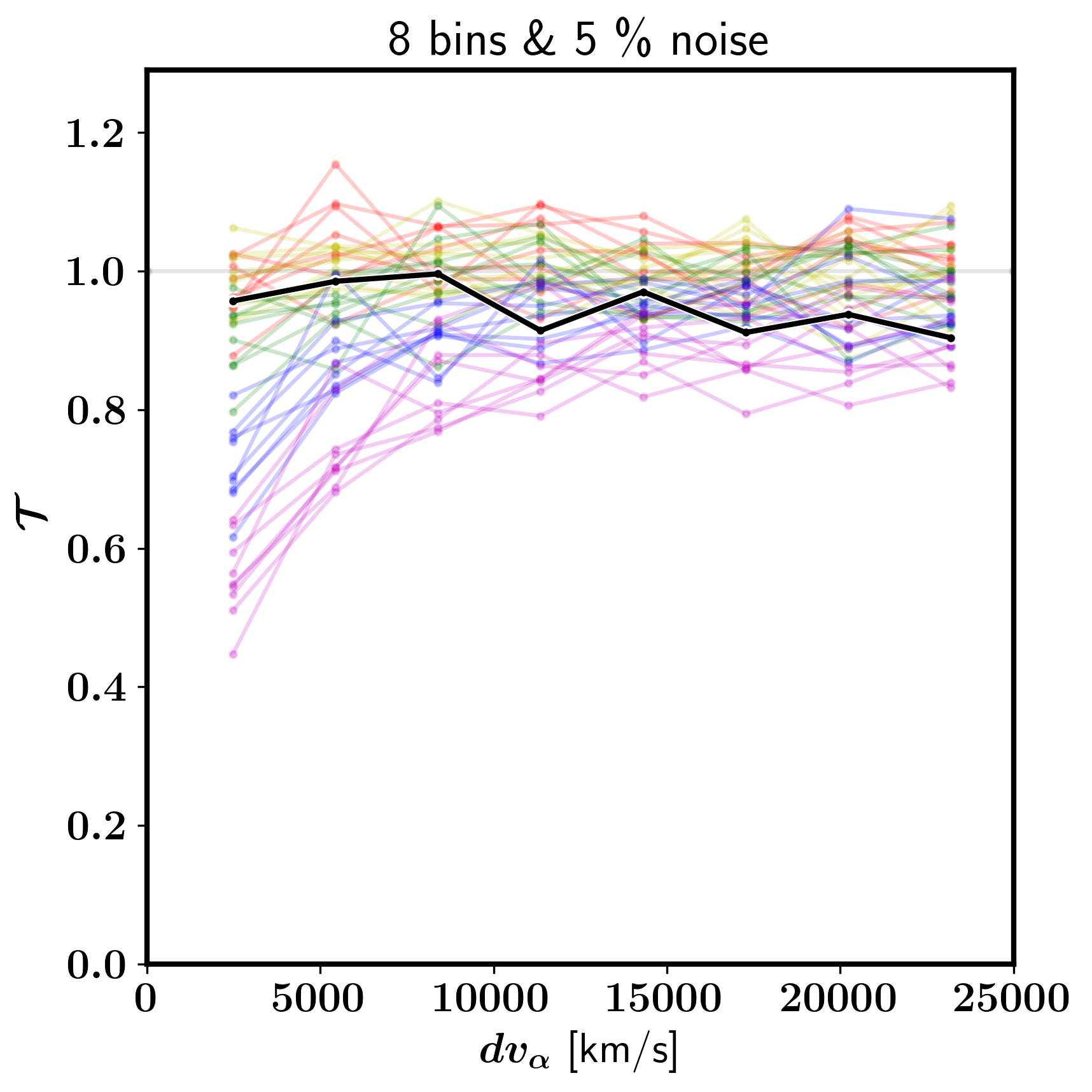}
    \includegraphics[scale=0.4]{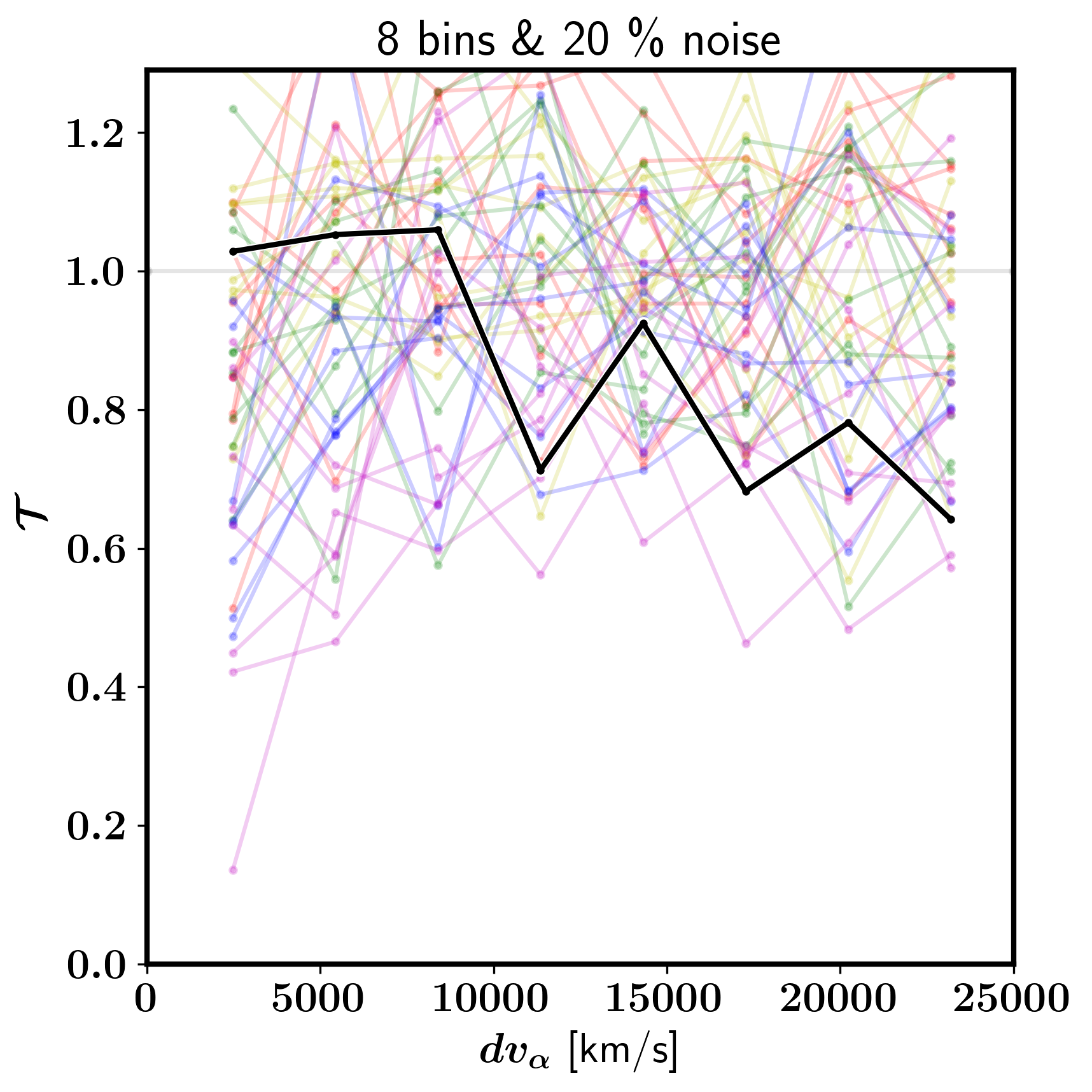}
    \includegraphics[scale=0.4]{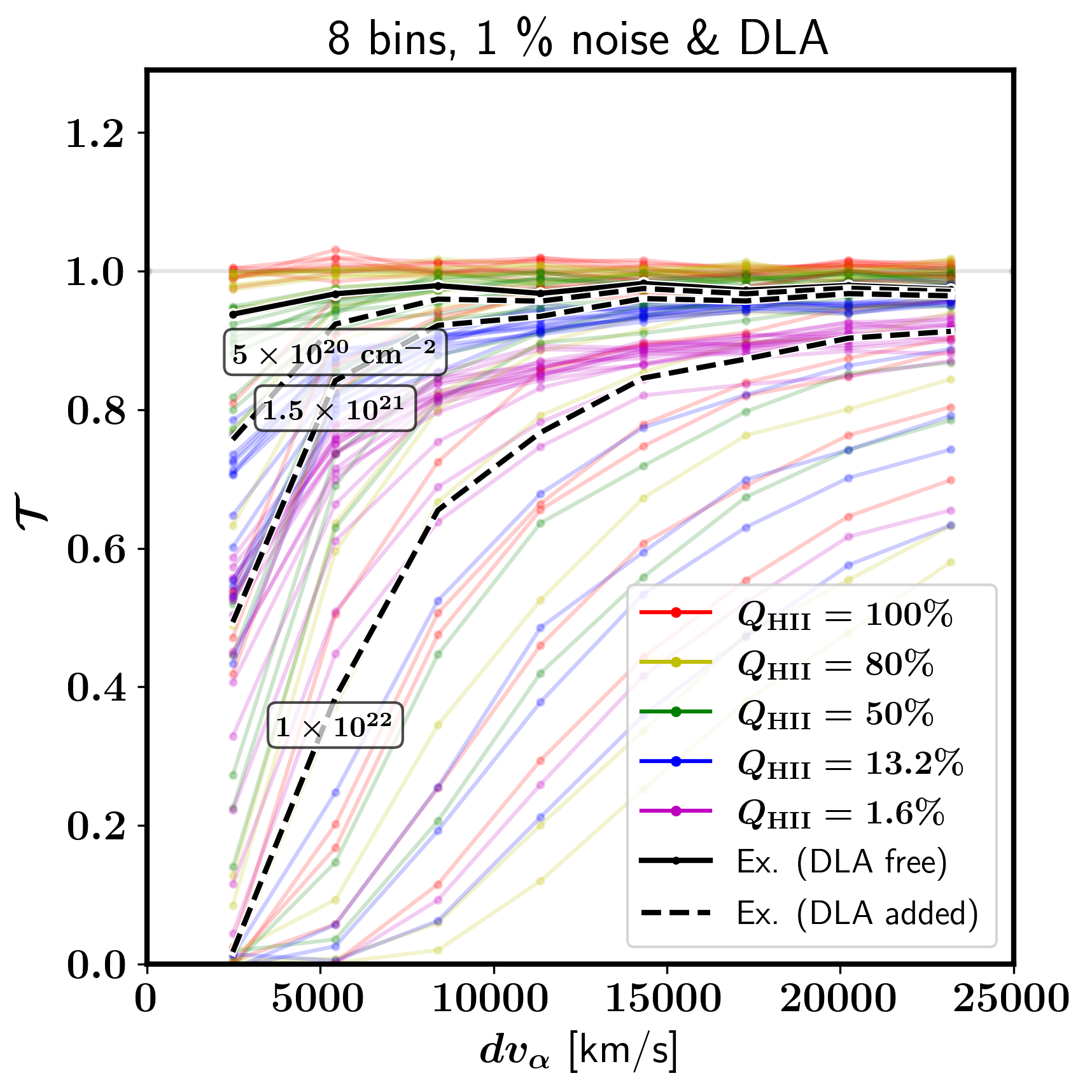}
    \includegraphics[scale=0.4]{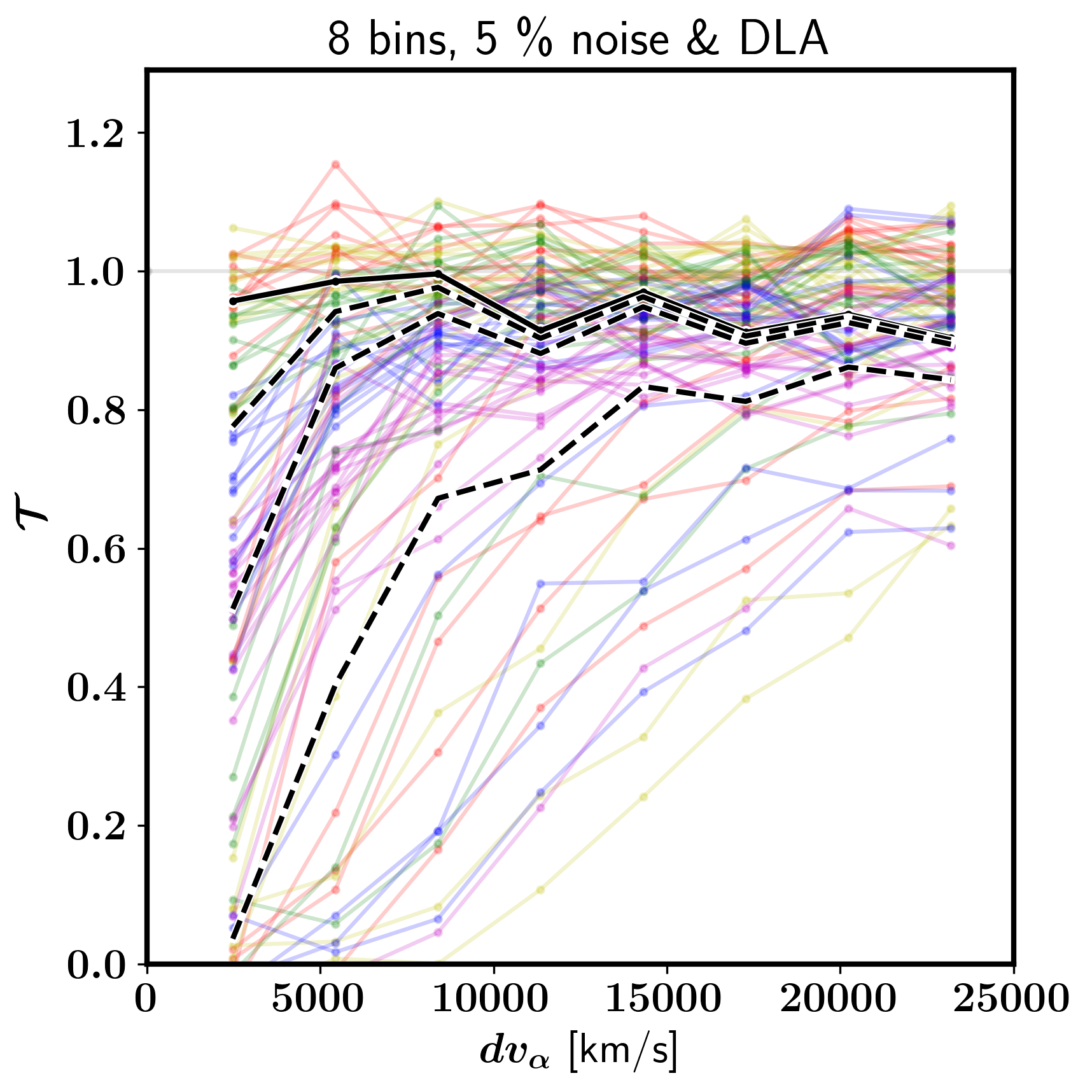}
    \includegraphics[scale=0.4]{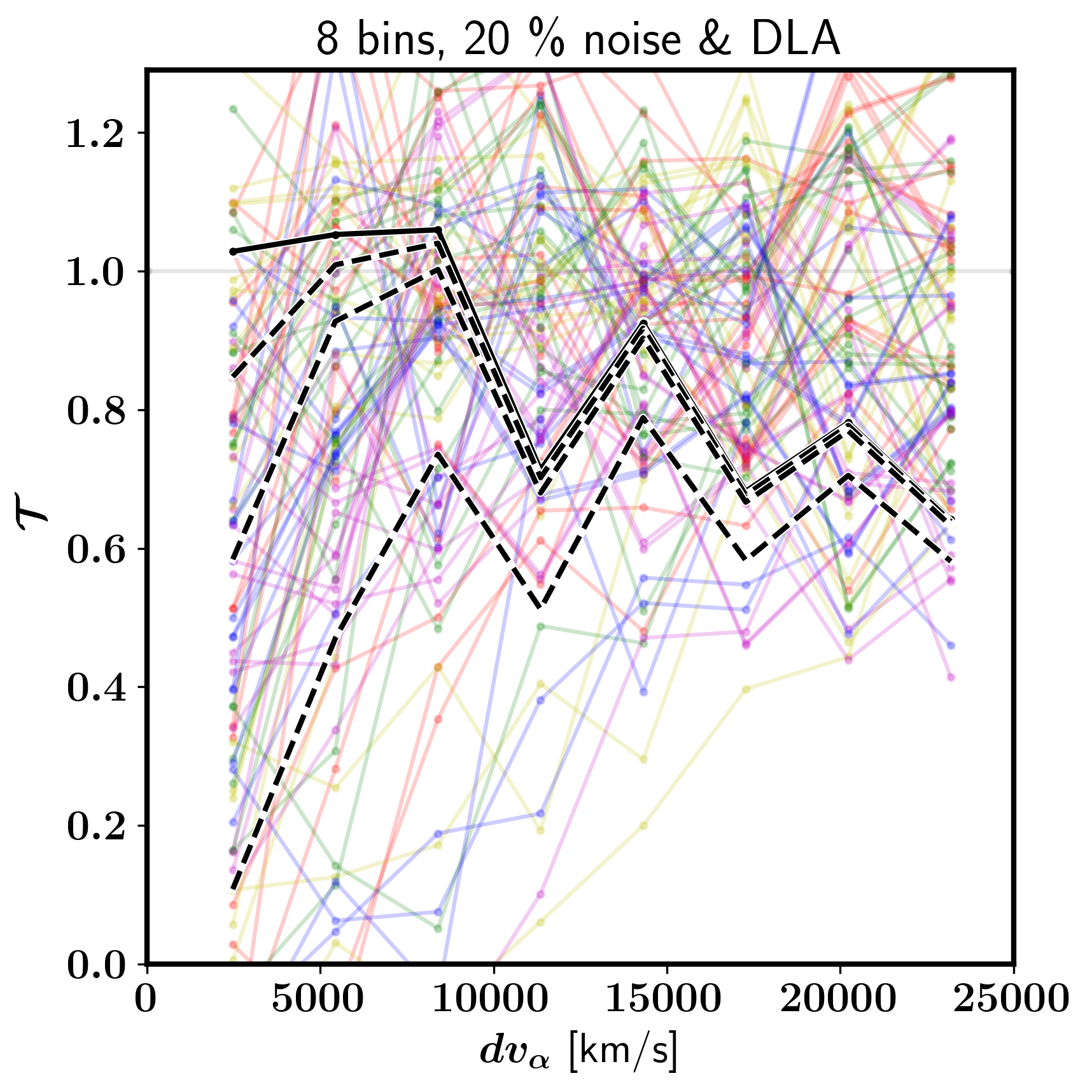}
    \caption{Binned transmission curves (i.e., normalized SEDs) with added random Gaussian photon noise only (upper panels) and those with additional DLA opacity in addition to the photon noise (lower panels). The left, middle, and right panels show samples with photon noise of 1, 5, and 20\% of the continuum flux, respectively. The black solid line is a randomly selected example from the $Q_{\rm HII}=50\%$ samples without DLA opacity, and the three black dashed lines represent cases where a DLA opacity of $N_{\rm HI}=5\times 10^{20},~1.5\times 10^{21},$ and $1\times 10^{22}~{\rm cm}^{-2}$ is added to the example case.}
   \label{fig:noisy}
  \end{center}
\end{figure*}

To calculate the Ly$\alpha$ opacity by the IGM, we follow the general workflow outlined in \cite{2021ApJ...922..263P} to conduct the calculation. The major difference in this work is that the calculation is done for a wider range of wavelength ($1216 - 1317$ \r{A}) to model the DW feature in low-resolution spectra, while \cite{2021ApJ...922..263P} covered only from $1212-1220$ \r{A} to model Ly$\alpha$ emission lines. 

The Ly$\alpha$ opacity is obtained by integrating the product of HI density and the Ly$\alpha$ cross section along the line of sight: 
\bea \label{eq:Tr} 
\tau_\alpha(\nu_e) = \int^{r_{\rm max}}_{r_{\rm min}} n_{\rm HI}(\bold{r})\sigma_\alpha(T(\bold{r}),\nu(\bold{r})) a(z) dr, 
\eea 
where $\bold{r}$ is the proper position vector originating from the source galaxy, $a$ is the cosmic scaling factor, $\nu_e$ is the frequency in the rest frame of the source, $n_{\rm HI}$ is the neutral hydrogen density, and $T$ is the gas temperature. Additionally, \bea \nu(\mathbf{r})=\nu_{\rm e} \left[1-\frac{\hat{\mathbf{r}}\cdot\mathbf{v}_{\rm pe}(\mathbf{r})+a(z)H(z)r}{c}\right] \nonumber \eea accounts for the peculiar velocity of the IGM ($\bold{v}_{\rm pe}$) and the cosmic redshift of photons ($aHr$) to calculate the Ly$\alpha$ cross section in the IGM frame. We integrate from the virial radius of the source galaxy (i.e., $r_{\rm min}=r_{\rm vir}$) to the end of reionization ($r_{\rm max} = r(z=6)$) to calculate the IGM opacity\footnote{We assume matter inside the virial radius belongs to the circumgalactic medium and interstellar medium.}. In the case of the $z=6$ snapshot, where reionization has ended in the simulation (i.e., $Q_{\rm HII} = 100\%$), we simply set $\tau_\alpha = 0$ for all of the samples.

For other snapshots, we first integrate up to $r=24~h^{-1}~{\rm Mpc}$ using the density, velocity, and temperature fields in the snapshot, assuming the fields remain static during the propagation of the radiation. 
This assumption is justified by the fact that the light travel time over that distance (e.g., 13.5 Myr at $z = 7$) is much smaller than the Hubble time (e.g., $\sim 800$ Myr at $z = 7$) or the duration of reionization ($\sim 500$ Myr between $z = 10$ and $6$). We then evaluate the rest of the integral using the globally averaged HII fraction interpolated from averaged values across $\sim 50$ snapshots: 
\bea 
\tau_{\alpha,{\rm LS}} = \int^{r(z = 6)}_{24h^{-1}{\rm Mpc}} (1 - Q_{\rm HII}(z)) \bar{n}_{\rm H}(z) \sigma_\alpha a(z) dr. 
\eea 
This approach ignores fluctuations in $n_{\rm HI}$ at distances beyond $24~h^{-1}~{\rm Mpc}$, which should not significantly affect the DW opacity. This large-scale opacity, $\tau_{\alpha,{\rm LS}}$, contributes up to $\sim 0.1$ in optical depth (see \cite{2021ApJ...922..263P} and \cite{2022MNRAS.512.3243S} for detailed analyses). For this term, we also account for the asymmetry in the DW cross section due to contributions from higher Lyman series damping wings \citep{2013ApJ...772..123L}: \bea \sigma_{\alpha,{\rm Lee}}(d\nu) = \sigma_{\alpha}(d\nu)(1 - 1.792d\nu/\nu_\alpha), \eea where $\nu_\alpha = 2.47 \times 10^{15}~{\rm Hz}$ is the Ly$\alpha$ frequency.

We calculate the Ly$\alpha$ transmission, $\mathcal{T}\equiv\exp{(-\tau_\alpha)}$, for 2000 sight lines from each of the 100 brightest galaxies of each snapshot, across 1000 equally spaced Ly$\alpha$ velocity offsets between $dv_\alpha = 1000$ and $25000~{\rm km}{\rm s}^{-1}$, corresponding to wavelengths between 1220 and 1317 \r{A}. In total, we produce $2000\times100\times5=10^6$ transmission curves from the five snapshots. 

A subset of the transmission curves is shown in Figure~\ref{fig:original}. The transmission shows a stronger dependence on $Q_{\rm HII}$ below 50\%, with well-segregated transmission for $Q_{\rm HII} = 50$, $13.2$, and $1.6\%$ and no boundary cases—instances where the transmission values are near the overlap between different $Q_{\rm HII}$ levels. Thus, these cases are easily classifiable, visually. However, the $Q_{\rm HII}$ dependence becomes much weaker for higher global ionization cases (80 and 100\%) as the transmission converges to 100\%. Although these cases remain segregated from one another with certain boundaries, it is clear that noise from realistic observations (to be introduced below) would make the classification challenging. A similar trend can be seen from Figure 4 of \cite{2024MNRAS.528L..33C}, which shows results based on a different reionization simulation. 


\subsection{Mock Damping-wing Spectra} \label{sec:mock}

We process the Ly$\alpha$ transmission curves to generate realistic mock galaxy spectra near the Ly$\alpha$ resonance. LBGs typically exhibit near power-law rest-frame UV continua, where the normalization and slope can be accurately fitted given sufficient spectral coverage at wavelengths longer than the Ly$\alpha$ resonance (up to $\sim 3000~\text{Å}$). The transmission curve can then be recovered by normalizing the spectrum with the fitted continuum extrapolated to Ly$\alpha$. Thus, the Ly$\alpha$ transmission curves calculated from the simulation effectively serve as normalized spectral energy distributions (SEDs) from observations.

Given that the continuum SEDs of high-redshift galaxies are generally very faint, low-resolution spectrographs, such as NIRSpec/PRISM, are typically used to measure the continuum with sufficient SNR for each spectral bin. To model such low-resolution spectra, we coarsen the transmission curves into eight bins, each with a width of $3000~{\rm km~s}^{-1}$ in the rest frame, corresponding to the spectral bin size of such low-resolution instruments. We randomly select an example from the $Q_{\rm HII}=50\%$ cases to visually demonstrate how an SED is modified through the following processes. While this specific case has a relatively higher flux than the average of the $Q_{\rm HII}=50\%$ cases, it remains within the typical range, falling at the 24th percentile.

Next, we add random Gaussian noise to simulate different SNR levels. The standard deviation of the noise is set to 1\%, 5\%, and 20\% of the continuum flux in each bin, corresponding to SNRs at continuum of 100, 20, and 5, respectively. A large fraction of JWST NIRSpec spectra are detector noise-limited\footnote{\url{https://jwst-docs.stsci.edu/jwst-near-infrared-spectrograph/nirspec-observing-strategies/nirspec-background-recommended-strategies}}. Thus, we assume SNR scales linearly with flux, with constant noise in the wavelength range where the flux is significantly reduced by the DW opacity. While a continuum SNR of 100 would be feasible only for rare bright objects, an SNR of 20 is achievable for moderately UV-bright sources with $M_{\rm UV} \lesssim -22$ at $z \sim 7-8$ with a 1 hr integration of NIRSpec PRISM observations, and for exceptionally bright sources such as GN-$z$11, even at $z > 10$. The noise-added spectra are illustrated in the upper panels of Figure~\ref{fig:noisy}. With 1\% noise (upper left panel), it is relatively easy to visually determine that the example belongs to the $Q_{\rm HII}=50\%$ samples. However, this classification becomes more challenging with 5\% noise (upper middle panel) and nearly impossible with 20\% noise (upper right panel).

In addition to photon noise, we simulate samples with additional DLA absorption. More than half of galactic continuum spectra from $z \gtrsim 6$ appear to have extra absorption by a DLA with a column density of up to $\sim 10^{23}~{\rm cm}^{-2}$, in addition to the absorption by the IGM \citep{2023ApJ...949L..34H,2025A&A...693A..60H,2024ApJ...971..124U}. The DLA component is generally considered to be dense gas clumps in the ISM or CGM associated with the source galaxy. There have also been instances where a DLA appears to be associated with a remote absorber with a redshift offset of $\Delta z \sim 0.5$ from the source \citep{2024A&A...690A..70T}, though such cases appear to be rare. Since the spatial resolution of CoDa II is insufficient for resolving structures in the CGM and ISM, we opt to artificially add DLA opacity to the IGM transmission curves. 

Based on recent measurements \citep[e.g.,][]{2025A&A...693A..60H},  we simulate additional DLA components by duplicating the existing transmission curves and randomly assigning column densities between $N_{\rm HI} = 2\times 10^{20}$ and $1\times 10^{23}{~\rm cm}^{-2}$, drawn uniformly. We assume the DLA is a pointlike absorber associated with the source galaxy. We apply the DLA opacity to the transmission before adding noise to avoid suppressing the noise, which should remain constant, by the DLA opacity.

The ``DLA-added" samples are merged with the original ``DLA-free" samples. A subset of the DLA-added mock spectra is shown in the lower panels of Figure~\ref{fig:noisy}. We also duplicate the example case into three additional cases, where weak, moderate and strong DLA opacity with $N_{\rm HI}=5\times10^{20},~1.5\times10^{21},~{\rm and}~1\times10^{22}~{\rm cm^{-2}}$ are added to sample spectra. The DLA-added cases present a much greater challenge in determining $Q_{\rm HII}$ at all noise levels, as the attenuation by DLA opacity removes the boundary between different $Q_{\rm HII}$ values on the transmission.

\subsection{IGM vs. DLA Opacity in the Data Space} \label{sec:dataSpace}

\begin{figure*}
  \begin{center}
    \includegraphics[scale=0.47]{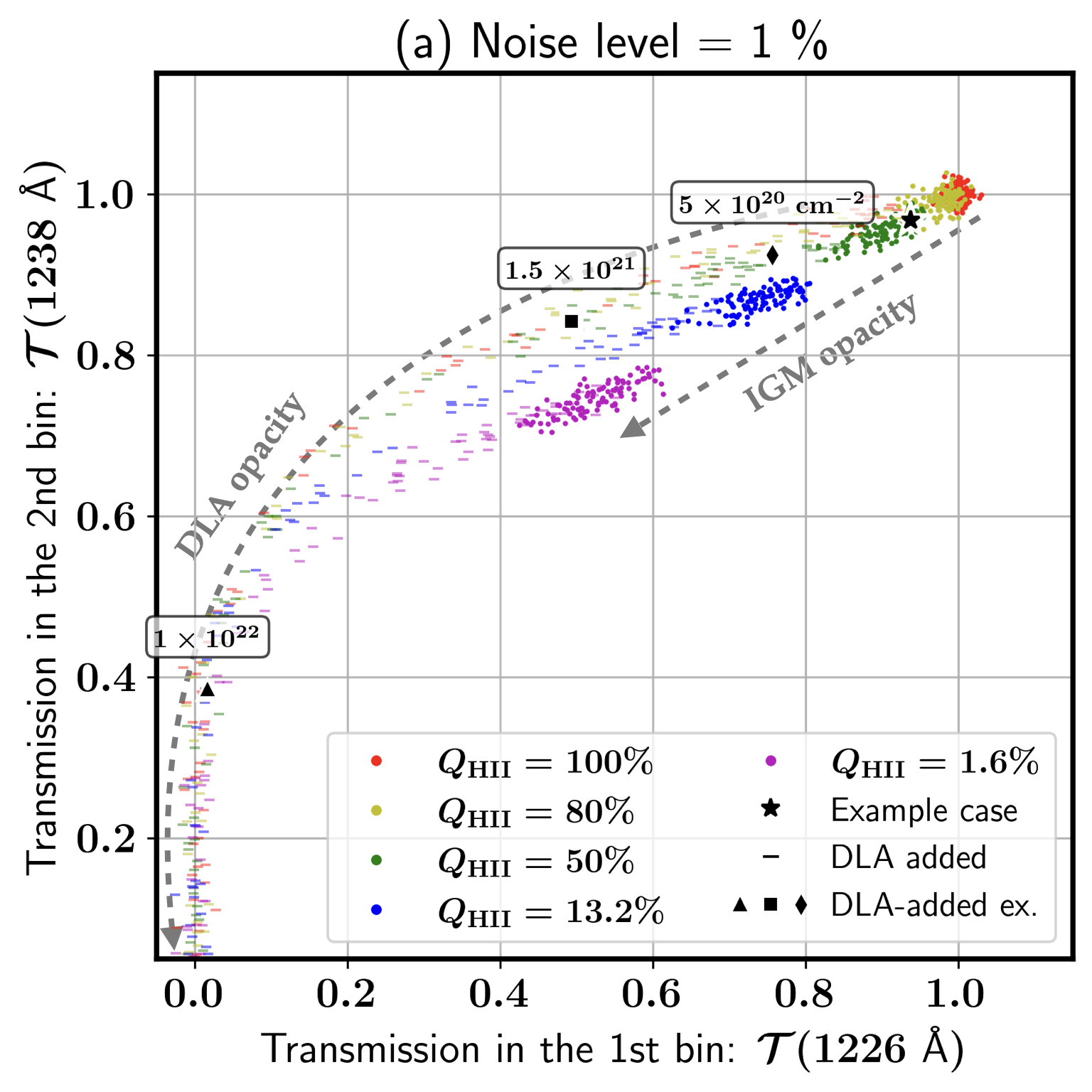}
    \includegraphics[scale=0.61]{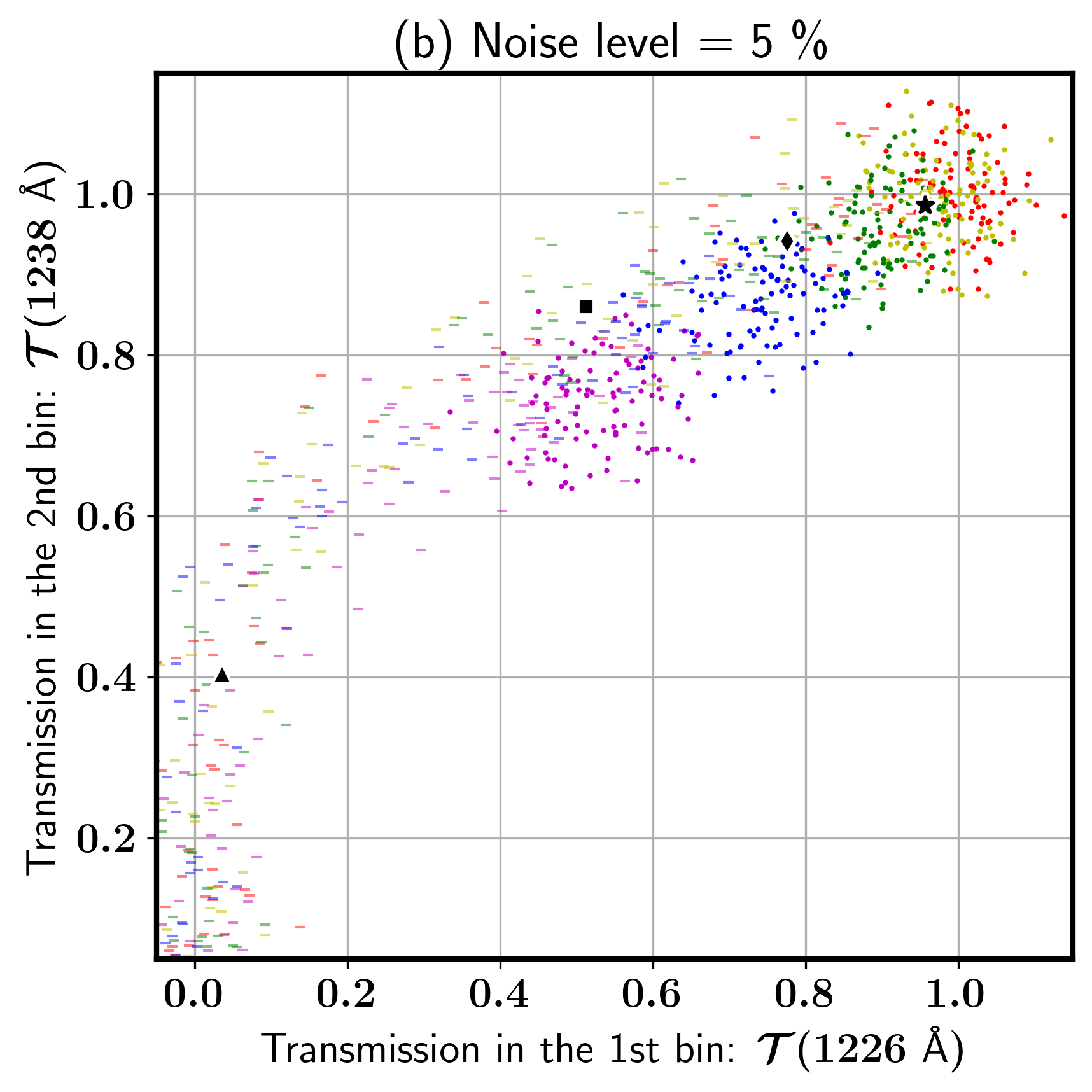}
    \includegraphics[scale=0.61]{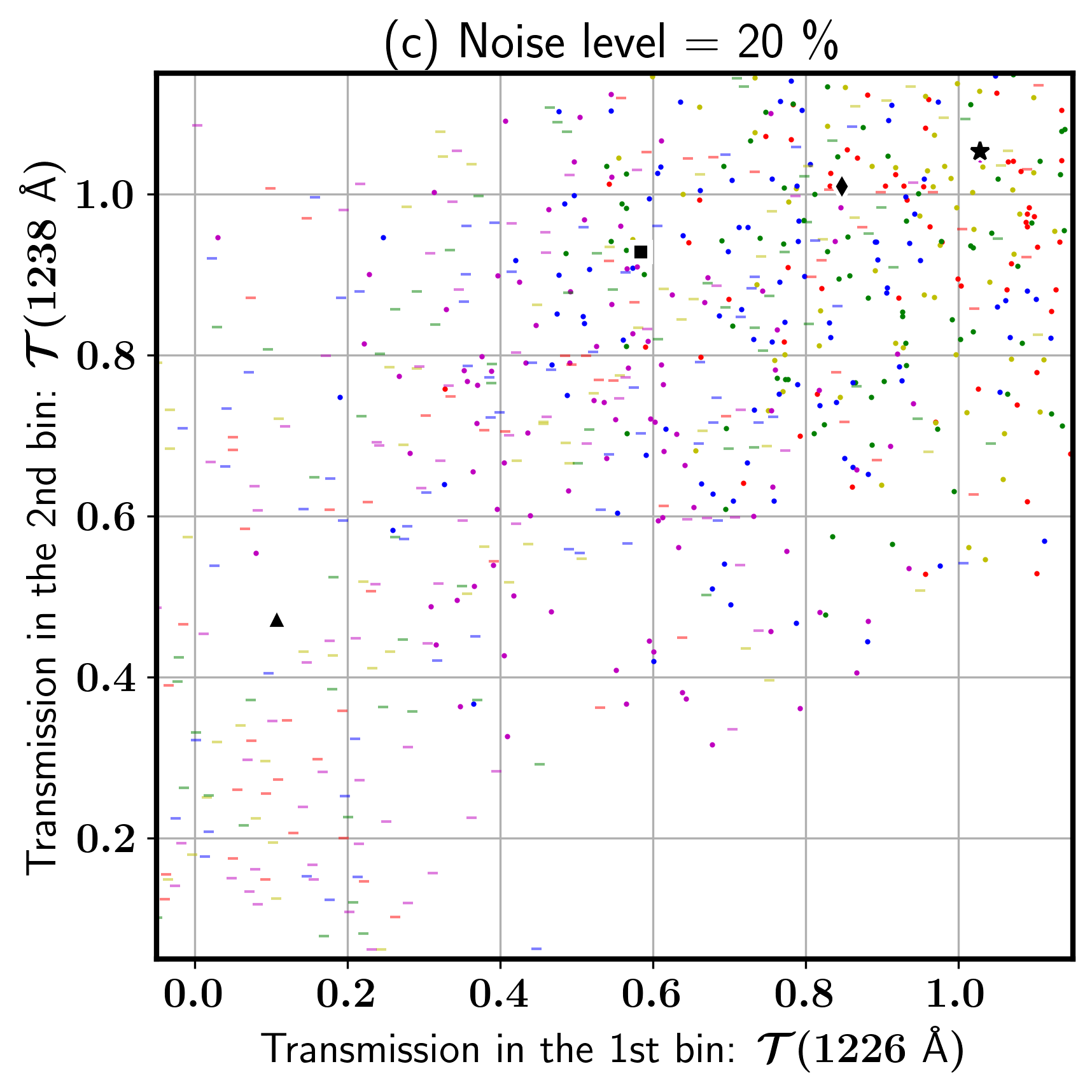}
    \caption{Scatter plot of $\mathcal{T}(1238\text{\AA}$) vs. $\mathcal{T}(1226\text{\AA}$) or transmission in the first vs. second shortest wavelength bins for the mock spectra shown in Fig.~\ref{fig:noisy}. The circular dots are for the DLA-free cases and the minus symbols are for the DLA-added cases. Red, yellow, green, blue, and magenta colors correspond to $Q_{\rm HII}=100\%,~80\%,~50\%,~13.2\%,$ and 1.6\%, respectively. The star symbol denote the example DLA-free case from Fig.~\ref{fig:noisy}, and the diamond, square, and triangle symbols denote the DLA-added versions of the example case with $N_{\rm HI}=5\times10^{20},~1.5\times10^{21},~{\rm and}~1\times10^{22}{\rm cm^{-2}}$, respectively. 
    Panel $(a)$, $(b)$, and $(c)$ are the results for the flux noise levels of 1\%, 5\%, and 20\%, respectively.}
   \label{fig:TrTr}
  \end{center}
\end{figure*}

Then, how do we deal with the contamination by DLA opacity? The difference in wavelength dependence can help us distinguish the DLA opacity from the IGM opacity. The Ly$\alpha$ cross section in DW scales as the inverse square of the wavelength offset. Since the DLA opacity comes from a pointlike absorber, it scales in the same way as the cross section itself does (i.e., $\tau_{\rm DLA}\propto dv_\alpha^{-2}$) while the large-scale IGM opacity comes from the optical depth integrated from the edge of an HII region to a semi-infinite distance (several hundred Mpcs to the end of EoR), thereby approximately scaling as the inverse of the offset (i.e., $\tau_{\rm IGM} \propto dv_\alpha^{-1}$).\footnote{In reality, the IGM opacity would scale differently from $dv_\alpha^{-1}$ because the mean HI density falls as reionization progresses and the Universe expands.}

Motivated by this fact, we make a scatter plot of transmissions at two different Ly$\alpha$ velocity offsets, $dv_\alpha=3500$ and $5000~{\rm km}~{\rm s}^{-1}$, corresponding to 1226 and 1238 Å in wavelength in Figure~\ref{fig:TrTr} to distinguish DLA-added and DLA-free spectra based on the wavelength dependence of the opacity. Those two quantities plotted, $\mathcal{T}(1226~\text{Å})$ and $\mathcal{T}(1238~\text{Å})$, correspond to the two data points with the shortest wavelengths in Figure~\ref{fig:noisy}. The example cases from Figure~\ref{fig:noisy} are shown as black symbols to show their location in the data space. 

The 1\% noise case shown in Figure~\ref{fig:TrTr}a clearly reveals the distinct distribution of DLA-added vs. DLA-free cases in the data space. As marked by two the dashed arrows, the DLA-added cases have a relatively higher flux at 1238 Å when the flux at 1226 Å is similar due to the steeper decline of DLA opacity with increasing wavelength. Also, the DLA-added cases have a much wider range of flux inherited from the large dynamic range of the DLA opacity. 

The black square in Figure~\ref{fig:TrTr}a is a $Q_{\rm HII}=50\%$ case with a DLA opacity of $N_{\rm HI}=1.5\times10^{21}~{\rm cm}^{-2}$, and it exemplifies the difference between DLA-added vs. DLA-free groups. It has $\mathcal{T}(1226~\text{Å})=0.5$, which falls within the range occupied by DLA-free samples with $Q_{\rm HII}=1.6\%$ (magenta dots), $0.45\lesssim \mathcal{T}(1226~\text{Å}) \lesssim 0.65$. However, its flux in the second bin, $\mathcal{T}(1238~\text{Å})=0.86$, is outside the range of that group, $0.75\lesssim \mathcal{T}(1238~\text{Å}) \lesssim 0.82$, thereby allowing us to determine that it does not belong to the DLA-free $Q_{\rm HII}=1.6\%$ cases.

However, the difference in the wavelength dependence between DLA and IGM opacity does not appear large enough to allow for an easy determination of $Q_{\rm HII}$ for DLA-added samples, which have both opacity components. Unlike the DLA-free sample, which forms compact and distinct clusters by $Q_{\rm HII}$ in the data space, DLA-added samples form extended trails with much shorter distances between different $Q_{\rm HII}$ values. Moreover, the trails converge to a single line with increasing DLA opacity, as $\mathcal{T}(1226~\text{Å})$ is attenuated to near zero, exemplified by the black triangle symbol in Figure~\ref{fig:TrTr}a, a case with $Q_{\rm HII}=50\%$ and $N_{\rm HI}\gtrsim 10^{22}~{\rm cm}^{-2}$. Obviously, it would be nearly impossible to determine $Q_{\rm HII}$ for these strong-DLA samples.

Increasing the noise level increases scatter on the data space, mixing samples with different labels. In the 5\% noise case shown in Figure~\ref{fig:TrTr}b, DLA-added samples are mostly mixed with DLA-free cases at $\mathcal{T}(1226~\text{Å})>0.5$. The example DLA-added cases with $N_{\rm HI}= 5\times10^{20}$ (square) and $1.5\times10^{21}~{\rm cm}^{-2}$ (diamond) are now visually indistinguishable from DLA-free cases with $Q_{\rm HII}=13.2$ (blue dots) and 1.6\% (magenta dots), respectively. Also, DLA-free cases with $Q_{\rm HII}=50\%$, $80\%$ and $100\%$ are largely mixed with each other due to the small difference in the IGM opacity. When the noise level is increased to 20\% (Fig.~\ref{fig:TrTr}c), the grouping of samples by $Q_{\rm HII}$ is mostly disrupted due to the increased scatter, severely hindering visual classification. 

We note that other combinations of spectral bins at longer wavelengths also exhibit similar trends but are more vulnerable to noise as the Ly$\alpha$ opacity is weaker for those bins. Thus, the first two spectral bins are most useful for determining $Q_{\rm HII}$ and DLA presence. 

\section{ML Model for Constraining $Q_{\rm HII}$} \label{sec:mlModel}
\subsection{Random Forest Classifier}

We use the Random Forest Classifier (hereafter RFC) to create an ML model that constrains $Q_{\rm HII}$ from input galactic rest-frame UV spectra with DW feature. RFC is a supervised ML algorithm widely used for classification and regression tasks involving labeled data. It is an ensemble learning algorithm, which combines the predictions of multiple base estimators to improve predictive accuracy and reduce the risk of overfitting \citep{RFCpaper}.

\begin{figure}
  \begin{center}
    \includegraphics[scale=0.35]{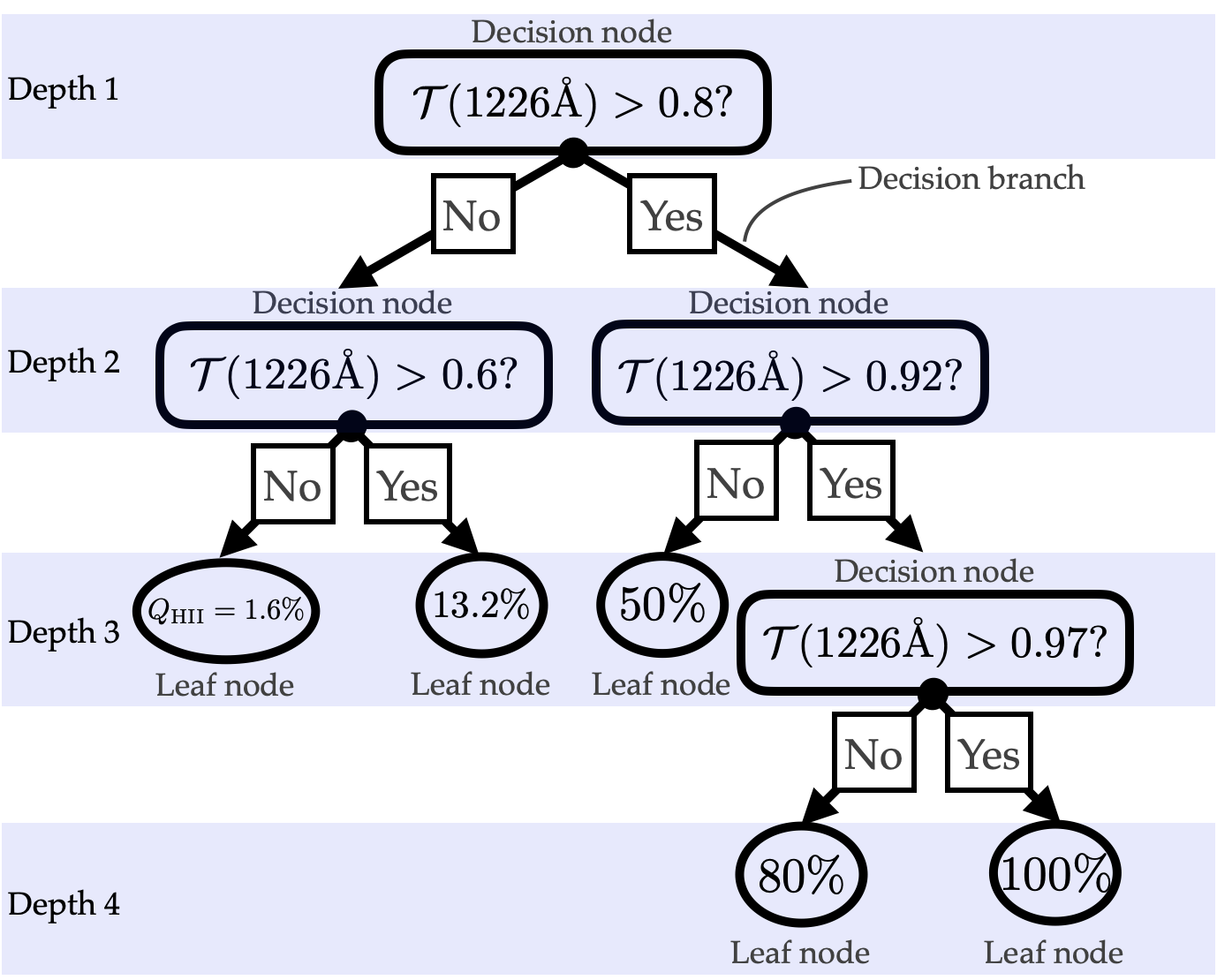}
    \caption{Example decision tree for predicting $Q_{\rm HII}$ based on Ly$\alpha$ transmission at $\lambda=1226$ Å. This is a simplified case using data from only one wavelength, while a practical model would utilize all 8 wavelengths. The tree starts at depth 1 with the first decision node, branching into subsequent nodes based on transmission thresholds. Each decision is made by evaluating whether $\mathcal{T}$(1226 Å) exceeds a specific threshold.}
   \label{fig:decisionTree}
  \end{center}
\end{figure}

Each base estimator in RFC is a decision tree (see Fig.~\ref{fig:decisionTree} for an example). A decision tree is a flowchart-like structure where each internal node represents a test on a feature, each branch represents the outcome of the test, and each leaf node represents a class label (in classification) or a continuous value (in regression). The paths from the root to the leaf represent classification rules.

An individual decision tree is limited in modeling complex boundaries in datasets when used alone. By aggregating the results of many such decision trees, the random forest algorithm significantly enhances generalizability and robustness. This ensemble approach leverages the diversity of the individual trees to make more accurate and stable predictions.

\subsection{Model Training} \label{sec:model}

\begin{figure*}
  \begin{center}
    \includegraphics[scale=0.9]{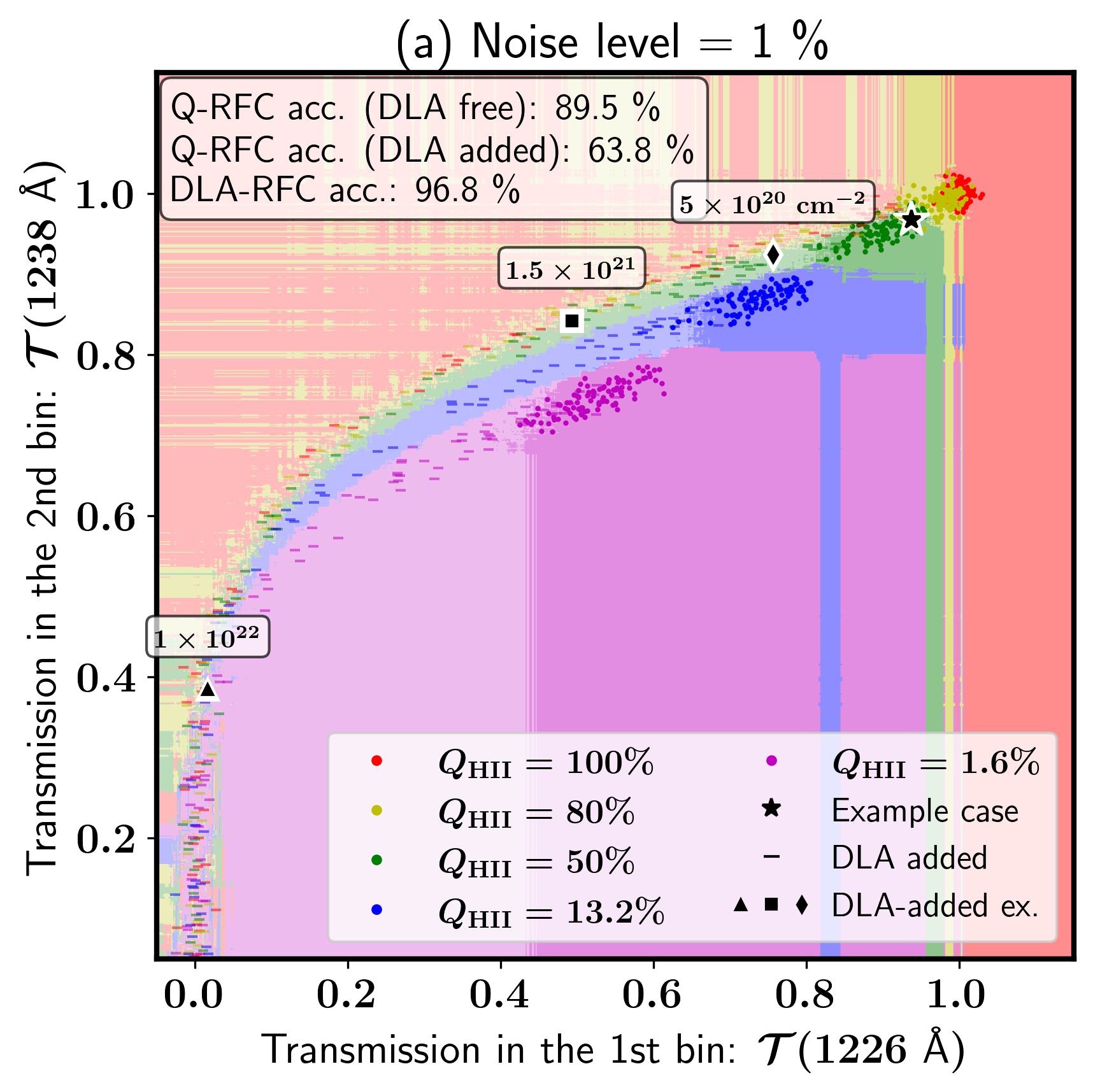}
    \includegraphics[scale=0.61]{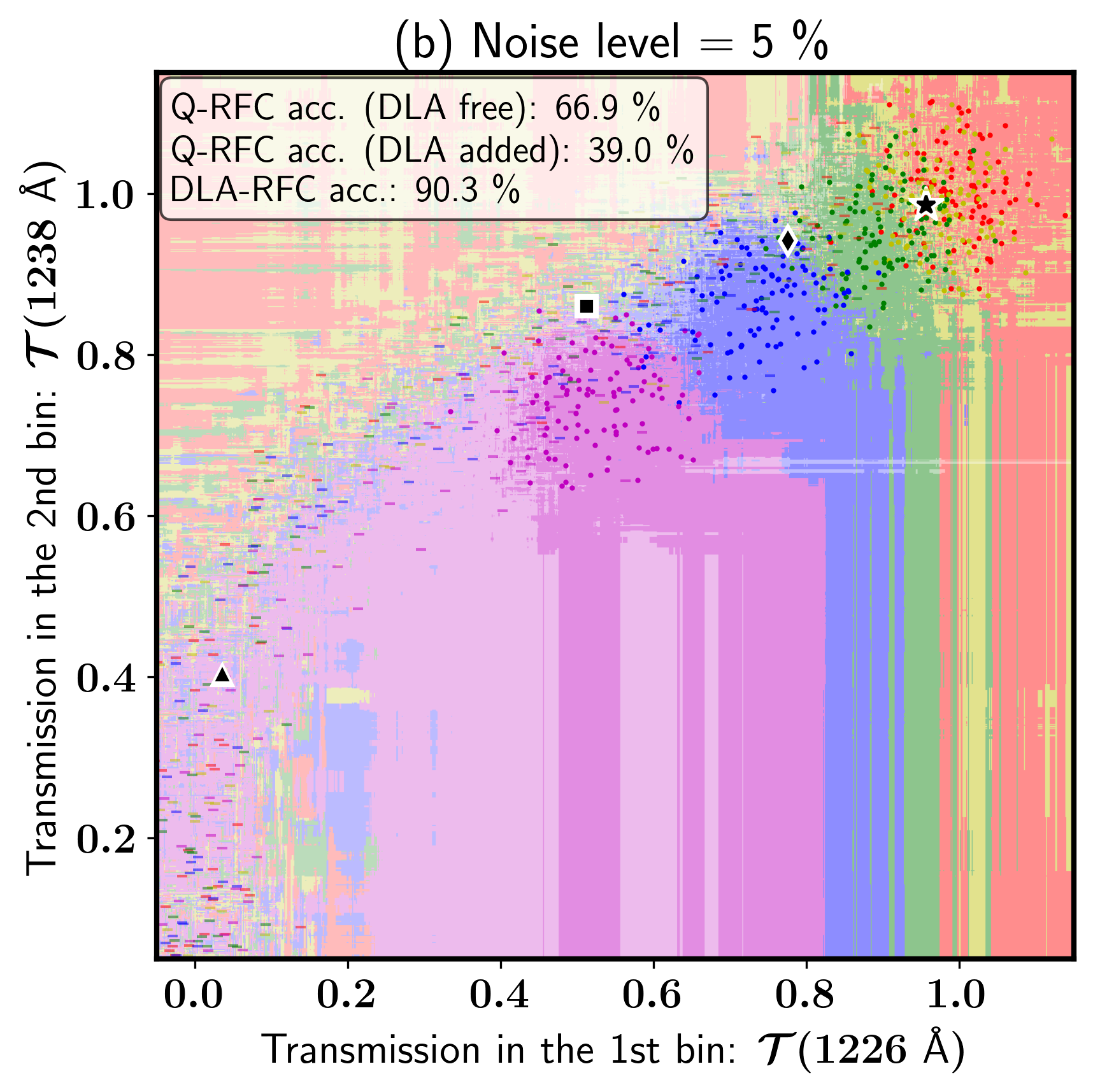}
    \includegraphics[scale=0.61]{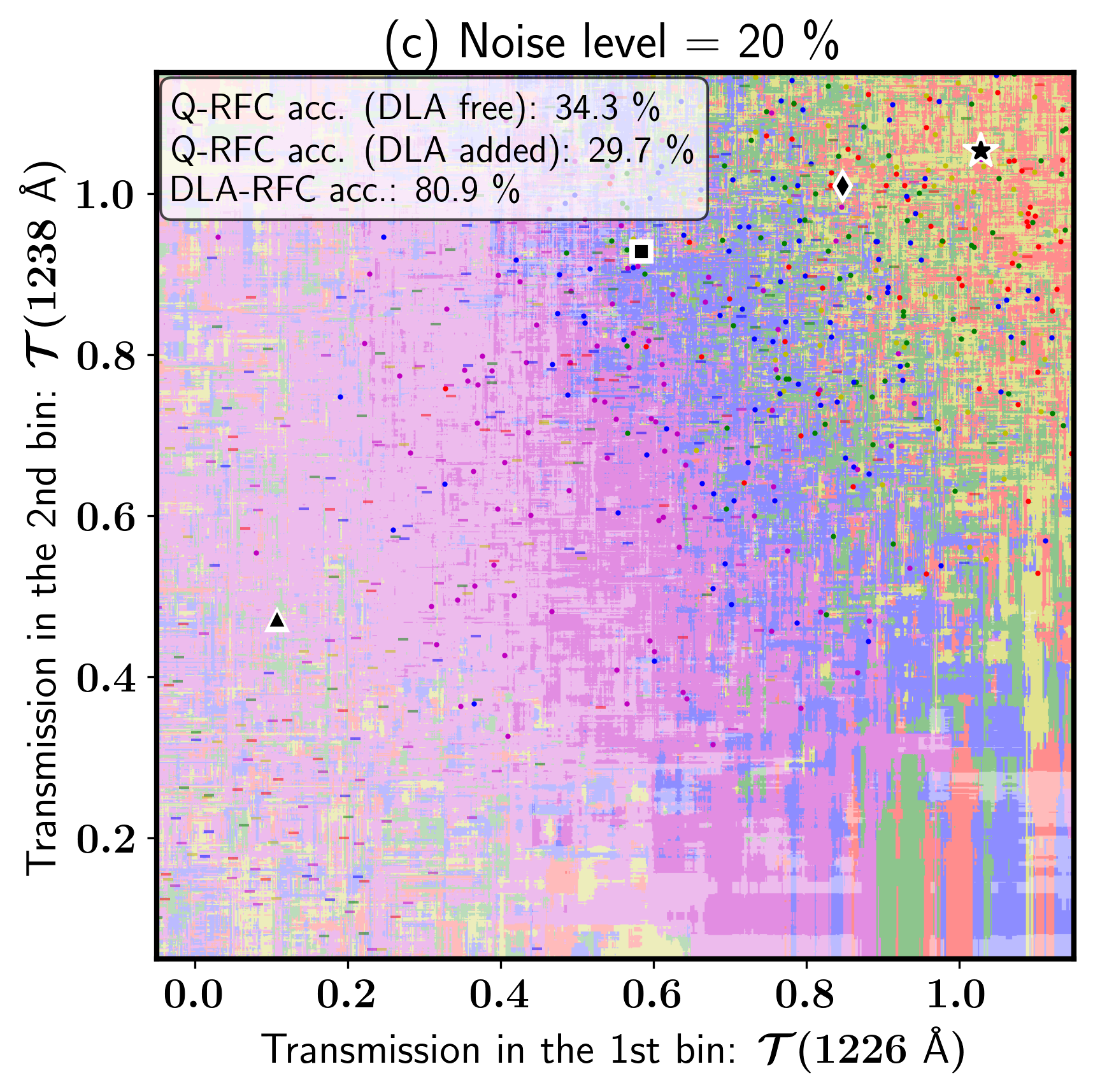}
    \caption{The 2D RFC results for the first and second spectral bins overlaid to the Fig.~\ref{fig:TrTr} as colored background. The hue of the background visualizes the Q-RFC results, where red, yellow, green, blue, and magenta corresponds to the same $Q_{\rm HII}$ values as the data points do: 100\%, 80\%, 50\%, 13.2\%, 1.6\%, respectively. The brightness of the colors visualizes the DLA-RFC results, where the brighter/darker background marks the region where the samples are classified DLA-added/DLA-free.}
   \label{fig:TrTrClf}
  \end{center}
\end{figure*}

In our study, the input data for our ML model comprises $10^6$ mock Ly$\alpha$ DW spectra, each consisting of eight flux data bins, with labels corresponding to the $Q_{\rm HII}$ values. We utilize the `RandomForestClassifier' class from the `sklearn.ensemble' module in the SciPy package to train an RFC model. Although the `RandomForestRegressor' class would typically be more appropriate for predicting continuous $Q_{\rm HII}$ values, we opted for the classifier due to the discrete nature of our labels, limited to five distinct $Q_{\rm HII}$ values in this work, and for easier visualization of the results. `RandomForestRegressor' essentially uses an ensemble of decision trees similar to RFCs for regression tasks. Thus, the underlying principles of our RFC model remain fundamentally aligned with those of the regressor.

As well as the RFC for determining $Q_{\rm HII}$ (Q-RFC, hereafter), we also train another RFC to determine whether the spectrum has a DLA opacity component (DLA-RFC, hereafter). The DLA-RFC is trained on the same training data, but with another set of binary labels containing whether the spectrum was added an additional DLA opacity. As demonstrated in Figure~\ref{fig:TrTr}, DLA-added cases appear more difficult to determine their $Q_{\rm HII}$. Thus, we aim for improving the classification accuracy by sorting out DLA-added cases using this DLA-RFC. 

The Q-RFC can be trained both with and without the DLA-added mock spectra. We use the Q-RFC trained with the DLA-added spectra in Section~\ref{sec:clfResult} to estimate the Q-RFC accuracy for both DLA-added and DLA-free sample. In Section~\ref{sec:application}, we use the Q-RFC trained with the DLA-free spectra only, as we will not apply our model to DLA-added samples based on the analysis from Section~\ref{sec:clfResult} except for one exceptionally high-SNR case analyzed in Section~\ref{sec:JADES-58975}. 

In addition to our main eight-parameter RFC models, we also generate two-parameter models using transmission values for the two shortest wavelengths, $\mathcal{T}(1238\text{\AA}$) and $\mathcal{T}(1226\text{\AA}$), to visualize the decision boundaries in the transmission-transmission plots of Figure~\ref{fig:TrTr}. Although the two-parameter models yield less accurate classification results and are not used in the main analysis, the decision boundaries visualized on a 2D plane for different noise levels provide useful insights into how the model is affected by noise. The decision boundary from this 2D data space offers a useful perspective on the full 8D data space because these two bins provide the most constraining information in practice due to the much weaker Ly$\alpha$ opacity in the other six bins with longer wavelengths. These six bins are more susceptible to noise and provide weaker constraints, thus only offering a `minority report' that slightly improves the model's accuracy.

The number of data points per sample defines the dimensionality of the data space on which the RFC will train. For continuous quantities such as SED, it is often advisable to reduce data dimensionality using methods like principal component analysis (PCA) to avoid the complexities of training ML models in high-dimensional spaces. However, in our case, the mock spectra are already binned into just eight wavelength bins, making further compression unnecessary. We experimented with PCA to reduce the data dimensions even further, but this did not enhance the classification results. Therefore, we use the Ly$\alpha$ DW spectra in their original binned form without any additional processing.

For the hyperparameters of RFC, we set the number of base estimators to 30 (i.e., `n\_estimators=30') and the maximum depth of individual estimator to 20 (i.e., `max\_depth=20'). Setting these parameters too high can result in overfitted models, which create unnecessarily complex decision boundaries and increase computational time without meaningful gains in classification accuracy. This choice of hyperparameters leads to overfitted models in some cases, giving a significantly higher accuracy for training data set than for the test set, but the computation time for querying the RFCs stays below a millisecond and does not consume a significant computing time. Thus, we apply the above-mentioned parameter choice (n\_estimators=30 \& max\_depth=20) across all the cases for consistency.

The accuracy of the model is defined by the fraction of correctly classified cases from the test data set. Owing to the large number of our mock spectra ($10^6$), the accuracy of the models converges after training with only 5\% of the entire mock samples ($5\times10^4$). Thus, we use the remaining 95\% as the test set to calculate the accuracy of the RFCs. 

As will be shown below, a mock spectrum with a weak DLA ($N_{\rm HI} \lesssim 10^{21} {\rm cm}^{-2}$) is much more prone to misidentification to a DLA-free case with a lower $Q_{\rm HII}$ value than its correct $Q_{\rm HII}$ label than stronger DLA cases. This implies that both the DLA-RFC and Q-RFC accuracies is subject to the assumed probability distribution of the column density of added DLA, and the accuracy calculated in this work should not be considered an absolute value.

\begin{figure*}
  \begin{center}
    \includegraphics[scale=0.38]{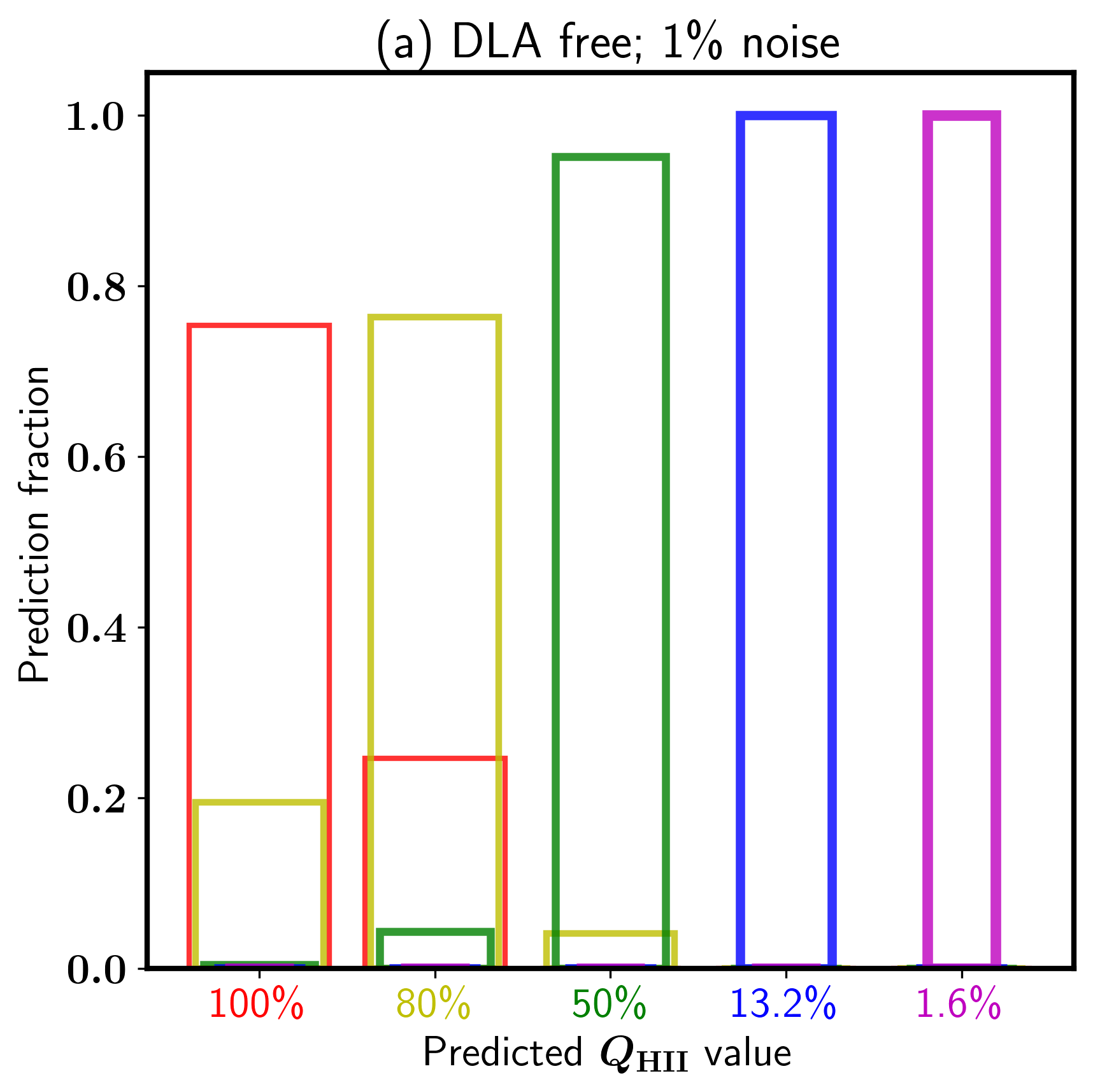}
    \includegraphics[scale=0.38]{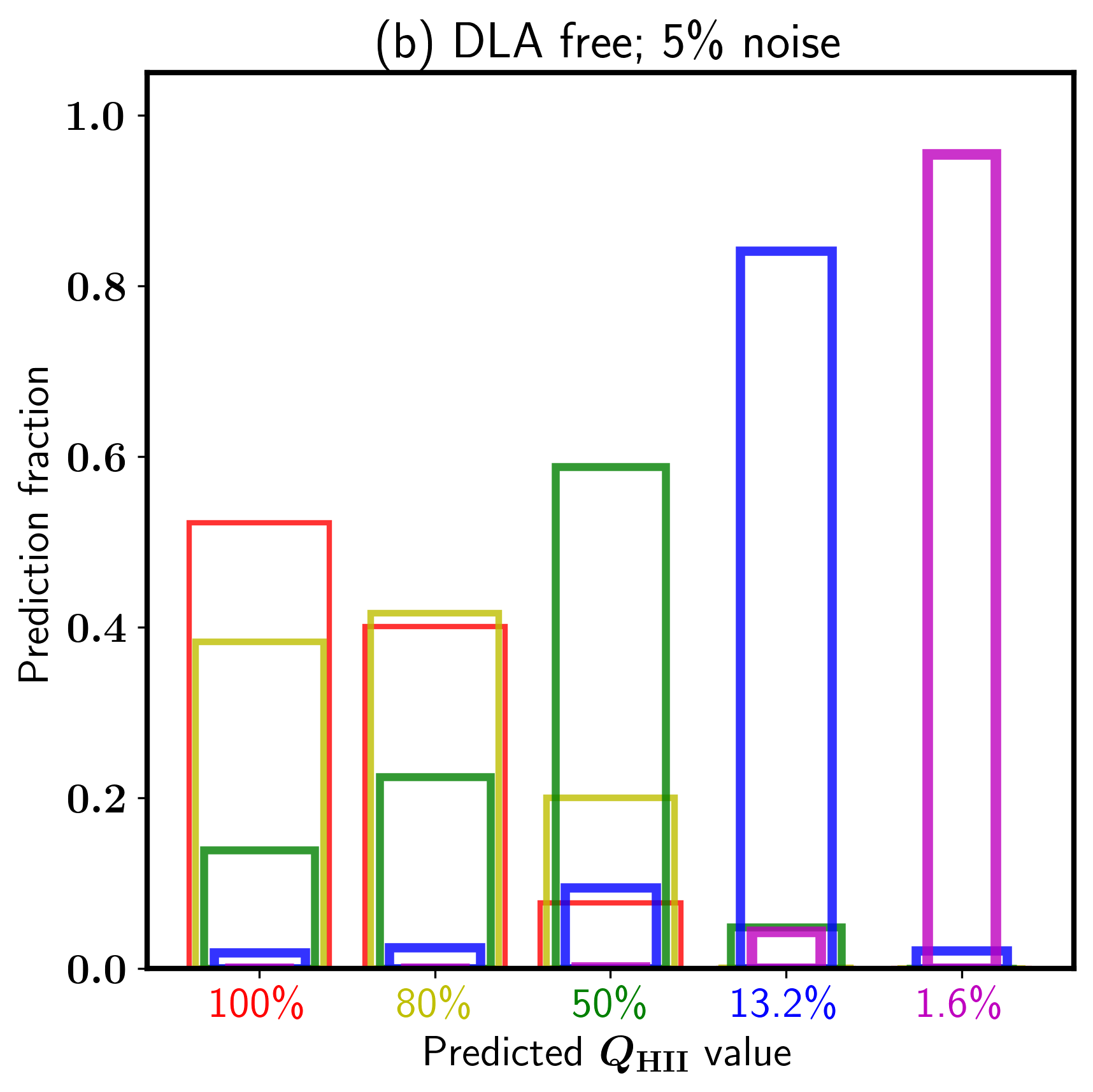}
    \includegraphics[scale=0.38]{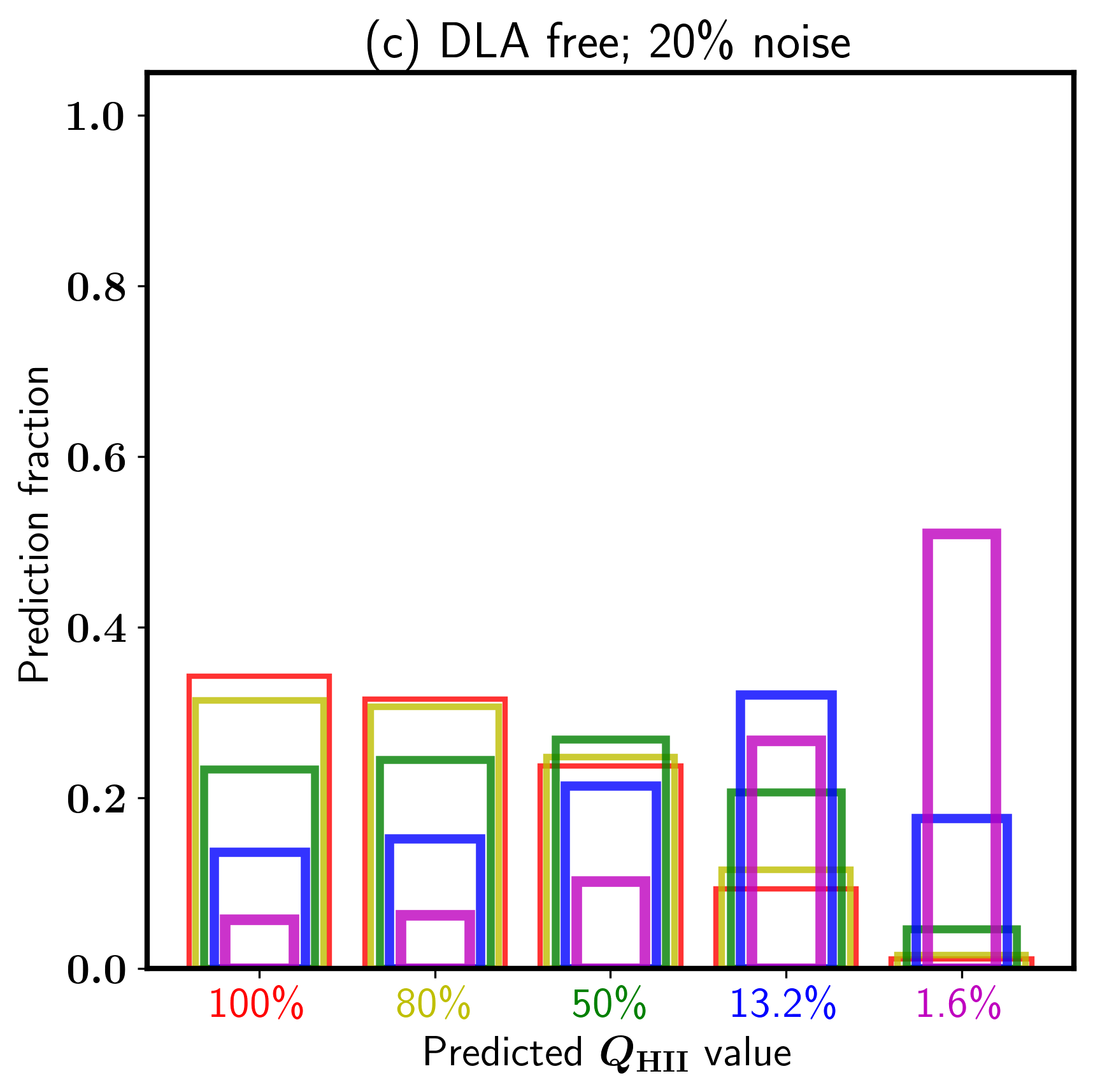}
    \includegraphics[scale=0.38]{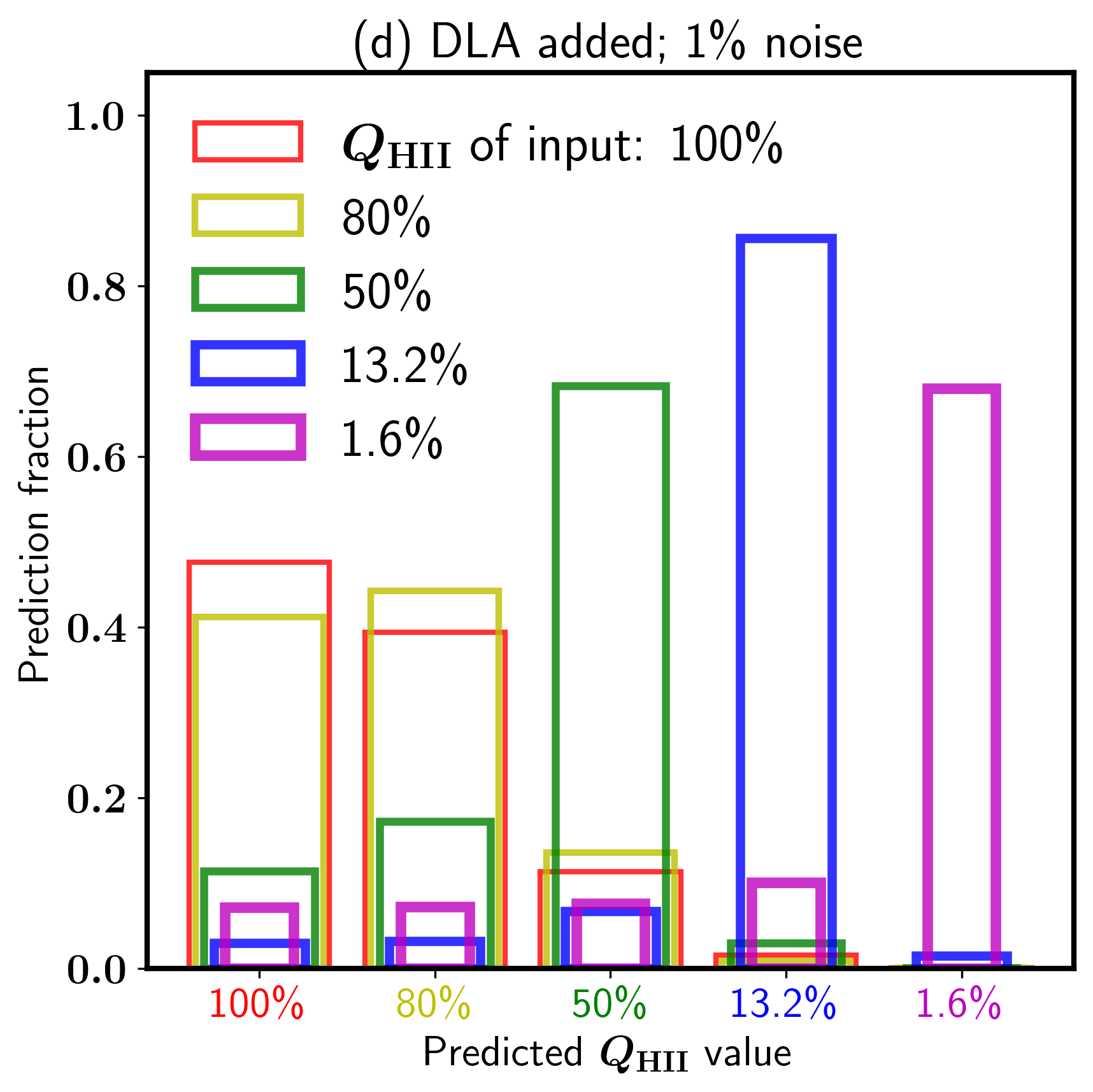}
    \includegraphics[scale=0.38]{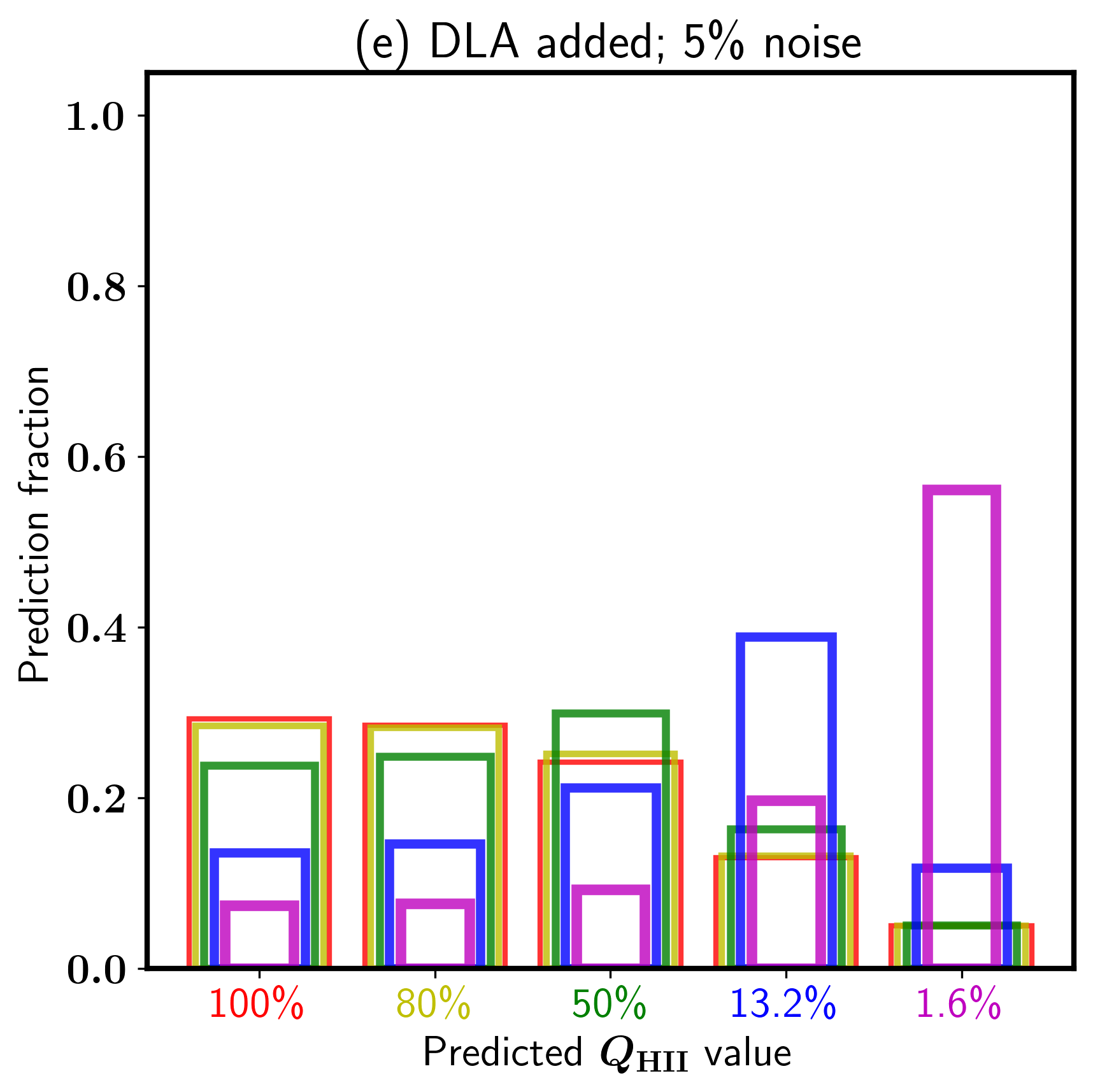}
    \includegraphics[scale=0.38]{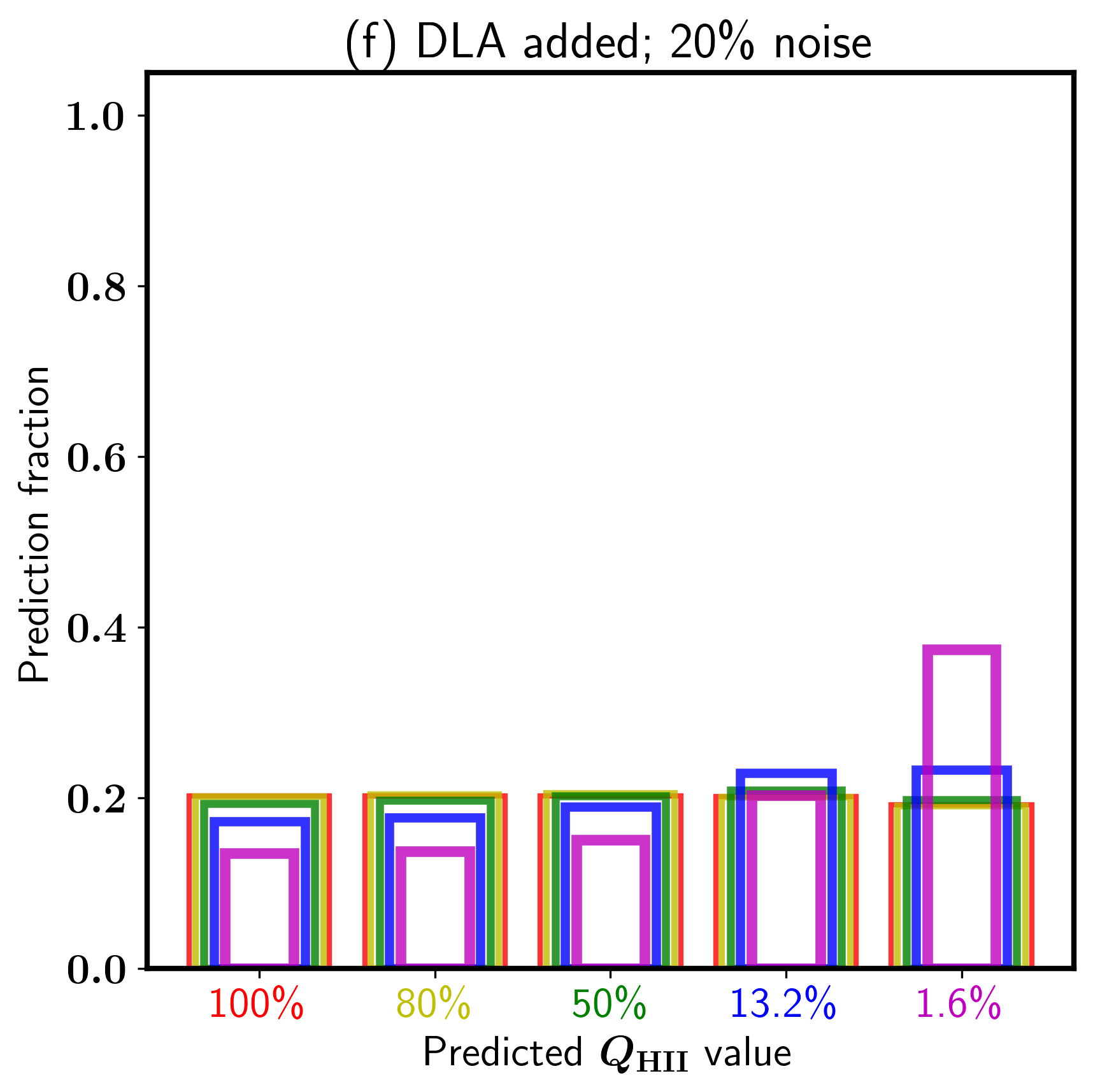}
\caption{Histogram of Q-RFC prediction results by the $Q_{\rm HII}$ values of the input sample, $100\%,~80\%,~50\%,~13.2\%$ and $1.6\%$ shown as red, yellow, green, blue, and magenta histogram, respectively. The predicted $Q_{\rm HII}$ value on the $x$-axis is colored in the same way so that the histogram matching the $x$-label color is given the fraction of corrected classified for sample (i.e., Q-RFC accuracy) for the corresponding $Q_{\rm HII}$ value. The DLA-free (upper panels) and DLA-added cases (lower panels) are shown for three different noise levels, 1\%, 5\%, and 20\%, in the left, middle, and right panels, respectively. Here, the DLA-free and DLA-added samples are based on the DLA-RFC results, not on the true labels. }
   \label{fig:accHist}
  \end{center}
\end{figure*}

\subsection{Classification Results for the Test Samples}\label{sec:clfResult}

In Figure~\ref{fig:TrTrClf}, we visualize the decision boundaries of our 2D RFC models for the three noise levels (1\%, 5\%, and 20\%) as colored backgrounds overlaid on the transmission-transmission plots from Figure~\ref{fig:TrTr}. The decision boundaries are generated by querying the trained 2D Q-RFC and DLA-RFC models on a $1000 \times 1000$ grid in the 2D data space.\footnote{As noted above, these 2D models and decision boundaries are used for visualization purposes only.} The hue of the background represents the decision boundary for $Q_{\rm HII}$ based on the Q-RFC model, while the brightness indicates the DLA-RFC results, with brighter/darker regions corresponding to DLA-added/DLA-free classifications, respectively.

Each panel displays the classification accuracy values for DLA-RFC and Q-RFC, derived from the primary 8D models, in the upper-left corner. The Q-RFC accuracy is shown separately for DLA-free and DLA-added groups, where we use the DLA-RFC result—rather than the true DLA presence label—to separate these groups. This approach provides a conservative estimate of the Q-RFC accuracy for DLA-free samples, assuming that the true DLA presence is unknown and must be determined by the DLA-RFC. Consequently, any misclassified DLA-free samples would lower the classification accuracy compared to a scenario where the true DLA presence is known, as Q-RFC tends to be less accurate for DLA-added samples.

Figure~\ref{fig:accHist} shows the Q-RFC results for $5\times5$ combinations of input and predicted $Q_{\rm HII}$ values as colored histograms, separately for the DLA-free and DLA-added groups. The $x$-labels represent the predicted $Q_{\rm HII}$ values, and the color of the bar indicates the true $Q_{\rm HII}$ value of the input spectra. The bar whose color matches the $x$-label color shows the correctly classified fraction (i.e., classification accuracy) of the Q-RFC for the corresponding $Q_{\rm HII}$ label. 

\subsubsection{One Percent Noise; SNR=100}
The RFC results for 1\% noise or continuum SNR of 100 is shown in Figure~\ref{fig:TrTrClf}$a$. The Q-RFC correctly classifies 90\% of the DLA-free samples, but is less accurate (64\%) for the DLA-added samples. The DLA-RFC performs with high accuracy, achieving a 97\% success rate in distinguishing DLA-added from DLA-free samples.

The Q-RFC decision boundary effectively separates different $Q_{\rm HII}$ groups of the DLA-free samples, owing to the strong clustering of these groups in the data space. It is also effective for the DLA-added samples with $Q_{\rm HII}=1.6$, 13.2, and 50\% samples with $\mathcal{T}(1226~\text{Å})> 0.2$, as indicated by the bright magenta, blue, and green regions that distinguish these samples. The DLA-free example case (star symbol) and the cases with moderate ($1.5\times10^{21}~{\rm cm}^{-2}$; square symbol) and weak ($5\times10^{20}~{\rm cm}^{-2}$; diamond symbols) DLAs are all correctly classified as having $Q_{\rm HII}=50~\%$.

Misclassification occurs mostly in regions with fragmented decision boundaries, where the RFC struggles to find clear separations between labels and overfits to the training data. The boundary between $Q_{\rm HII}$ = 100\% (red) and 80\% (yellow) is highly fragmented for both DLA-free and DLA-added samples, as the 1\% noise is comparable to the opacity difference between these $Q_{\rm HII}$ values. The decision boundary is also poorly defined at $\mathcal{T}(1226~\text{Å})<0.2$ and $\mathcal{T}(1238~\text{Å})<0.6$ (lower left corner in the figure), a region populated by strong DLA samples. As pointed out in Section~\ref{sec:dataSpace}, classification is highly ineffective for these samples, across all noise levels, as the IGM and DLA opacity become highly degenerate, causing the samples to converge to the same region in the data space. The example case with a strong DLA ($N_{\rm HI} = 10^{22}~{\rm cm}^{-2}$; triangle symbol) is located in this region and is incorrectly classified as having $Q_{\rm HII}=100\%$.

The DLA-RFC decision boundary between the dark and bright background is well defined, reflected in the 97\% classification accuracy. Misclassifications involve weak-DLA cases with $N_{\rm HI}\sim 2\times10^{20}~{\rm cm}^{-2}$. The DLA-added examples with $N_{\rm HI}=5\times10^{20},~1.5\times10^{21},$ and $1\times10^{22}~{\rm cm}^{-2}$ are all correctly classified at this noise levels. 

The Q-RFC prediction statistics in Figure~\ref{fig:accHist} also show that the Q-RFC performs better for DLA-free samples (Fig.~\ref{fig:accHist}$a$) compared to DLA-added sample (Fig.~\ref{fig:accHist}$d$). Misclassification primarily occurs between adjacent $Q_{\rm HII}$ values with unclear decision boundaries. In the DLA-free samples (Fig.~\ref{fig:accHist}$a$), most misclassifications involve confusing around $20\%$ of $Q_{\rm HII}=80\%$ and $100\%$ samples with each other, with about $3\%$ of the $Q_{\rm HII}=50\%$ samples misclassified as $Q_{\rm HII}=80\%$. Conversely, the $Q_{\rm HII}=13.2\%$ and $1.6\%$ samples are nearly perfectly classified. While the accuracy decreases for DLA-added samples, it remains relatively high for $Q_{\rm HII}=13.2\%$ and $1.6\%$, indicating that low-resolution DW spectra are more effective at constraining reionization history during the early stages ($Q_{\rm HII} \lesssim 50\%$) than during the later stages.

These results demonstrate the impressive performance of RFC models in constraining $Q_{\rm HII}$ using LBG spectra observed with SNRs of 100 or higher. However, achieving such high SNRs from reionization-era galaxies is feasible only for rare bright objects, making it more practical to consider scenarios with higher noise levels.

\subsubsection{Five Percent Noise; SNR=20}

SNR of 20 is achievable for moderately bright galaxies with $M_{\rm UV}\lesssim -22$ at $z\sim7$ with $\sim 1$ hr exposure by NIRSpec PRISM observation. When the noise level is increased to 5\% (see Fig.~\ref{fig:TrTrClf}b), the increased mixing between different labels results in significantly more fragmented decision boundaries. The Q-RFC accuracy drops to 67\% for DLA-free samples and 39\% for DLA-added samples. Given that a complete random classification would yield a 20\% accuracy, the Q-RFC is particularly ineffective for DLA-added samples at this noise level. Meanwhile, the DLA-RFC maintains relatively high accuracy at 90.3\%.

Even with the increased noise, the Q-RFC still identifies well-defined boundaries between $Q_{\rm HII}=1.6\%$ (dark magenta) and $13.2\%$ (dark blue), and between $Q_{\rm HII}=13.2\%$ and $50\%$ (dark green) for DLA-free samples. However, for DLA-added samples across all $Q_{\rm HII}$ values and DLA-free samples with $Q_{\rm HII} \ge 50\%$, the boundaries are poorly defined due to noise surpassing the opacity differences. The DLA-free example case (star symbol) with a true $Q_{\rm HII}$ value of 50\% is incorrectly classified as $Q_{\rm HII} = 80\%$. Moreover, the DLA-added examples with $N_{\rm HI} = 5 \times 10^{20}$ (diamond symbol), $1.5 \times 1\times10^{21}$ (square symbol), and $10^{22}~{\rm cm}^{-2}$ (triangle symbol) are misclassified as $Q_{\rm HII} = 13.2\%$, $1.6\%$, and $1.6\%$, respectively. 

In case of the DLA-RFC, the samples with $\mathcal{T}(1226\text{Å})<0.5$ are classified as DLA-added with nearly 100\% accuracy. The triangle symbol (a case with $N_{\rm HI} = 10^{22}~{\rm cm}^{-2}$) falls into this category and is correctly identified as DLA-added with 100\% confidence. However, DLA-added samples with $\mathcal{T}(1226)~\text{Å}>0.5$ are largely mixed with DLA-free samples, as the noise far exceeds the difference in wavelength dependence between DLA and IGM opacity. Many DLA-added samples with $\mathcal{T}(1226\text{Å})>0.5$ are misclassified as DLA-free cases with a lower $Q_{\rm HII}$ than its true value. The $Q_{\rm HII}=50\%$ examples with a weak ($N_{\rm HI}=5\times10^{20}$; diamond symbol) and a moderate DLA ($N_{\rm HI}=1.5\times10^{21}$; square symbol), which are misclassified as DLA-free cases with $Q_{\rm HII}=13.2\%$, fall into this category. The fraction of such confused cases is only 10\% in this case, but it could be higher if the fraction of weak DLAs is higher than assumed in this work. 

For the DLA-RFC, samples with $\mathcal{T}(1226~\text{Å}) < 0.5$ are classified as DLA-added with nearly 100\% accuracy, including the triangle symbol (representing $N_{\rm HI} = 10^{22}~{\rm cm}^{-2}$), which is correctly identified with 100\% confidence. However, DLA-added samples with $\mathcal{T}(1226\text{Å}) > 0.5$ are frequently confused with DLA-free samples due to the noise overwhelming the differences in the wavelength dependence of DLA and IGM opacity and are often misclassified as DLA-free cases with a lower $Q_{\rm HII}$ than their true values. For example, the $Q_{\rm HII} = 50\%$ cases with a weak ($N_{\rm HI} = 5 \times 10^{20}$; diamond symbol) and moderate DLA ($N_{\rm HI} = 1.5 \times 10^{21}$; square symbol) are misclassified as DLA-free cases with $Q_{\rm HII} = 13.2\%$. Although only 10\% of the DLA-added samples fall into this category, this fraction could be higher if weak DLAs are more prevalent than assumed.

The Q-RFC prediction statistics in Figure~\ref{fig:accHist}$e$ highlight that DLA-added samples are much less effective in constraining $Q_{\rm HII}$, except for $Q_{\rm HII} = 1.6\%$, which still achieves 50\% accuracy. Samples with other $Q_{\rm HII}$ values slightly exceed 20\% accuracy, with $Q_{\rm HII} = 13.2\%$ reaching 40\%. This is because the highly fragmented decision boundary still roughly captures the statistical differences in the data distribution. However, given the low accuracy and the lack of clear visual confirmation, we do not consider the Q-RFC effective in this regime. Figure~\ref{fig:accHist}$b$ indicates that the Q-RFC can reach meaningful accuracy if DLA-added cases are filtered using the DLA-RFC, with accuracy particularly high for $Q_{\rm HII} = 13.2\%$ and $1.6\%$, exceeding 85\%. The accuracy is slightly lower for $Q_{\rm HII}=50\%$, 80\%, and 100\% cases, but still significantly higher than the random baseline of 20\%.

\subsubsection{Twenty Percent Noise; SNR=5}

In 20\% noise case, shown in Figure~\ref{fig:TrTrClf}$c$, the Q-RFC accuracy degrades to 34\% for DLA-free samples and 30\% for DLA-added samples (see Fig.~\ref{fig:TrTrClf}c).  The highly fragmented decision boundaries between different colors illustrate the ineffectiveness of the model. The Q-RFC prediction statistics in Figure~\ref{fig:accHist}$c$ and $f$ also demonstrate that the Q-RFC is unable to correctly determine $Q_{\rm HII}$ regardless of DLA presence.

\begin{figure}
  \begin{center}
    \includegraphics[scale=0.5]{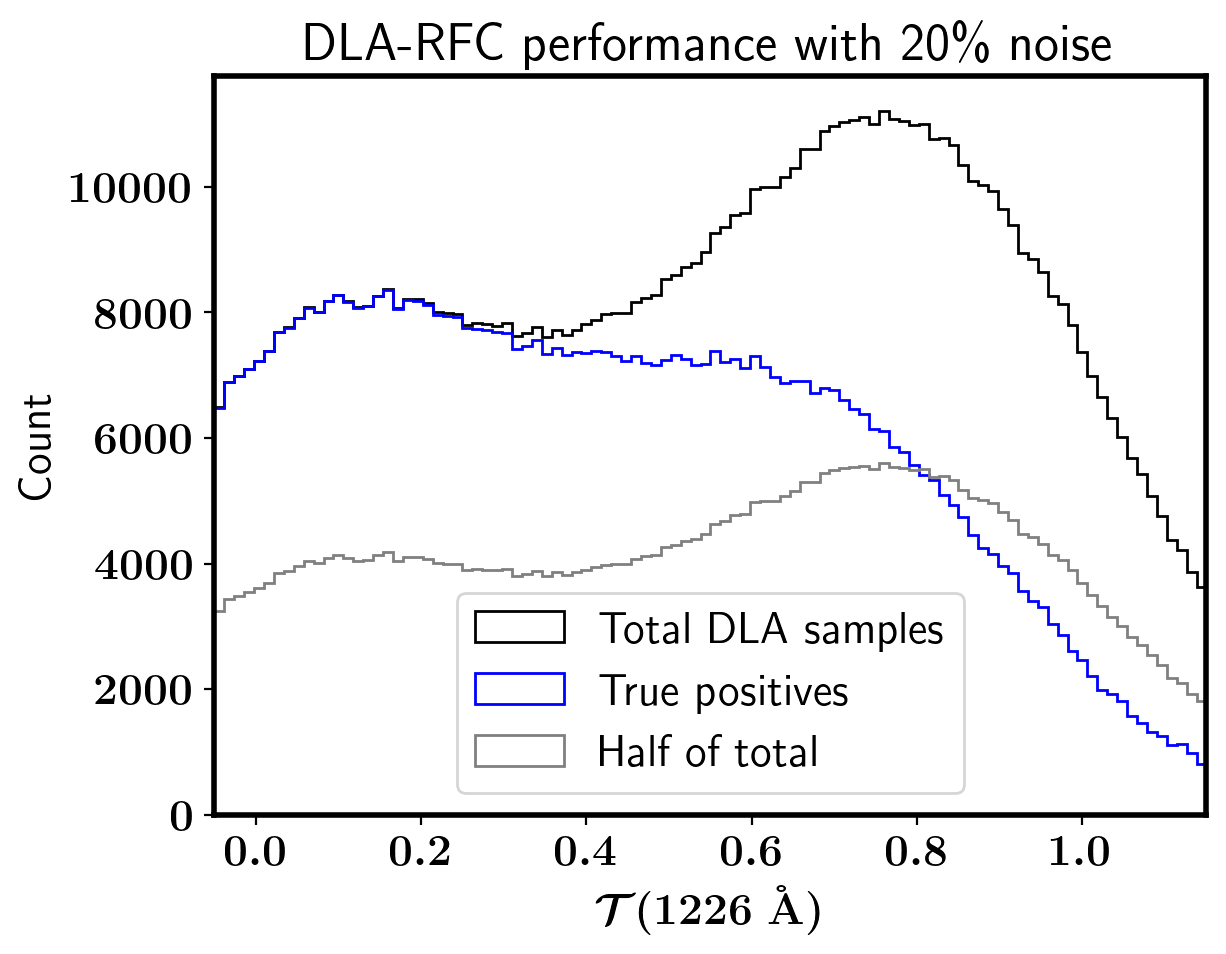}
    \caption{Histogram of total DLA samples (black) compared to the correctly classified DLA samples by our DLA-RFC model (blue) with 20\% noise as a function of $\mathcal{T}(1226~\text{Å})$, the transmission in the shortest wavelength bin. The gray histogram represents the limit of a completely random 50-50 selection.}
   \label{fig:dlaRfcPerformance}
  \end{center}
\end{figure}

The DLA-RFC establishes an approximate boundary between DLA-free and DLA-added samples at $\mathcal{T}(1226~\text{Å})\sim 0.5$ (indicated by the bright vs. dark background in Fig.~\ref{fig:TrTrClf}$c$) and correctly determine DLA presence for more than 80\% of the samples. Figure~\ref{fig:dlaRfcPerformance} shows that most of the correctly classified DLA-added cases (i.e., true positives) are distributed in the low-transmission region in the data space. Given that DLA-RFC performs a binary classification between DLA-free versus DLA-added, its 80\% accuracy in this noise level is modest at best. However, it can still serve as a screening tool for higher-SNR observations.

\section{Application to Existing Observations} \label{sec:application}

We train and apply our RFC models to existing high-redshift spectra to assess their effectiveness in constraining $Q_{\rm HII}$ at the redshift of the source galaxy, thereby providing insights into reionization models. To ensure the RFC models are trained accurately for the specific conditions of each spectrum, we generate mock Ly$\alpha$ DW spectra tailored to each input spectrum. This involves matching the wavelength range of the spectral bins, the SNR of the flux data in each bin, and the spectral resolution of the instrument.

First, we smooth the Ly$\alpha$ transmission curves from the CoDa II simulation using a Gaussian line-spread function, 
\bea \label{eq:lsp}
f_{\rm LSF}(dv_\alpha)= \frac{1}{\sigma \sqrt{2\pi}} \exp{\left(-\frac{dv^2_\alpha}{2\sigma}\right)},
\eea
where we set $\sigma = c/(2.355R)$ so that the full width at half maximum (FWHM) is given by the speed of light, $c$, divided by the spectral resolution\footnote{The spectral resolution of PRISM can be obtained from \url{https://jwst-docs.stsci.edu/jwst-near-infrared-spectrograph/nirspec-instrumentation/nirspec-dispersers-and-filters}.}, $R$. Next, we rebin the smoothed transmission curves to match the wavelength range of the spectral bins and add Gaussian noise based on the SNR of each bin. We use spectral bins whose entire extent falls within the range of $dv_\alpha = 1000 - 25000~{\rm km/s}$ to sample the wavelength range where DW opacity is significant, while avoiding modeling uncertainties related to the velocity offset of the Lyman break. We then train both the DLA-RFC and Q-RFC models using these tailored mock spectra to determine the presence of a DLA and the value of $Q_{\rm HII}$ for each spectrum.

The continuum in the galaxy spectrum is fitted using a power-law between 1350 and 2600 Å, and the intrinsic UV continuum level is extrapolated to 1215.67 Å to estimate the DW opacity of the IGM. Additionally, we subtract any Ly$\alpha$ emission-line leakage into other spectral bins caused by the low-resolution spectroscopy, provided the exact location and strength of the line are known from higher-resolution spectroscopy. As with the smoothing of the transmission curves, we apply the line-spread function from Equation~(\ref{eq:lsp}) to estimate the leakage across spectral bins. 

We inspect publicly available spectra from the JWST Advanced Deep Extragalactic Survey\footnote{https://jades-survey.github.io} \citep[JADES, GTO 1180, GTO 1181, PI: D. Eisenstein, GTO 1210, PI: N. Lützgendorf, GO 3215, PI: D. Eisenstein \& R. Maiolino;][]{2023arXiv230602465E,2024A&A...690A.288B,2025ApJS..277....4D} and the Cosmic Evolution Early Release Science program\footnote{https://ceers.github.io} \citep[ERS 1345, PI: S. Finkelstein;][]{2024ApJ...969L...2F} to identify suitable spectra for applying our model. We select galaxies with continuum SNRs of 10 or above, along with measured Ly$\alpha$ line offsets and EW.  As demonstrated below, accurate Ly$\alpha$ line measurement is crucial for our analysis, as leakage from the Ly$\alpha$ line can significantly contaminate flux in other spectral bins. We then use our DLA-RFC to select three galaxies that are likely DLA-free (GN-$z$11, CEERS-1026, and JADES-12637) for analysis. Additionally, we select JADES-58975 for analysis despite the presence of apparent DLA opacity, due to its high continuum SNR of 40. The Jupyter notebook detailing these procedures is publicly available on GitHub. \footnote{\url{https://github.com/hcosmosb/RFC4LBGDW}}

\begin{figure*}
  \begin{center}
    \includegraphics[scale=0.4]{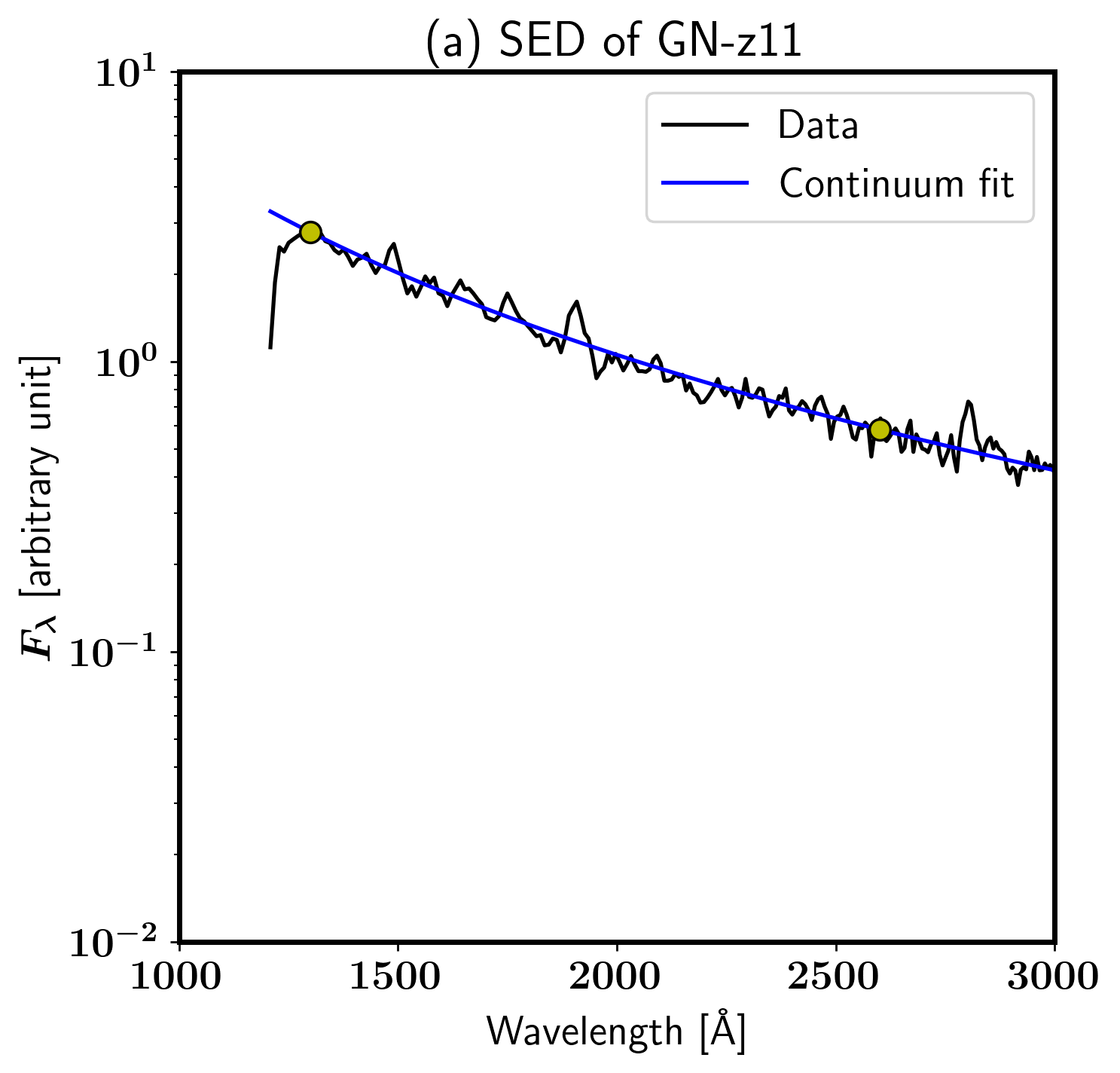}
    \includegraphics[scale=0.4]{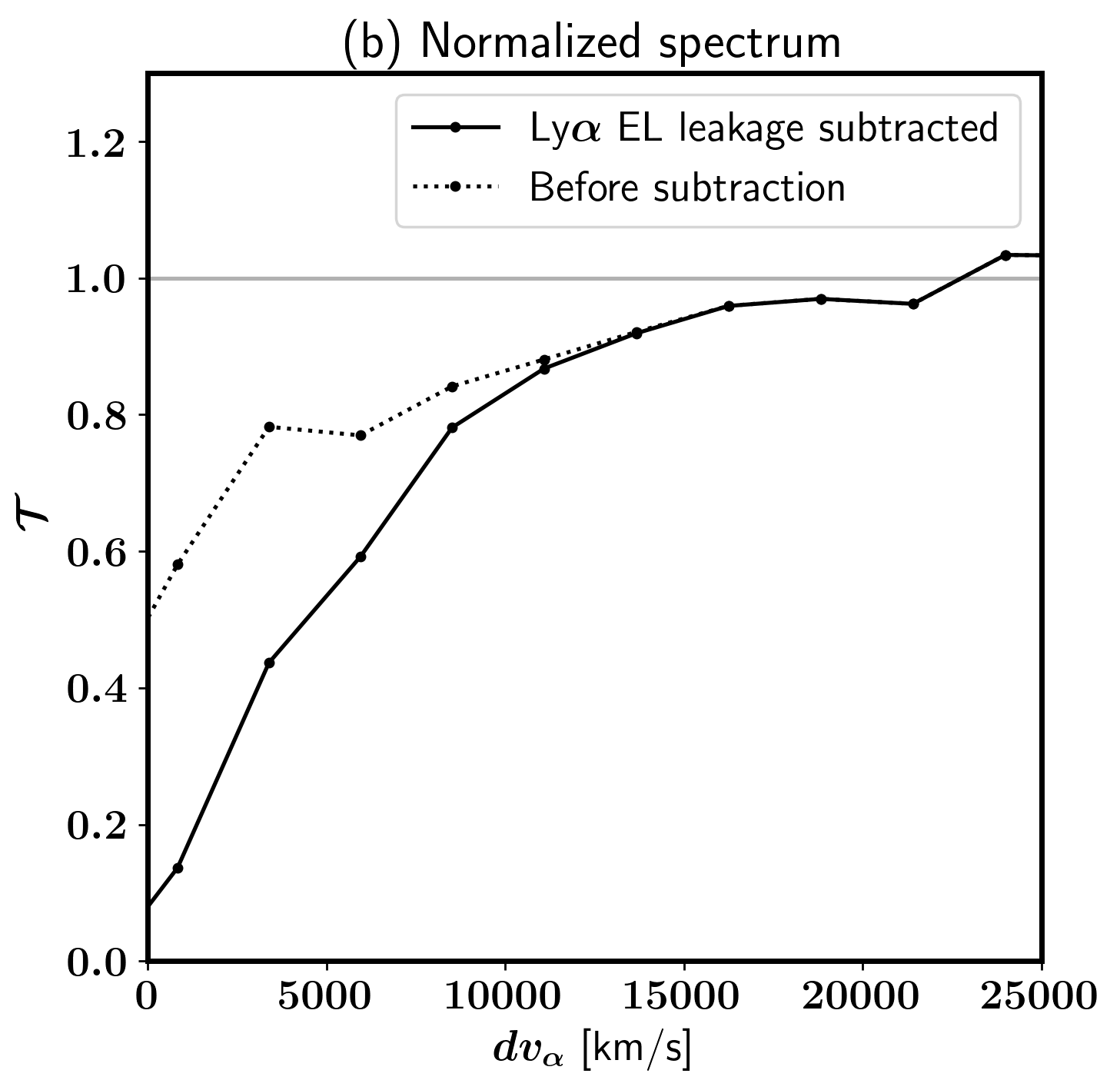}
    \includegraphics[scale=0.4]{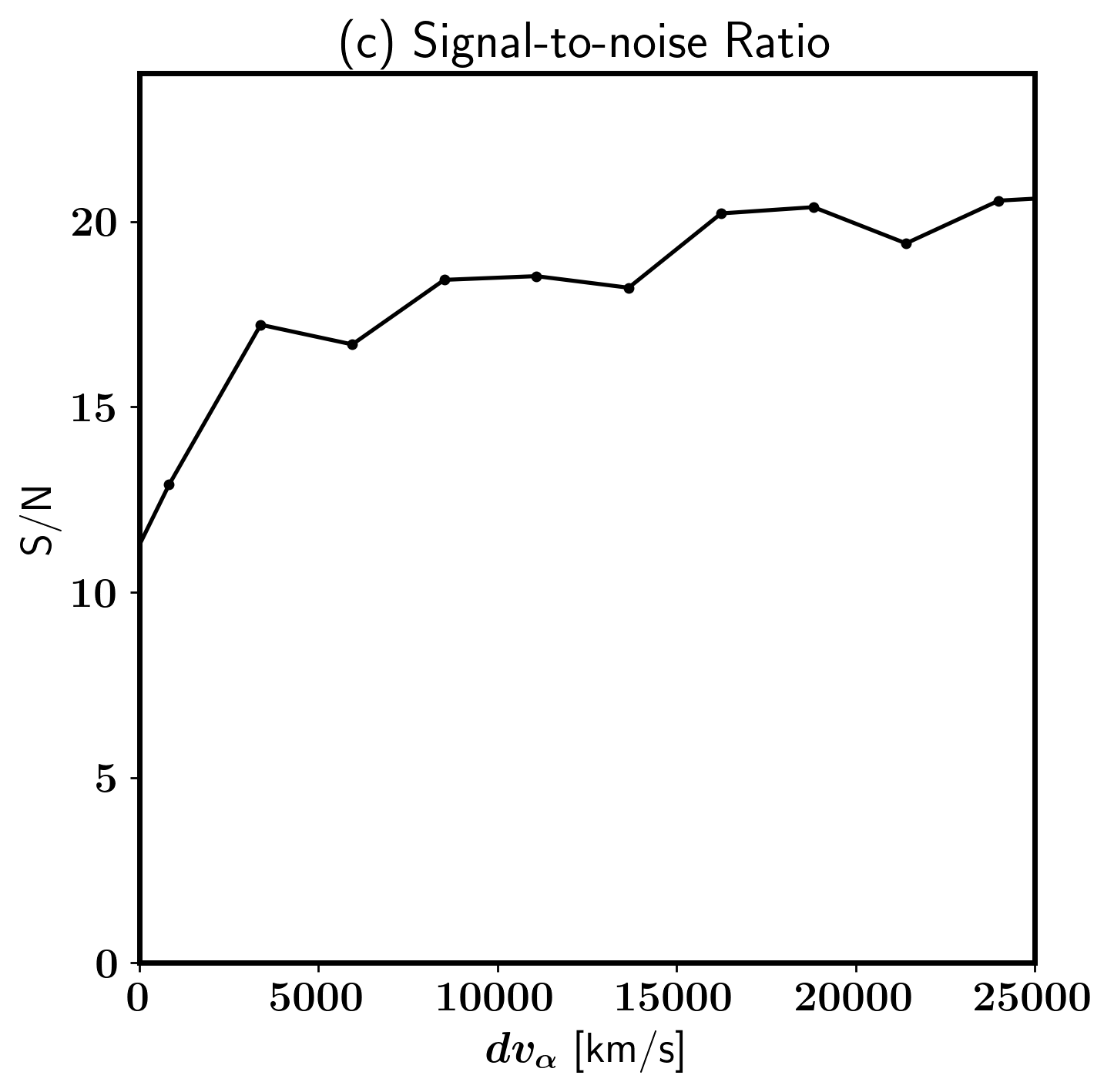}
    \includegraphics[scale=0.4]{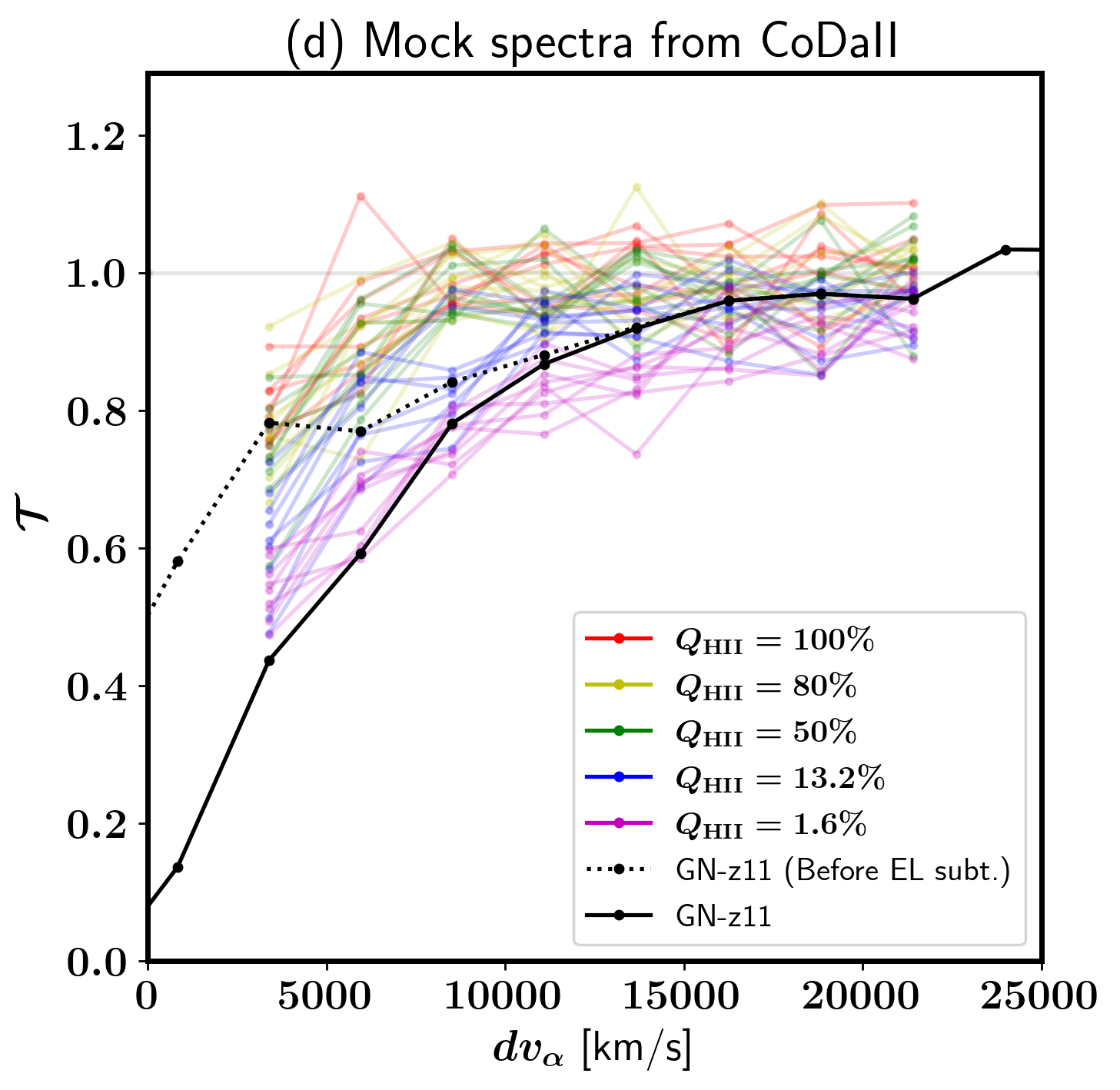}
    \includegraphics[scale=0.4]{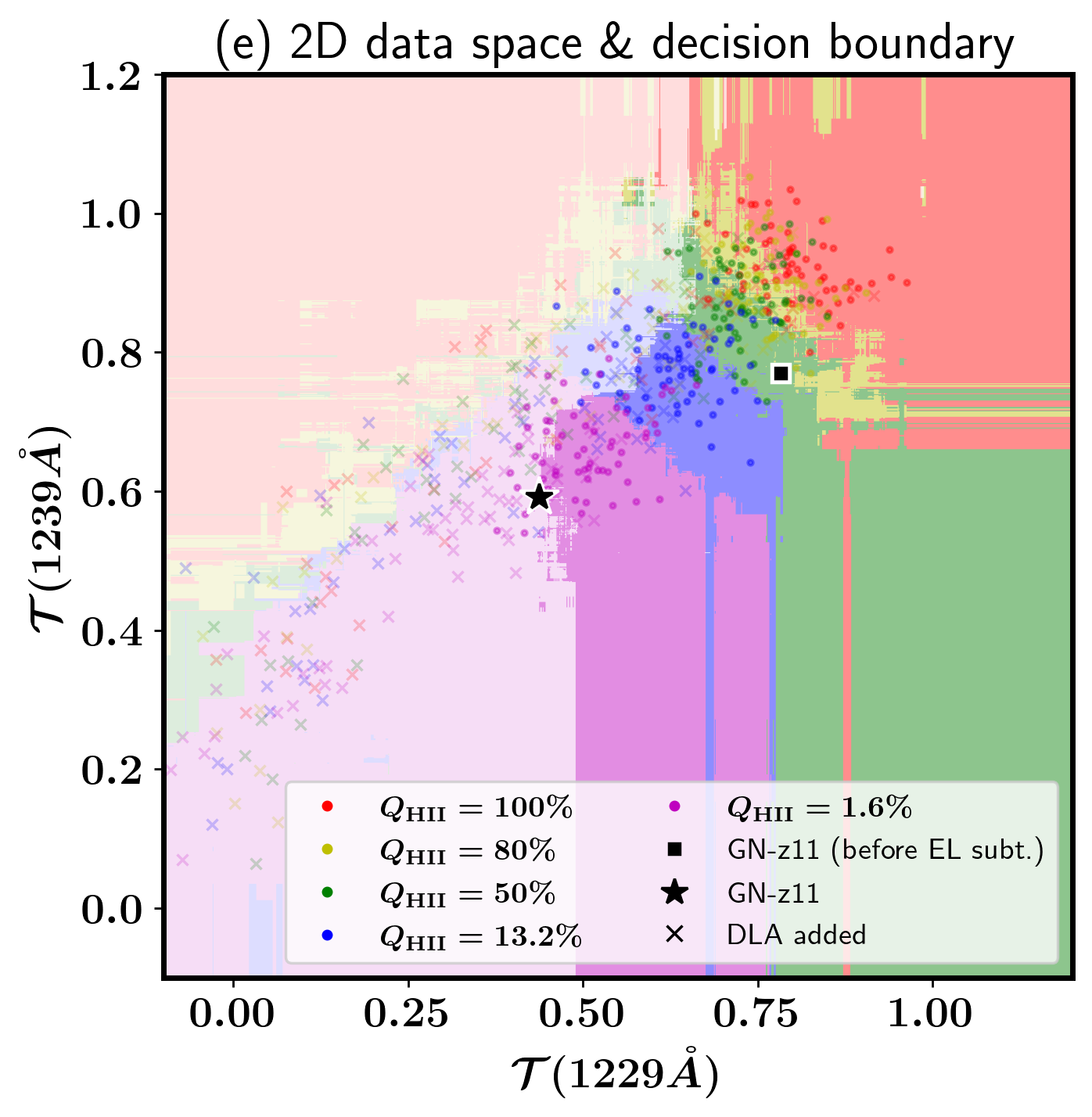}
    \includegraphics[scale=0.4]{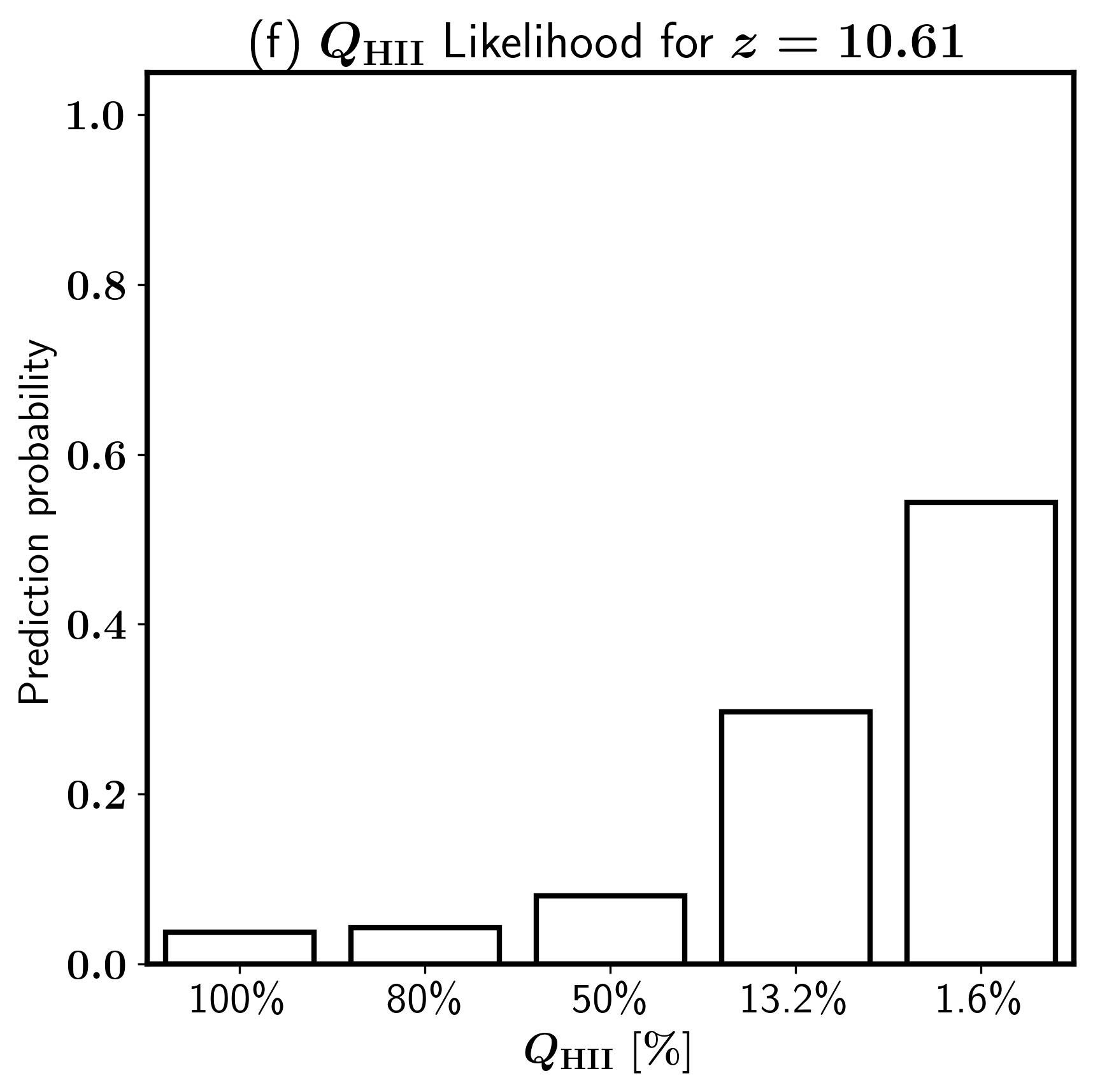}
\caption{ This figure summarizes the workflow of constraining $Q_{\rm HII}$ at $z=10.6$ with the spectrum of GN-$z$11. Panel (a): the raw spectral energy distribution (SED) at the rest frame is shown as the black line, and the power-law curve, which best fits the data between 1350 and 2600 \r{A}, is shown as the blue line. The fitting range is enclosed by two yellow circular dots.
Panel (b): Ly$\alpha$ transmission as the function of $dv_\alpha$ between 0 and $25,000~{\rm km}~{\rm s}^{-1}$, obtained by normalizing the SED with the fitting function. The dashed and solid lines represent the normalized flux before and after correcting for the leakage from the emission line, respectively. 
Panel (c): signal-to-noise ratio (SNR) of each data point in panel (b). 
Panel (d): mock Ly$\alpha$ spectra from CoDa II generated according to the flux range of the data bins and SNR of the observed spectrum. The spectrum of GN-$z$11 is also shown as a black line. 
Panel (e): transmission-transmission plot of the two shortest wavelength bins, with the results of the 2D Q-RFC and DLA-RFC shown as colored backgrounds as in Fig.~\ref{fig:TrTrClf}. The GN-$z$11 spectrum after subtracting the emission-line contribution is shown as the black star symbol, and that before the subtraction is shown as the black square symbol.
Panel (f): prediction probability of $Q_{\rm HII}$ values for $z=10.6$ from our Q-RFC model.}
   \label{fig:GNz11}
  \end{center}
\end{figure*}

\subsection{GN-$z$11}

GN-$z$11 is an extremely bright galaxy with $M_{\rm UV}=-21.5$, spectroscopically confirmed at $z>10$ by JWST. It exhibits a clear DW feature in its NIRSpec PRISM spectrum \citep{2023A&A...677A..88B}. This galaxy hosts an AGN with a broad Ly$\alpha$ line extended to $dv_\alpha\approx1000~{\rm km}~{\rm s}^{-1}$ \citep{2024Natur.627...59M,2024A&A...691A.145M}. The emission line is centered at $dv_\alpha=555~{\rm km}~{\rm s}^{-1}$ with EW of 18 Å.
Two publicly available spectra of GN-$z$11 from JADES were combined by averaging them, weighting each spectrum by the inverse of its SNR. The SNR of the combined spectrum was calculated using the Pythagorean summation of the SNRs from each observation. We use eight spectral bins at $dv_\alpha=3400$, 5900, 8500, 11,100, 13,700, 16,200, 18,800 and $21,400~{\rm km}~{\rm s}^{-1}$ for the analysis.

Figure~\ref{fig:GNz11} illustrates the workflow used to estimate $Q_{\rm HII}$ at the redshift of GN-$z$11 ($z=10.6$). Panel (a) shows the raw SED near the rest-frame Ly$\alpha$ and the power-law fit to the continuum. Panel (b) displays the eight spectral bins near Ly$\alpha$ used to constrain $Q_{\rm HII}$.  We subtract the leakage from the Ly$\alpha$ emission line based on the spectral resolution of PRISM around the rest-frame Ly$\alpha$ ($R=32$). The emission-line subtraction results in a significant change in flux, with the normalized flux being altered by approximately 0.35 in the shortest wavelength bin with $dv_\alpha \approx 4000~{\rm km}{\rm s}^{-1}$. Panel (c) shows that the combined SNR of the two GN-$z$11 spectra is close to 20, comparable to the 5\% noise cases in Section~\ref{sec:mlModel}. 

Panel (d) presents the mock spectra for different $Q_{\rm HII}$ values alongside the GN-$z$11 spectrum. At the first spectral bin ($dv_\alpha=4000~{\rm km}~{\rm s}^{-1}$ or $\lambda_{\rm rest} = 1229\text{Å}$), the flux of GN-$z$11 is slightly below $Q_{\rm HII}=1.6\%$ samples after the emission line subtraction (0.44) while it is above $Q_{\rm HII}=13.2\%$ samples before subtraction (0.78).
It is evident that this flux change from the emission-line subtraction would significantly impact the classification result for GN-$z$11.

Panel (e) shows the mock and GN-$z$11 spectra in the data space of the two shortest wavelength bins at 1229 and 1239 Å, with the results of the 2D Q-RFC and DLA-RFC models represented by the colored background, similar to in Figure~\ref{fig:TrTrClf}. As described in Section~\ref{sec:clfResult}, all the data points are used for the final classification. GN-$z$11 is classified as having $Q_{\rm HII}=1.6\%$ after the emission-line subtraction (star symbol) and $Q_{\rm HII}=50\%$ before subtraction. The data point before emission-line subtraction is highly offset from the mock data points, indicating that it is invalid for classification without subtraction. Regarding DLA presence, GN-$z$11 is slightly outside the decision boundary for DLA-free samples due to its slightly lower flux compared to typical $Q_{\rm HII}=1.6\%$ cases. The DLA-RFC model determines GN-$z$11 has a 60\% probability of containing a DLA.  However, its location in the data space is still populated by other DLA-free $Q_{\rm HII}=1.6\%$ samples, suggesting that GN-$z$11 could still be DLA-free. Given that prediction probability depends on the DLA fraction and the assumed DLA column density distribution in the mock samples, it remains inconclusive whether GN-$z$11 indeed contains a DLA.

Panel (f) illustrates the ``prediction probability" for each $Q_{\rm HII}$ value, based on the number of votes from individual decision trees in the Q-RFC for the emission-line subtracted spectrum. We can regard the prediction probability for each label as the likelihood for $Q_{\rm HII}$. The Q-RFC strongly favors $Q_{\rm HII}=1.6\%$ and $13.2\%$ over other $Q_{\rm HII}$ values, with prediction probabilities of 54\% and 30\%, respectively. Thus, we conclude $Q_{\rm HII}< 31.6\%$ at $z=10.6$ with 85\% confidence, where $31.6\% = [(13.2+50)/2]\%$ is the midpoint between the $Q_{\rm HII}=13.2$ and 50\% bins. This result is unlikely to constrain reionization models effectively, as most predict the Universe was less ionized than $31.6\%$ at this redshift. Instead, this outcome serves as a test confirming the validity of our RFC method.

\begin{figure*}
  \begin{center}
    \includegraphics[scale=0.3]{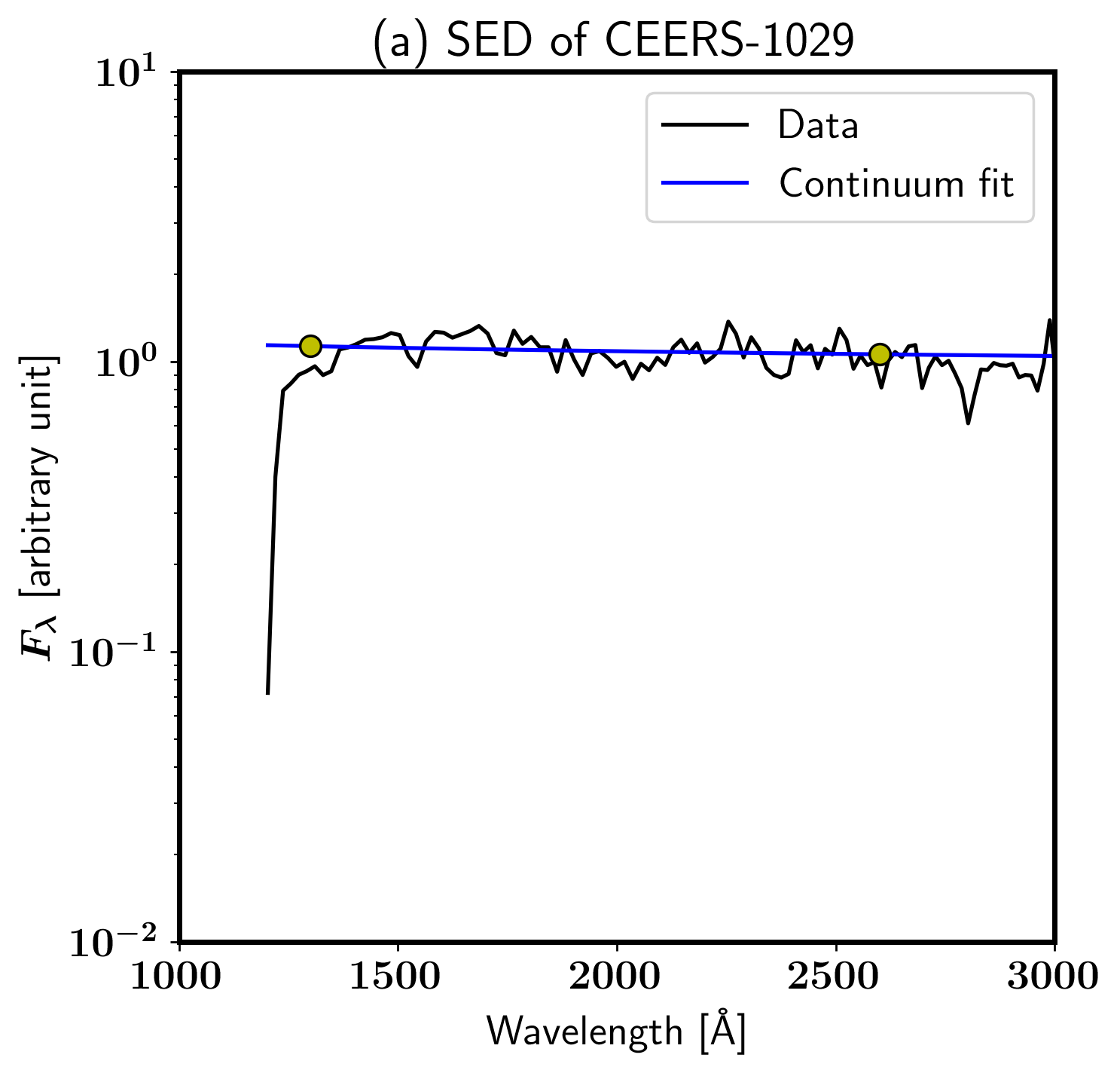}
    \includegraphics[scale=0.3]{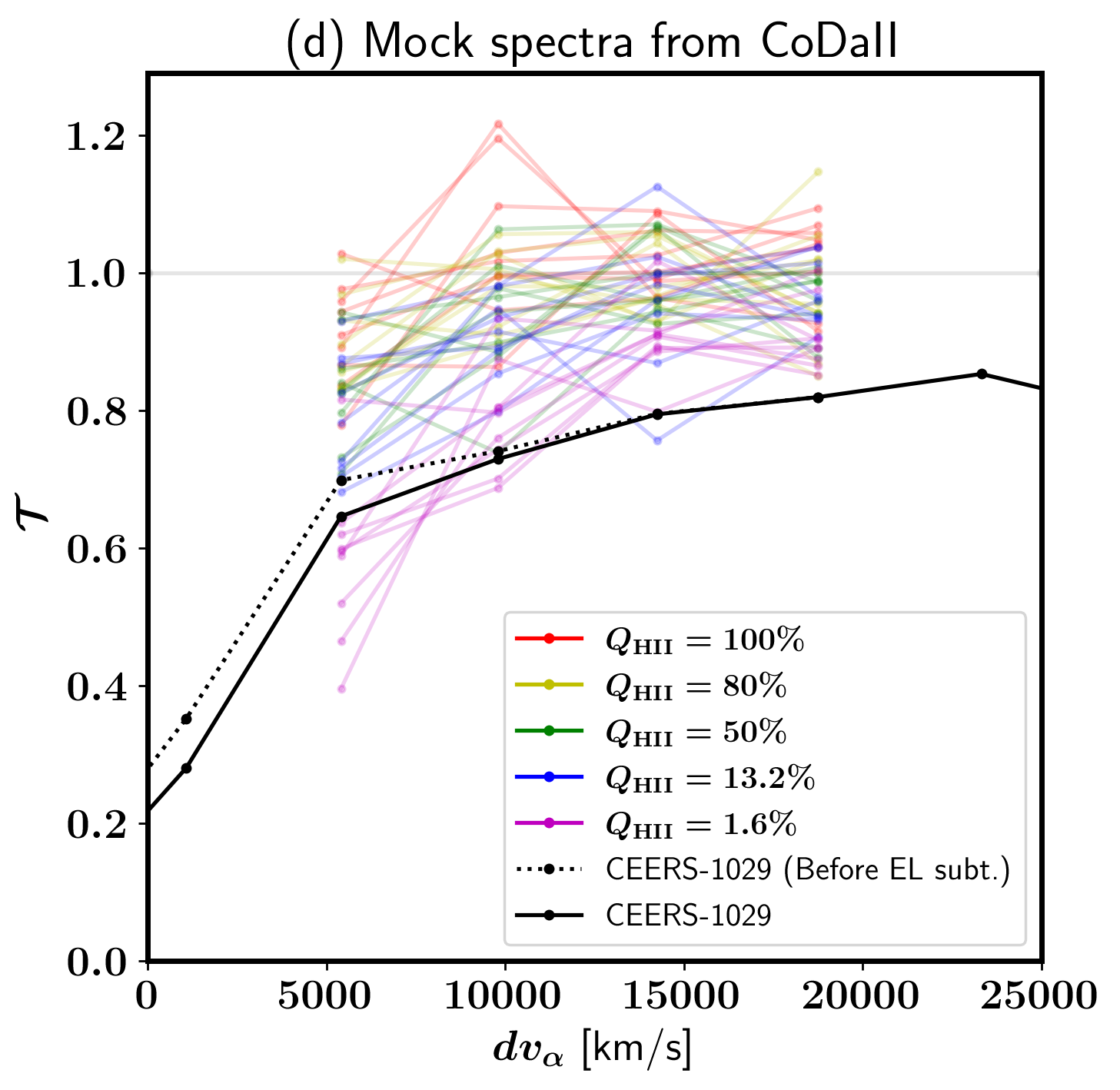}
    \includegraphics[scale=0.3]{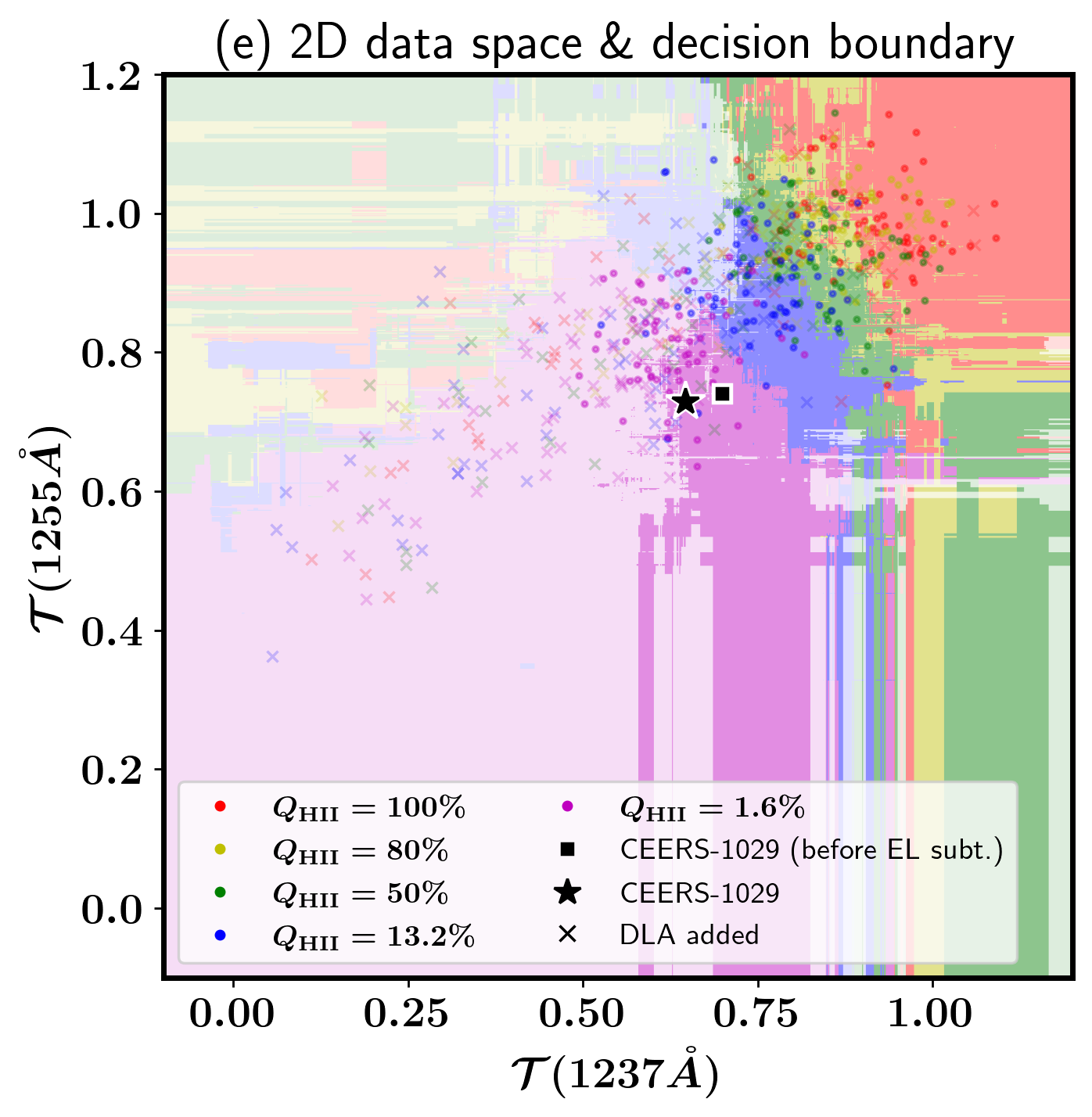}
    \includegraphics[scale=0.3]{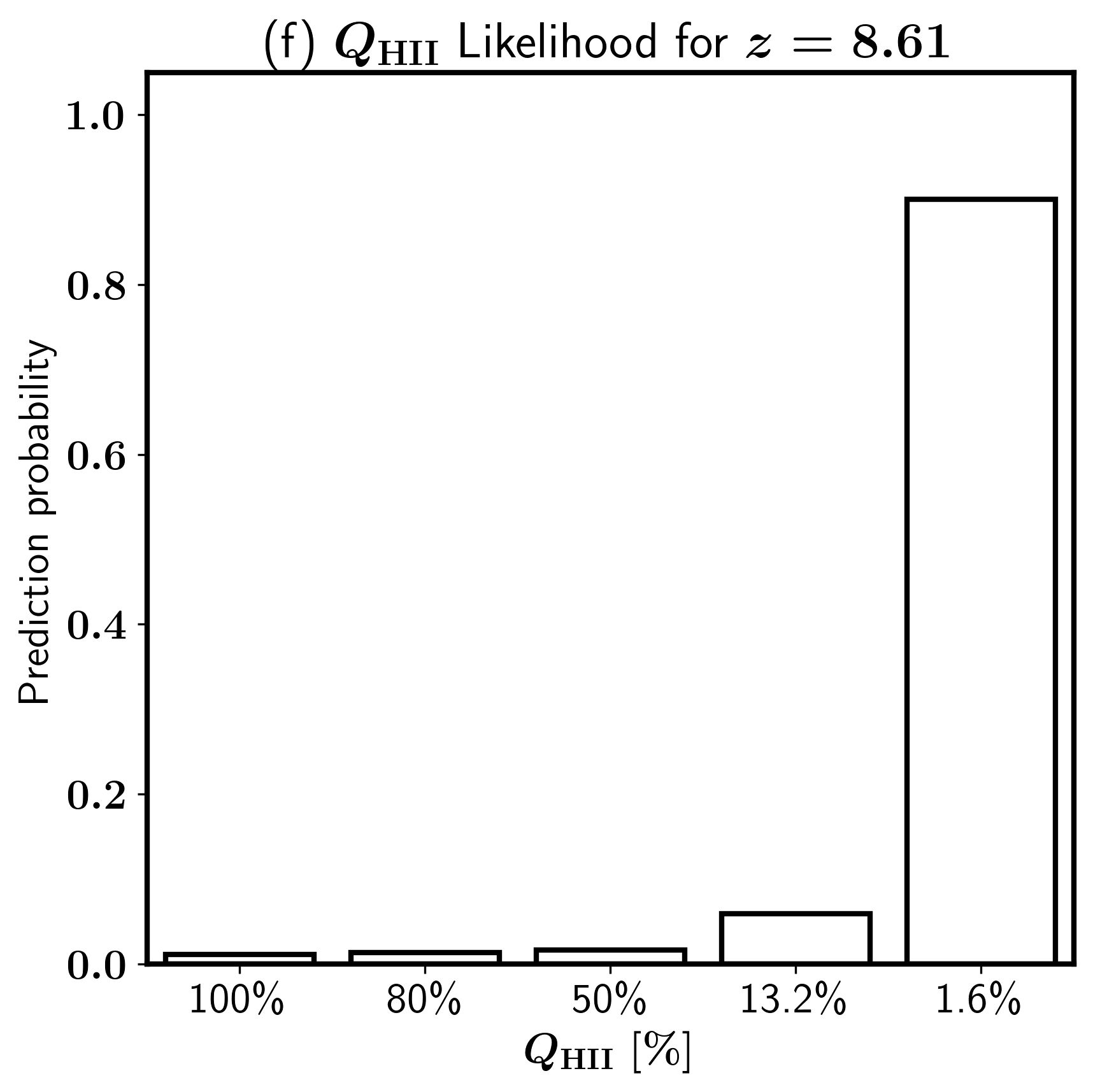}
\caption{Same as panels (a), (d), (e), and (f) of Fig.~\ref{fig:GNz11} for CEERS-1029 at $z=8.6$.}
   \label{fig:CEERS-1029}
  \end{center}
\end{figure*}

\subsection{CEERS-1029}

CEERS-1029 is one of the targets of the CEERS program, with $M_{\rm UV} = -21.53$ at $z = 8.61$, which was first discovered by \cite{2022ApJ...930..104L}. The EW and velocity offset of the Ly$\alpha$ emission are measured to be 3 Å and $dv_\alpha=2100~{\rm km}{\rm s}^{-1}$, respectively \citep{2024ApJ...975..208T}. The PRISM spectrum has a SNR of approximately $15$ in the continuum. We use four spectral bins at $dv_\alpha=5400$, 9800, 14,300, and $18,800~{\rm km}~{\rm s}^{-1}$ for the analysis.

Figure~\ref{fig:CEERS-1029} shows the raw flux data with the power-law fit, the normalized spectrum alongside the mock spectra for the target, a scatter plot of the two shortest wavelength bins, and the prediction probability of Q-RFC for the emission-line subtracted spectrum. The normalized flux in panel (d) show that subtracting the emission leakage results in only a minor change (few percent) in the shortest wavelength bin, due to the small EW of the Ly$\alpha$ emission. 

In the data space shown in panel (e), CEERS-1029 falls in the region for DLA-free $Q_{\rm HII}=1.6\%$ cases, regardless of the emission-line subtraction. The Q-RFC model predicts a 90\% probability for $Q_{\rm HII}=1.6\%$ and much lower probability for other $Q_{\rm HII}$ values (panel (f)). We interpret this result as an upper bound of $Q_{\rm HII}<7.4\%$ at $z=8.6$. This upper bound is quite tight for $z=8.6$ and would rule out some of the early reionization models.

However, we note that the spectrum of CEERS-1029 may not have been precisely fitted by a power law. The raw flux in panel (a) shows a long-range fluctuation around the fit, suggesting there may be another component in the continuum not captured by the power law. The normalized spectrum of CEERS-1029 near Ly$\alpha$ in panel (d) appears flatter than those of the mock $Q_{\rm HII}=1.6\%$ cases, with the flux at the two longer wavelength bins being lower than the typical range of the mock spectra for $Q_{\rm HII}=1.6\%$. If the actual continuum is lower by, for example, 10\%, than what the power-law fitting gave, the Q-RFC would have given significantly higher probabilities for $Q_{\rm HII}=13.2$ or $50\%$. 

\begin{figure*}
  \begin{center}
    \includegraphics[scale=0.3]{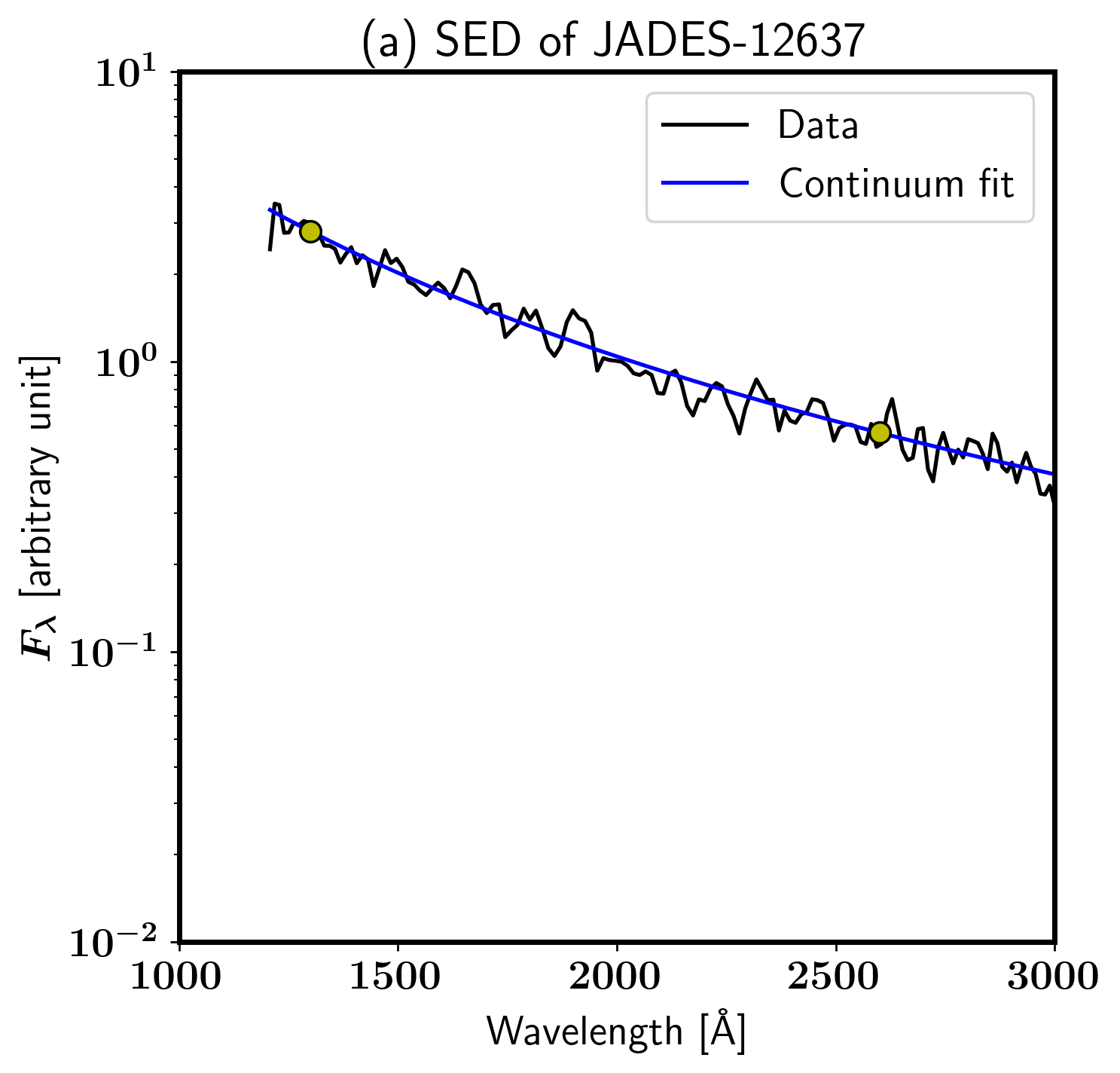}
    \includegraphics[scale=0.3]{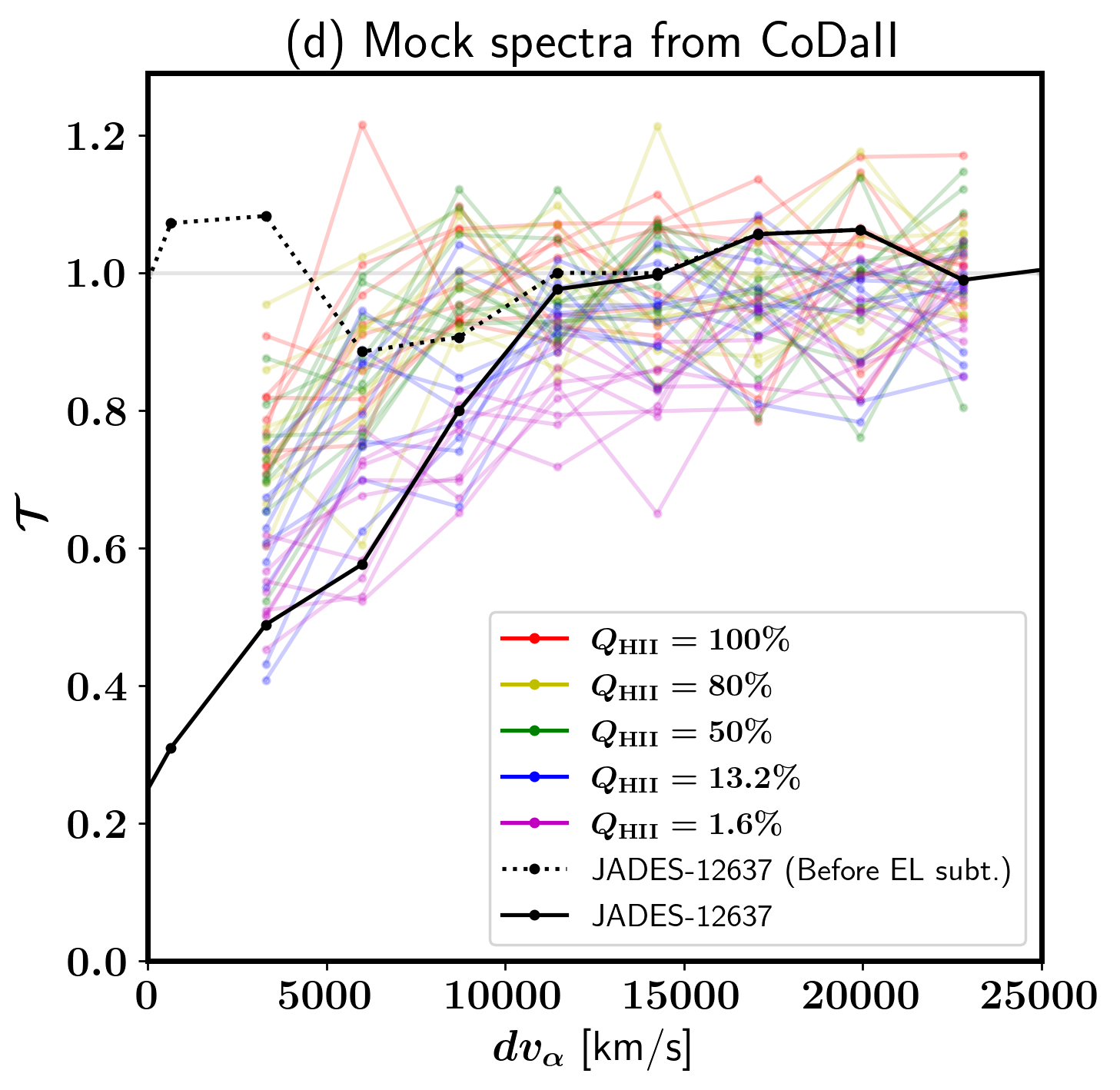}
    \includegraphics[scale=0.3]{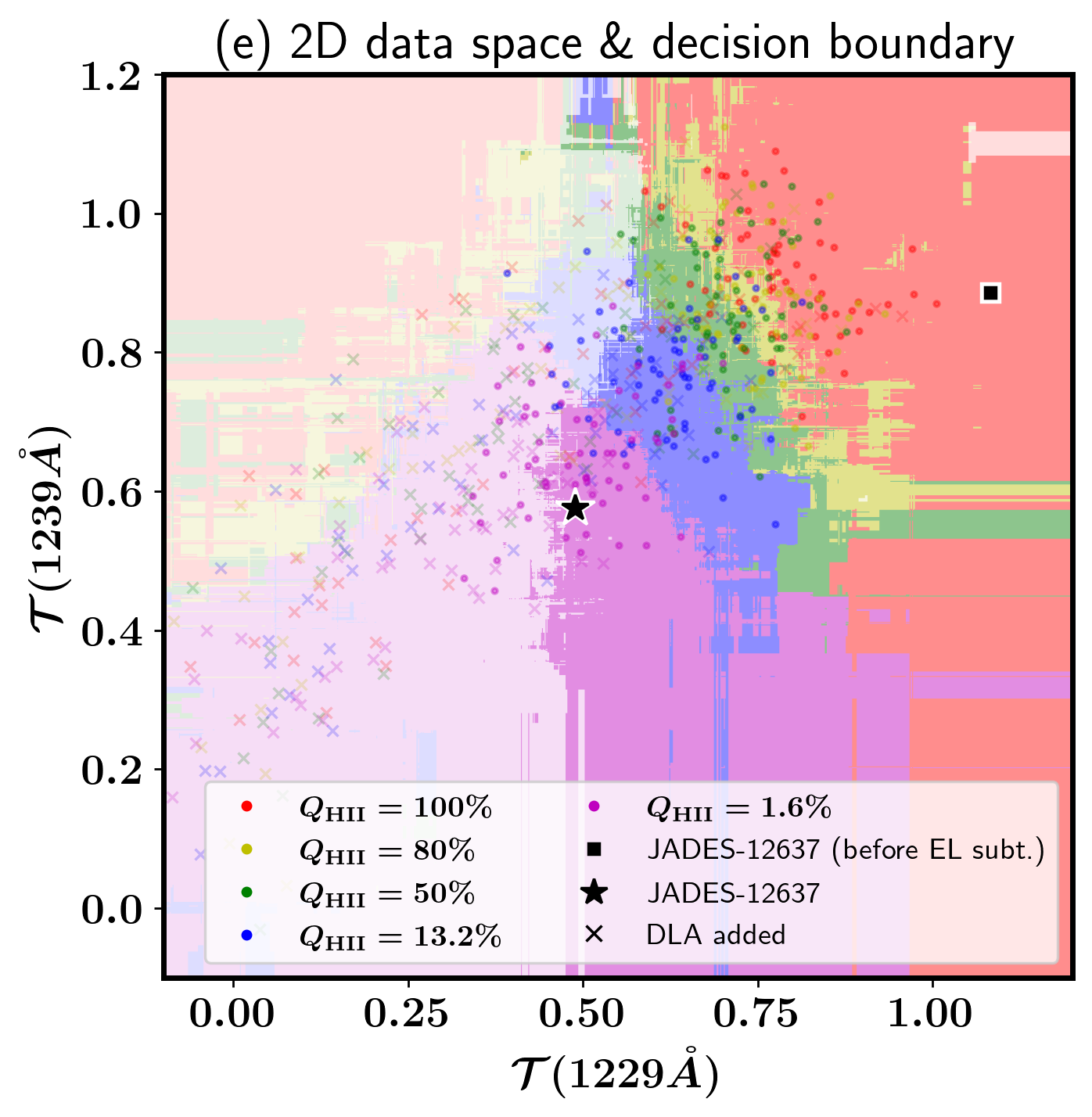}
    \includegraphics[scale=0.3]{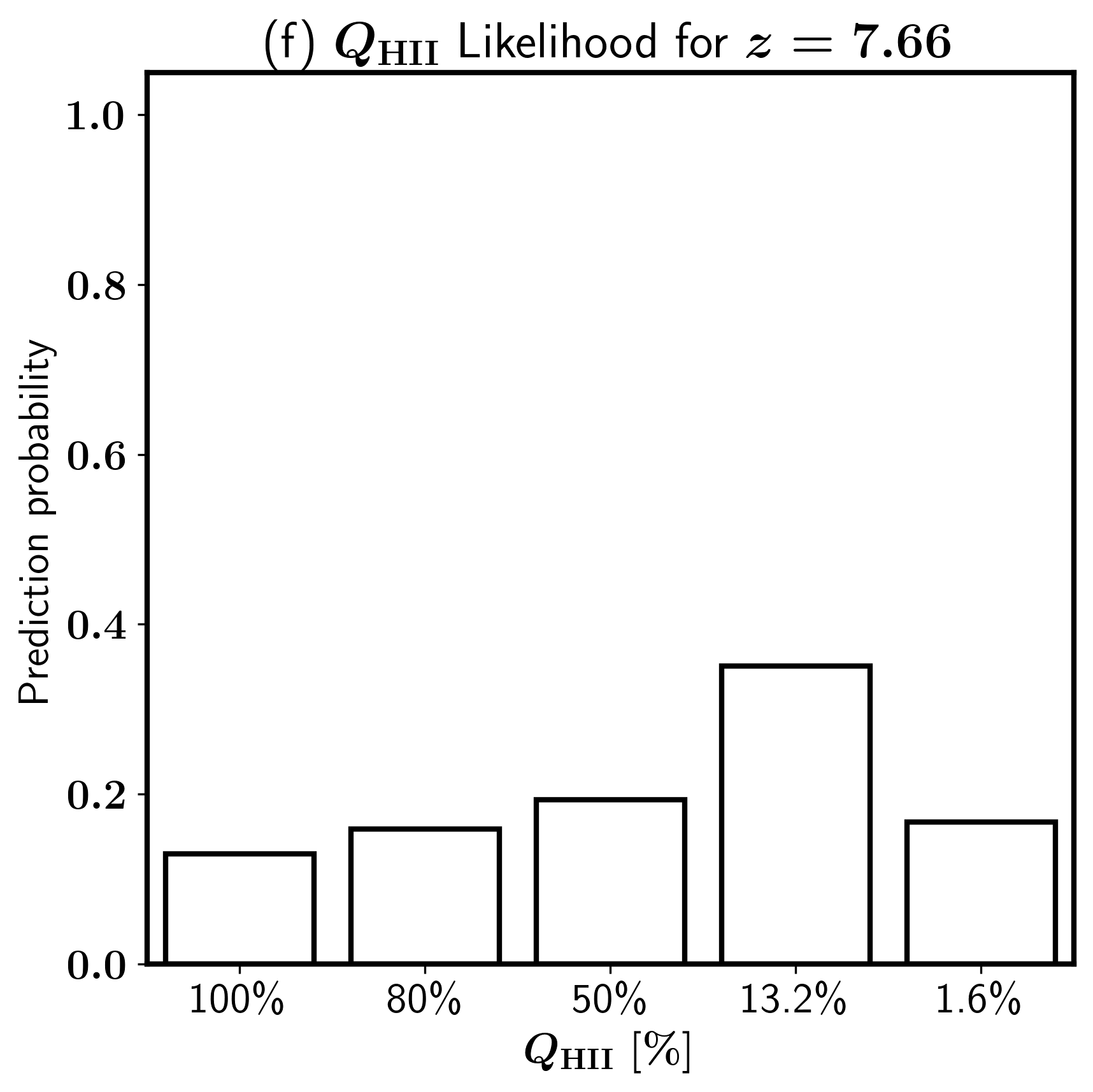}
\caption{Same as panels (a), (d), (e), and (f) of Fig.~\ref{fig:GNz11} for JADES-12637 at $z=7.7$.}
   \label{fig:JADES-12637}
  \end{center}
\end{figure*}

\subsection{JADES-12637}

JADES-12637 is one of the targets of the JADES program, with $M_{\rm UV}=-20.7$ at $z=7.7$. The Ly$\alpha$ equivalent width and $dv_\alpha$ of the Ly$\alpha$ emission are measured to be 33 Å and $277~{\rm km}{\rm s}^{-1}$, respectively \citep{2024ApJ...975..208T}. The PRISM spectrum has an SNR of approximately $10$ in the continuum. We use eight spectral bins at $dv_\alpha=3300$, 6000, 8700, 11,500, 14,200, 17,100, 19,900 and $22,800~{\rm km}~{\rm s}^{-1}$ for the analysis.

Figure~\ref{fig:JADES-12637} presents the raw flux with the power-law fit, the normalized spectrum of JADES-12637 and its corresponding mock spectra, a scatter plot of these spectra in the 2D data space of the two shortest wavelength bins, and the prediction probability for the Q-RFC model. The power-law fit in panel (a) appears to fit the data well. The normalized flux in panel (d) shows that subtracting the emission-line leakage, with EW = 33 Å, results in a significant change in the spectrum, nearly halving the flux in the first bin.

In the data space (panel (e)), JADES-12637 is positioned near the center of the DLA-free $Q_{\rm HII}=1.6\%$ group after the emission-line subtraction (indicated by the black star symbol). Without the subtraction (black square symbol), it falls entirely outside all of the mock data points. In panel (f), the Q-RFC weakly prefers $Q_{\rm HII}=13.2\%$ with a prediction probability of 35\%, with probabilities for other $Q_{\rm HII}$ values higher than 14\%. The apparent inconsistency with panel (e), appearing to prefer $Q_{\rm HII}=1.6\%$, is due to the five data points at longer wavelengths, not described in panel (e), preferring higher $Q_{\rm HII}$ due to their relatively higher flux. Combining probability for the three lower $Q_{\rm HII}$ values gives an upper bound of $Q_{\rm HII}<65\%$ with 71.2\% confidence, where 65 \% is obtained by taking the midpoint between $Q_{\rm HII}=80$ and 50\%. The weakness of the constraint from this galaxy is mainly due to the low SNR of the observation. While high-SNR observations would likely improve the result, the constraint can also be tightened by combining results from other galaxies at similar redshifts, treating the prediction probability as likelihood.

\begin{figure*}
  \begin{center}
    \includegraphics[scale=0.3]{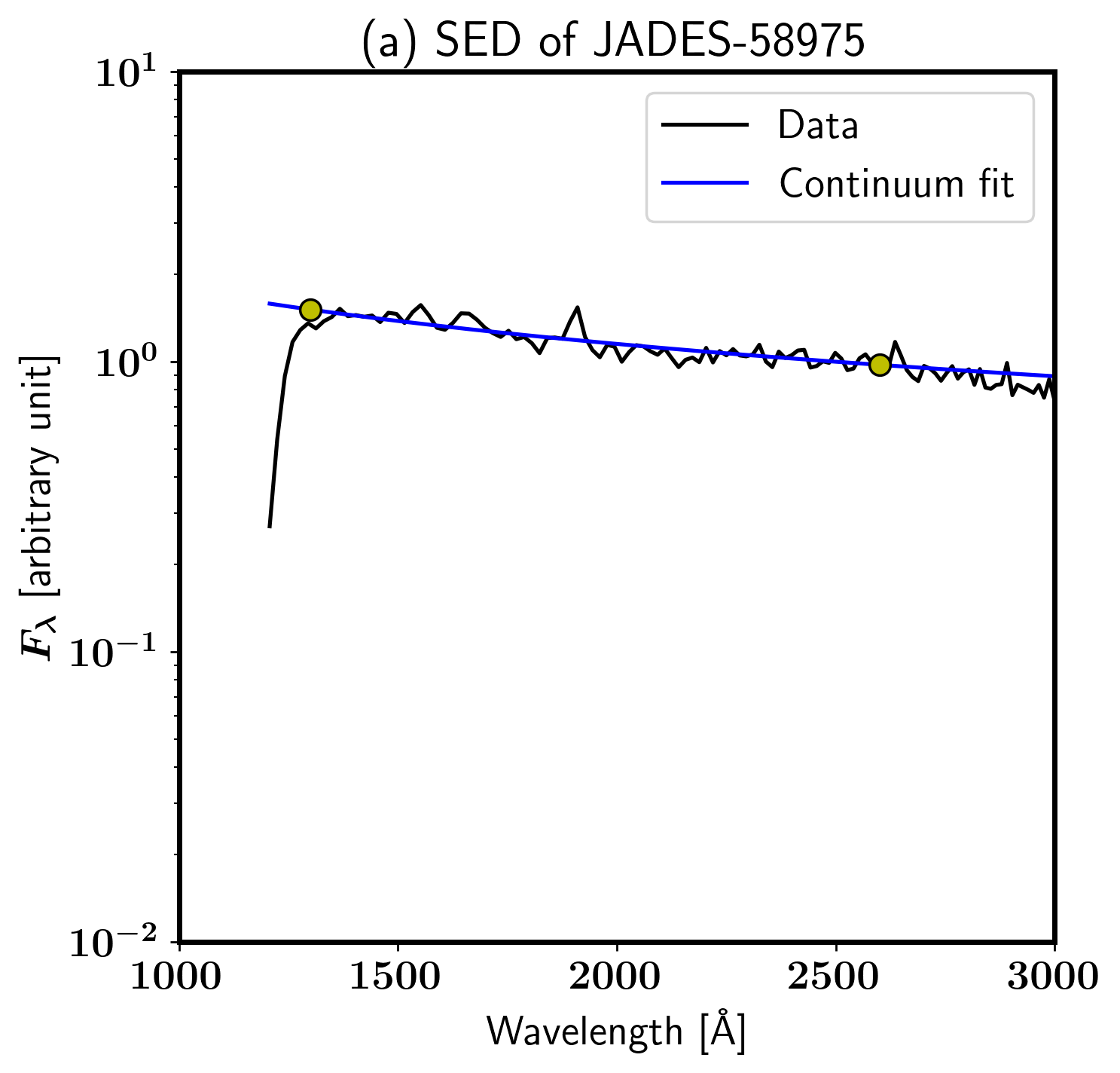}
    \includegraphics[scale=0.3]{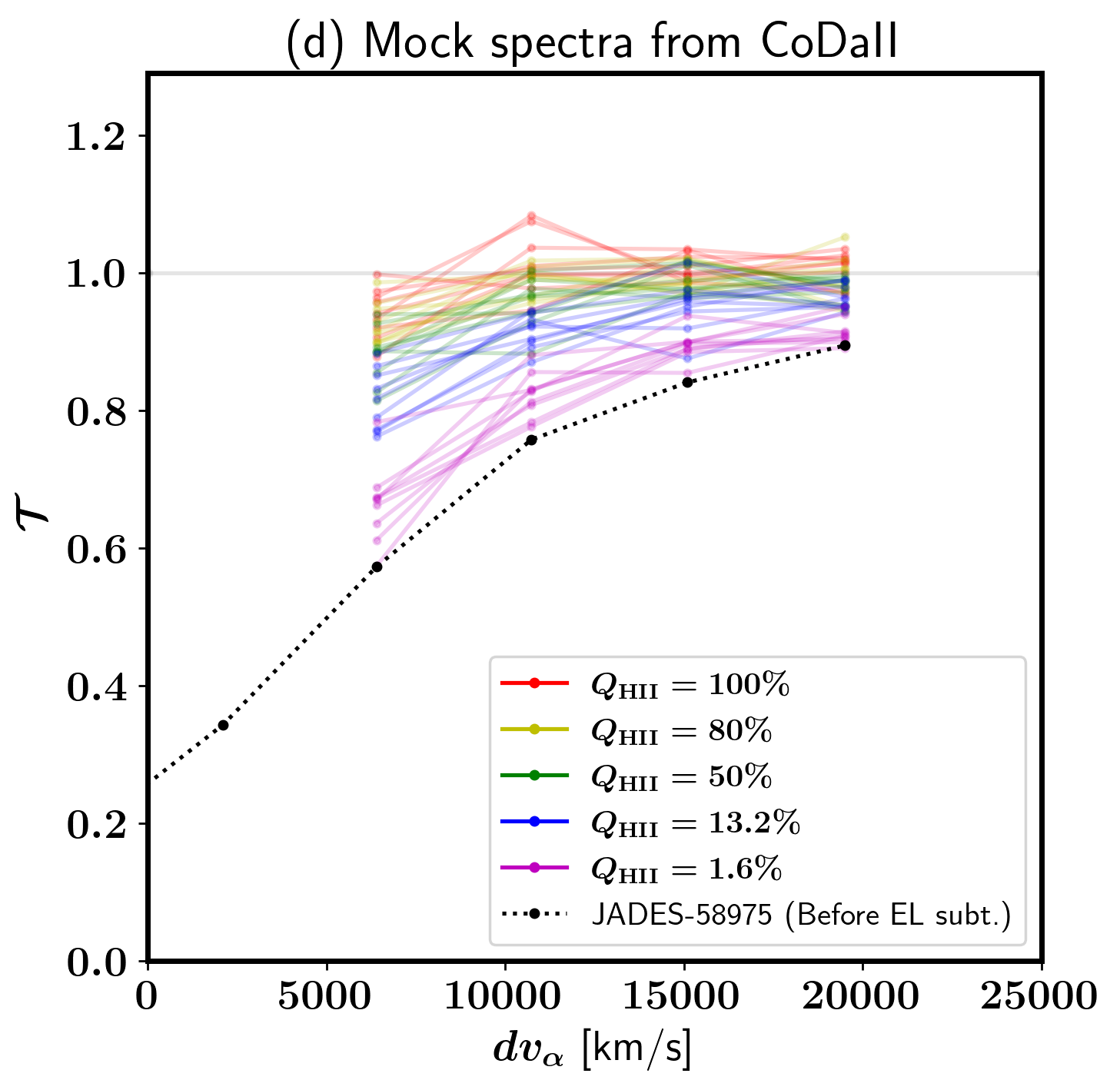}
    \includegraphics[scale=0.3]{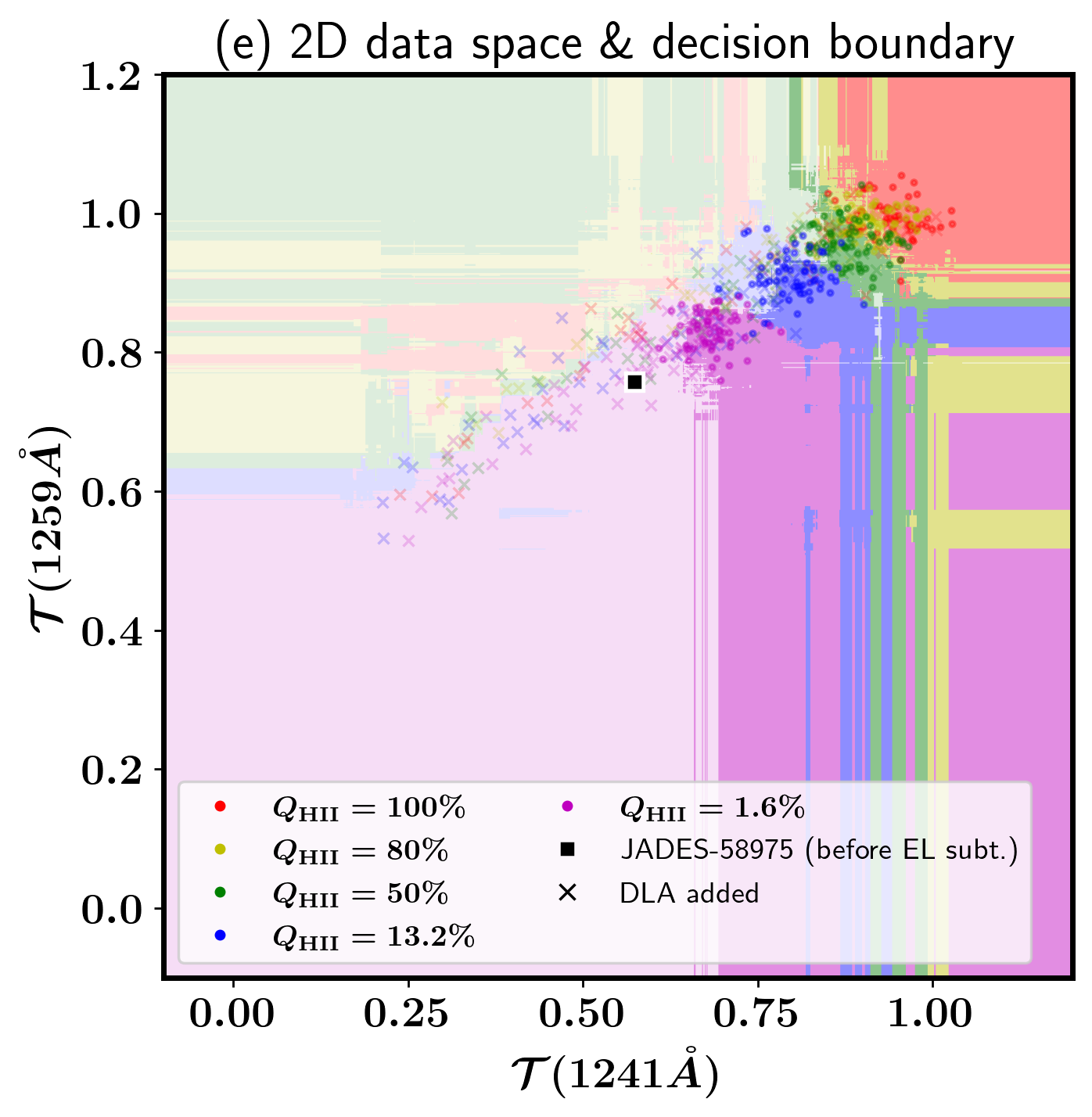}
    \includegraphics[scale=0.3]{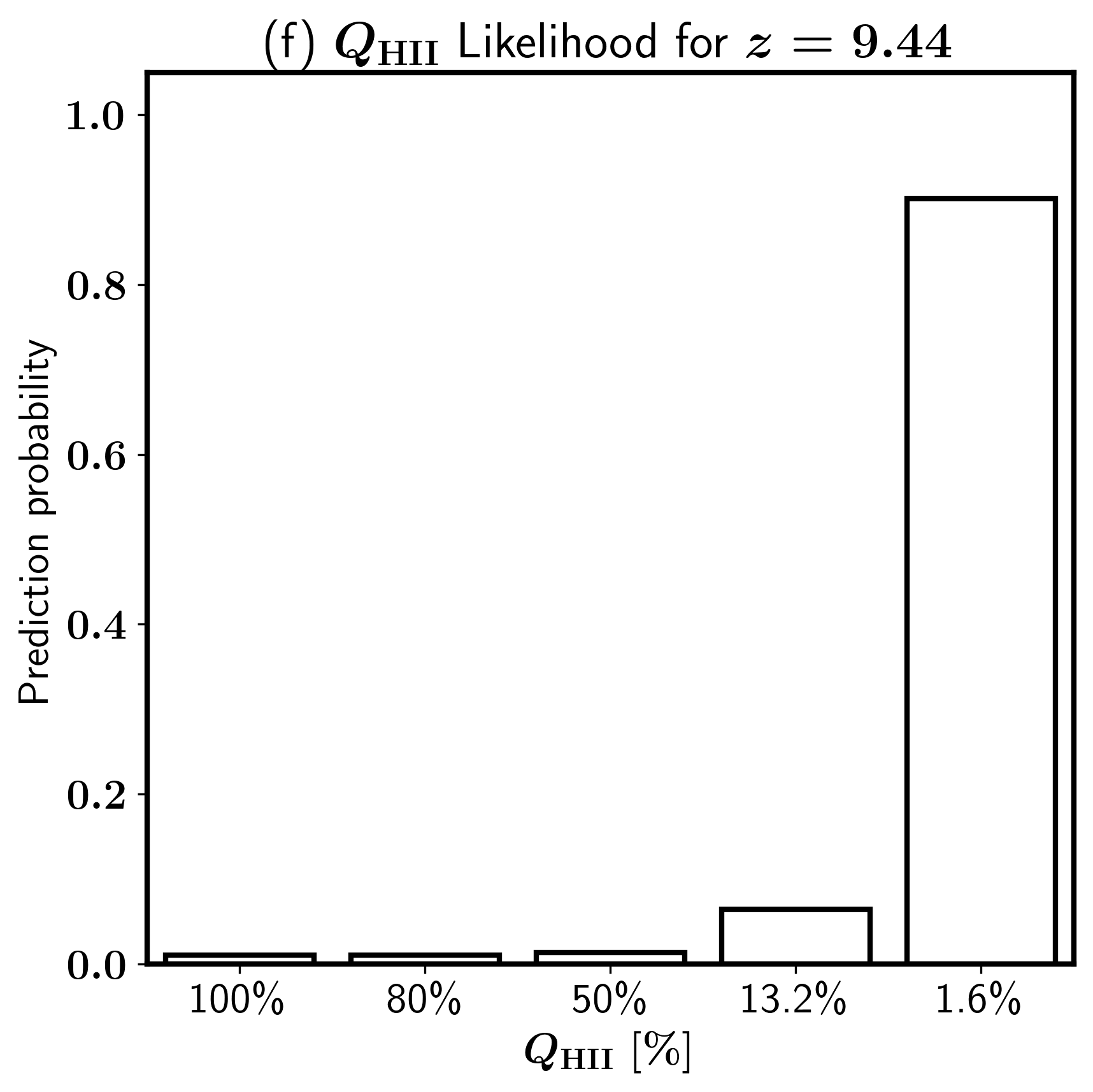}
\caption{Same as panels (a), (d), (e), and (f) of Fig.~\ref{fig:GNz11} for JADES-58975 at $z=9.4$.}
   \label{fig:JADES-58975}
  \end{center}
\end{figure*}

\subsection{JADES-58975} \label{sec:JADES-58975}

JADES-58975 is an $M_{\rm UV}=-20.3$ galaxy at $z=9$, with a publicly available spectrum showing an exceptionally high continuum SNR of $\sim40$. This galaxy is considered to be an obscured AGN \citep{2023arXiv231118731S}. The Ly$\alpha$ emission of this galaxy has not been measured. For the analysis, we use four spectral bins at $dv_\alpha=6400$, 10,700, 15,100, and $19,500~{\rm km}~{\rm s}^{-1}$. One additional bin, at $dv_\alpha=2100~{\rm km}~{\rm s}^{-1}$, is excluded because its wavelength range, $dv_\alpha\approx0 - 4200~{\rm km}~{\rm s}^{-1}$, includes uncertain flux around the Lyman break. Figure~\ref{fig:JADES-58975} presents the raw spectrum, normalized mock spectra, decision boundary in the 2D data space, and the $Q_{\rm HII}$ prediction probability.

The spectrum is well fitted by a power law with small deviations (panel (a)). In panel (d), the normalized spectrum falls below all of the mock DLA-free spectra, indicating that this galaxy likely has DLA absorption. The DLA presence is further confirmed by its location in the data space (panel (e)), which is within the region populated by DLA-added mocks. The DLA-RFC detects DLA absorption with 100\% confidence. JADES-58975 is well positioned in the $Q_{\rm HII}=1.6\%$ region, with a 90\% prediction probability for $Q_{\rm HII}=1.6\%$ from the Q-RFC (panel (f)).

In Section~\ref{sec:clfResult}, we have shown that the Q-RFC accuracy for galaxies with a DLA is generally low (<40\%) for spectra with SNR=20, but $Q_{\rm HII}=1.6\%$ samples are classified with relatively higher accuracy ($\sim 55\%$) even when DLA-added mocks are used (Fig.~\ref{fig:TrTrClf}$b$). Additionally, the continuum SNR of JADES-58975 is higher than 20, suggesting that the Q-RFC results for this galaxy can be considered reliable. 

It is worth noting that the presence of a DLA makes strong Ly$\alpha$ emission, which we could not subtract for this galaxy, less likely, as DLA opacity significantly suppresses the flux. However, there are cases where galaxies exhibit both strong Ly$\alpha$ emission and DLA absorption \citep[e.g.,][]{2024arXiv240816608W, 2025A&A...693A..60H}, possibly due to Ly$\alpha$ emission originating from structures outside the DLA obscuration in some galaxies. Thus, it is possible that the results here may have been influenced by unsubtracted Ly$\alpha$ emission leakage. If such leakage exists and is subtracted, the resulting spectrum would show even lower flux, changing the Q-RFC result. 

\section{Summary and Discussion} \label{sec:summary}

Using the state-of-the-art CoDa II simulation of the reionization era, we have developed a methodology to constrain the reionization history through the Ly$\alpha$ DW absorption feature in the spectra of LBGs. Leveraging the simulated galaxies and HI map from CoDa II, we created approximately $10^6$ mock galactic Ly$\alpha$ DW spectra across five values of $Q_{\rm HII}$ and trained RFCs to estimate the most likely value of $Q_{\rm HII}$ and detect the presence of DLAs in input spectra.

We found that individual spectra can produce tight constraints on $Q_{\rm HII}$ when the IGM is highly neutral ($Q_{\rm HII}<50\%$), provided that the rest-frame flux data at wavelengths between 1220 and 1270 Å has a continuum SNR of $\gtrsim 20$, and no DLA is present in the sight line. We applied our model to four LBG spectra with high SNR ($\gtrsim 10$) to confirm the model's effectiveness. During this analysis, we also showed that accurately subtracting the leakage from Ly$\alpha$ emission is crucial for low-resolution spectra, such as those obtained using the NIRSpec PRISM. 

\subsection{Reionization History Constraint}

\begin{figure}
  \begin{center}
    \includegraphics[scale=0.55]{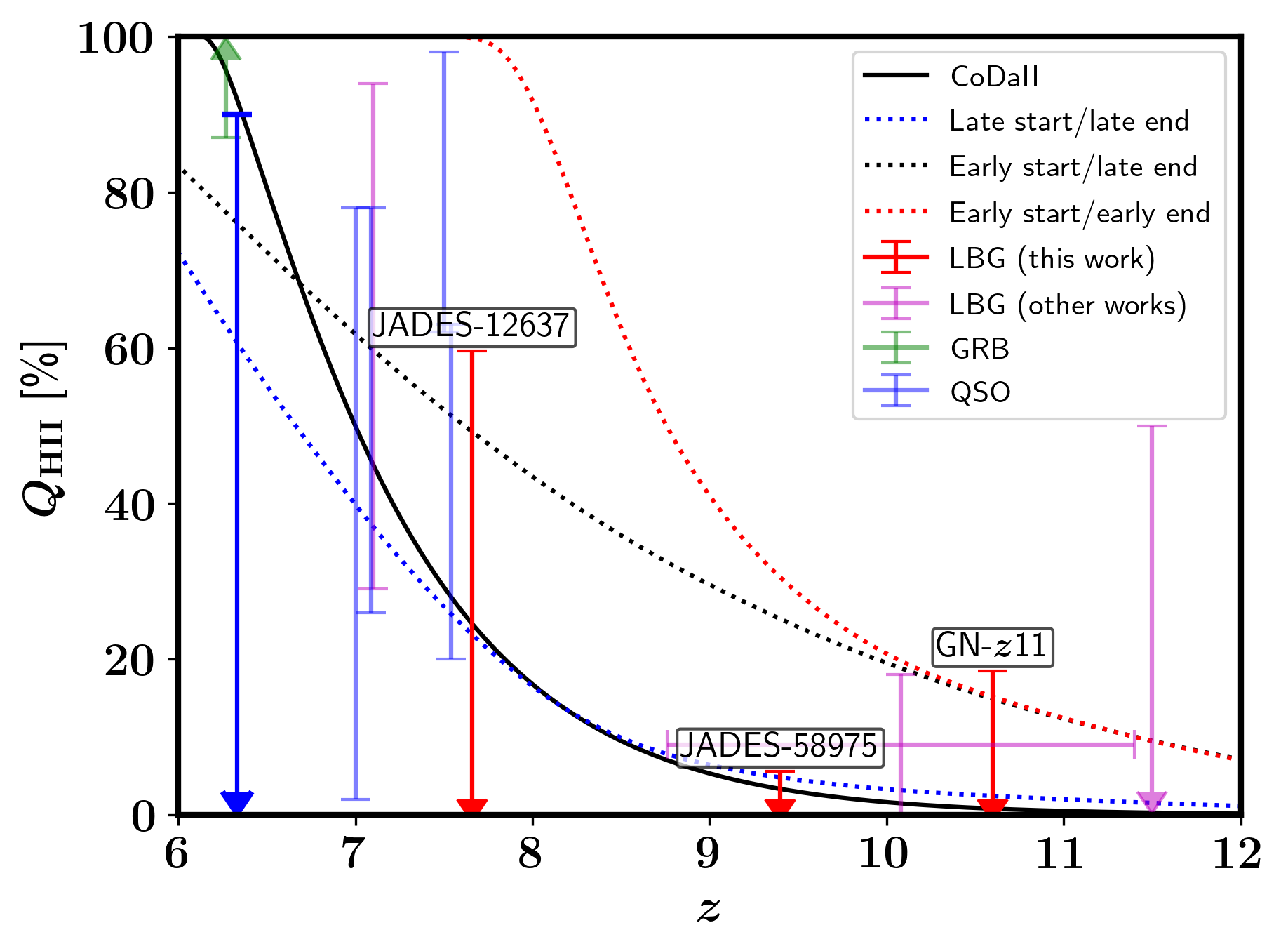}
\caption{Reionization history constraints derived from our RFC model applied to JWST NIRSpec/PRISM data on Lyman break galaxies (LBGs) are shown as red arrows, indicating the upper bounds with 68\% CL from JADES-12637, JADES-58975, and GN-$z$11, respectively. Constraints from another study using LBGs by \citet{2023NatAs...7..622C} and \citet{2024ApJ...971..124U} are shown as magenta error bars. Constraints from the DW feature in other observables are shown as a green error bar for GRBs \citep{2006PASJ...58..485T} and blue error bars for QSOs \citep{2013MNRAS.428.3058S,2018ApJ...864..142D,2019MNRAS.484.5094G,2020ApJ...896...23W}. These error bars are for 68\% CL except for the leftmost blue bar, with a thicker line from \citep{2013MNRAS.428.3058S}, calculated for 95\% CL. The reionization history of CoDa II is shown as a black solid line, and that of the \textsc{late start/late end}, \textsc{early start/late end}, and \textsc{early start/early end} models from \cite{2025ApJ...980...83C} are shown as blue, black, and red dotted lines, respectively.}
   \label{fig:reionHist}
  \end{center}
\end{figure}

Figure~\ref{fig:reionHist} summarizes the reionization history constraints from Section~\ref{sec:application} along with reionization history in the CoDa II simulation and that from the \textsc{late start/late end}, \textsc{early start/late end}, and \textsc{early start/early end} models of \cite{2025ApJ...980...83C}. The upper bounds from JADES-12637, JADES-58975, and GN-$z$11 are calculated for 68\% confidence limit (CL), assuming the five $Q_{\rm HII}$ values (100, 80, 50, 13.2, and 1.6 \%) represent intervals [100, 90]\%, [90, 65]\%, [65, 31.6]\%, [31.6, 7.4]\%, and [7.4, 0]\%, respectively, where we took midpoints between snapshots to determine the intervals. We assume the likelihood within the interval, as predicted by the Q-RFC, is uniformly distributed. The resulting upper bounds are 59.6, 5.6, and 18.5~\% at $z=7.7,~9.4$, and 10.6, respectively. We do not include the constraint from CEERS-1029 in the figure due to its seemingly inaccurate continuum fitting. Note that 68\% CL constraints from other studies using the DW feature of Gamma-ray bursts (GRBs), QSOs, and LBGs are also shown.

While the constraint from GN-$z11$ does not rule out any model show here, the upper bound from JADES-58975 disfavors the \textsc{early start/late end} and \textsc{early start/early end} models, where $Q_{\rm HII}$ rises above 20\% before $z=10$. The constraint from JADES-12637 is weak, but disfavors the \textsc{early start/early end} scenario. The CoDa II and \textsc{late start/late end} models reach $\approx 15\%$ ionization by $z=8$ and exhibit similar reionization histories down to $z=7$, both of which are consistent with the constraints from this work and with most of other DW constraints shown in the figure. The tight constraint from JADES-58975 highlights the potential effectiveness of high-SNR observations of DWs, analyzed using reionization simulations, which provides a prior on the possible HII bubble size distribution for a given $Q_{\rm HII}$. However, the presence of a DLA and the absence of a Ly$\alpha$ emission line measurement for this galaxy suggest higher uncertainty in the result. Excluding the constraint from this galaxy would also allow the \textsc{early start/early end} scenario, in which $\sim 40\%$ of the IGM is reionized by $z=8$.

We note that numerous constraints from other Ly$\alpha$ observables \citep[e.g.,][]{2018ApJ...856....2M,2019ApJ...878...12H,2020MNRAS.495.3602W,2020ApJ...904..144J,2020A&A...641A...6P,2021ApJ...919..120M,2022MNRAS.517.3263B,2023ApJ...949L..40B,2023ApJ...947L..24M,2024ApJ...975..208T,2024ApJ...967...28N} are not shown in the figure. While those constraints are often inconsistent with each other, most disfavor scenarios in which reionization ends early, such as the \textsc{early start/early end} model \cite[see also discussions in][]{2025ApJ...980...83C}. 

We also acknowledge that the reionization history constraints from this work still have remaining uncertainties, which could not be quantified in the scope of this work. Therefore, the constraints presented here are preliminary and subject to refinement in future studies. In the following sections, we discuss these uncertainties (Sec.~\ref{sec:challenges}) and outline future prospects (Sec.~\ref{sec:future}).

\subsection{Challenges and Possible Solutions} \label{sec:challenges}

\textbf{DLAs:} In theory, DLA opacity is distinguishable from IGM opacity due to the difference in wavelength dependence. Determining DLA presence in a target galaxy can be done even with low SNR ($\sim5$) spectra from NIRSpec PRISM-like instruments. However, constraining $Q_{\rm HII}$ from DLA-contaminated spectra requires high SNR ($\sim100$), which is only attainable from rare, bright objects such as JADES-58975 \citep[see also discussions in][]{2025arXiv250113899H}. A practical approach involves initially identifying galaxies without DLA signatures using low-SNR ($\lesssim 20$) observations. Once DLA-free candidates are identified, they can be followed up with higher-SNR ($\gtrsim 20$) observations for input into our model. Additionally, analyzing multiple spectra from galaxies at similar redshifts can help exclude DLA-contaminated results and improve overall reliability.

\textbf{Reionization model:} The mock Ly$\alpha$ spectra in this work are based on the reionization model of the CoDa II simulation, which may be inaccurate if the assumed ionizing emissivity of galaxies deviates from reality. The HII bubble size distribution, a key factor influencing DW strength, depends on whether ionizing photons predominantly originate from rare, bright galaxies or numerous, faint galaxies \citep[see, e.g.,][]{2024MNRAS.528.4872L}. Another highly uncertain aspect of reionization is the ionizing efficiency of minihalos below the atomic-cooling threshold ($\lesssim10^8 M_\odot$), which are unresolved in most cosmological simulations. These sources can dominate reionization at $z \gtrsim 10$ before the buildup of the Lyman-Werner (LW) background and lead to a substantially different reionization geometry \citep{2012ApJ...756L..16A,2021ApJ...914...44A}. After star formation is suppressed in these minihalos by the LW background, they can act as sinks for ionizing photons, altering the HII bubble size distribution in the later stages of reionization \citep{2020MNRAS.491.1600M,2021MNRAS.504.2443B}.

Furthermore, the DW strength depends not only on $Q_{\rm HII}$ but also on redshift, which, in principle, requires multiple simulations with different reionization histories to model the possible range of DW strength at a given redshift. This aspect was not accounted for in this work to perform the calculation with a single simulation. Additionally, the simulation volume may be insufficient to capture rare, bright galaxies such as GN-$z$11. While observational uncertainties currently dominate, these modeling uncertainties need to be addressed as data quality improves.

\textbf{Continuum shape:} This work fits the galactic continuum with a simple power law, not accounting for uncertainty in its shape. For instance, CEERS-1029 shows deviations from a power-law continuum, potentially biasing our $Q_{\rm HII}$ constraints. Highly star-forming galaxies during reionization can exhibit significant contributions from nebular continuum \citep{2023MNRAS.523.3516C,2024ApJ...976..193R}, which can masquerade a DW near Ly$\alpha$, adding degeneracy in constraining the reionization history \citep{2024arXiv240803189K}. A potential solution is to identify galaxies with nebular continuum signatures through detailed SED modeling and focus deep observations on galaxies less likely to have this component. As demonstrated, tight constraints can be achieved from a single galaxy with high-SNR data if there is no uncertainty in the continuum shape. Prioritizing galaxies with minimal continuum shape uncertainty and obtaining high-SNR data for these samples is recommended.

\textbf{Ly$\alpha$ emission:} Spectral bins closer to the Ly$\alpha$ resonance can provide stronger constraints on $Q_{\rm HII}$ due to increased DW absorption near the resonance. However, these bins are more susceptible to contamination from Ly$\alpha$ emission. As shown in the cases of GN-$z$11 and JADES-12637, strong emissions with EW greater than $10$ Å can render the analysis ineffective \citep{2023A&A...677A..88B}. Therefore, accurately measuring the Ly$\alpha$ EW or setting stringent upper limits using high-resolution spectroscopy is crucial for this method to be effective.

\subsection{Future Prospects} \label{sec:future}

JWST will continue to observe galactic Ly$\alpha$ damping wings (DW) throughout its operational lifetime. The CAPERS program (P.I.: Mark Dickinson) on JWST aims to observe approximately 2000 objects, dedicating 3 hr of exposure per target with NIRSpec/PRISM. This extensive dataset will be well suited for the methodologies developed in this work. CAPERS also plans to conduct medium-resolution spectroscopy on a subset of these targets, which we demonstrated to be crucial for accurately subtracting Ly$\alpha$ emission leakage. Additionally, EUCLID will be able to observe hundreds of Ly$\alpha$ DWs \citep{2022A&A...662A.112E}. By analyzing high-SNR spectra from selected targets at $z \sim 7 - 9$, we will be able to distinguish between different reionization scenarios with high confidence, enhancing our understanding of the star-formation rate during this epoch.

Future work will involve improving our model’s performance by incorporating the complete dataset from the new CoDa III simulation, which features higher spatial resolution and a reionization history that better aligns with recent observational constraints. This will help reduce modeling uncertainties. In the updated model, we plan to utilize a larger number of snapshots and adopt regression models to refine the prediction of $Q_{\rm HII}$, minimizing discretization effects. This approach will also enable us to regress the DLA column density directly from the spectra, allowing us to construct detailed DLA statistics and explore the conditions of the CGM and ISM in the high-redshift Universe.

\section*{Acknowledgement}
H.P. thanks Junhwan Choi and Charlot Mason for their insightful comments on this work and Christopher Cain for providing his reionization model data. H.P. thanks the organizers of the Cosmic Dawn at High Latitude conference held in Stockholm, Sweden between June 24-28, 2024, which allowed for helpful discussion with the participants. G.Y. acknowledges Ministerio de Ciencia e Innovación (Spain) for partial financial support under research grant PID2021-122603NB-C21. K.A. is supported by NRF-2021R1A2C1095136 and RS-2022-00197685. H.Y. is supported by MEXT/JSPS KAKENHI grant No. 21H04489 and JST FOREST Program, grant No. JP-MJFR202Z.
\bibliographystyle{apj}
\bibliography{reference}

\end{document}